\def\statusstring{Accepted for
                  IEEE Transactions on Information Theory. 
                  Manuscript received July 21, 2011;
                  date of current version October 20, 2012.}

\documentclass[journal, twocolumn]{IEEEtran}

\usepackage{amssymb} 
\usepackage{amsmath} 
\usepackage{color}

\usepackage{theorem} 

\usepackage{latexsym} 
\usepackage{epsfig} 
\usepackage{psfrag} 
\usepackage{bm, bbm} 
\usepackage{xspace} 
\usepackage{graphicx} 
\usepackage{cite} 
\usepackage{url} 
\usepackage{subfigure,afterpage, wrapfig, pifont}

\IEEEtriggeratref{58}

\renewcommand{\mathbf}[1]{{\bm{#1}}}

\newcommand{\ignore}[1]{}

\newcommand{\confer}{\emph{cf.}}
\newcommand{\eg}{\emph{e.g.}}

\newcommand{\ie}{\emph{i.e.}}
\newcommand{\etc}{\emph{etc.}}

\renewcommand{\leq}{\leqslant}
\renewcommand{\geq}{\geqslant}

\newcommand{\R}{\mathbb{R}}
\newcommand{\Rp}{\mathbb{R}_{\geq 0}}
\newcommand{\Rpp}{\mathbb{R}_{>0}}
\newcommand{\Z}{\mathbb{Z}}
\newcommand{\Zp}{\Z_{\geq 0}}
\newcommand{\Zpp}{\Z_{>0}}

\newcommand{\matr}[1]{\mathbf{#1}}
\newcommand{\vect}[1]{\mathbf{#1}}
\newcommand{\code}[1]{\mathcal{#1}}

\newcommand{\set}[1]{\mathcal{#1}}

\newcommand{\graph}[1]{\mathsf{#1}}

\newcommand{\defeq}{\triangleq}
\newcommand{\onestareq}{\overset{\text{(a)}}{=}}
\newcommand{\twostarseq}{\overset{\text{(b)}}{=}}
\newcommand{\threestarseq}{\overset{\text{(c)}}{=}}
\newcommand{\fourstarseq}{\overset{\text{(d)}}{=}}
\newcommand{\fivestarseq}{\overset{\text{(e)}}{=}}
\newcommand{\sixstarseq}{\overset{\text{(f)}}{=}}

\newcommand{\onestarleq}{\overset{\text{(a)}}{\leq}}
\newcommand{\twostarsleq}{\overset{\text{(b)}}{\leq}}
\newcommand{\threestarsleq}{\overset{\text{(c)}}{\leq}}
\newcommand{\fourstarsleq}{\overset{\text{(d)}}{\leq}}
\newcommand{\fivestarsleq}{\overset{\text{(e)}}{\leq}}

\newcommand{\onestargeq}{\overset{\text{(a)}}{\geq}}

\newcommand{\threestarsgeq}{\overset{\text{(c)}}{\geq}}

\newcommand{\fivestarsgeq}{\overset{\text{(e)}}{\geq}}

\newcommand{\onestar}{\text{(a)}}
\newcommand{\twostars}{\text{(b)}}
\newcommand{\threestars}{\text{(c)}}
\newcommand{\fourstars}{\text{(d)}}
\newcommand{\fivestars}{\text{(e)}}
\newcommand{\sixstars}{\text{(f)}}

\newcommand{\codeC}{\set{C}}

\newcommand{\cgamma}{\tilde \gamma}

\newcommand{\matrgamma}{\boldsymbol{\gamma}}
\newcommand{\hmatrgamma}{\boldsymbol{\hat \gamma}}
\newcommand{\cmatrgamma}{\boldsymbol{\tilde \gamma}}

\newcommand{\vgamma}{\boldsymbol{\gamma}}
\newcommand{\hgamma}{\hat \gamma}

\newcommand{\matrtheta}{\boldsymbol{\theta}}
\newcommand{\cmatrtheta}{\boldsymbol{\tilde \theta}}

\newcommand{\matrA}{\matr{A}}

\newcommand{\cmatrI}{\matr{\tilde I}}
\newcommand{\matrP}{\matr{P}}

\newcommand{\matrallone}[1]{\matr{1}_{#1 \times #1}}

\newcommand{\e}{\operatorname{e}}

\newcommand{\bel}{\beta}

\newcommand{\belivait}{\beta_{i,\va_i}^{(t)}}
\newcommand{\beljt}{\beta_j^{(t)}}
\newcommand{\beljvajt}{\beta_{j,\va_j}^{(t)}}

\newcommand{\vbel}{\boldsymbol{\bel}}

\newcommand{\setA}{\set{A}}

\newcommand{\setE}{\set{E}}

\newcommand{\setElefttime}[1]{\overleftarrow{\setE}^{(#1)}}
\newcommand{\setErighttime}[1]{\overrightarrow{\setE}^{(#1)}}
\newcommand{\setEfull}{\set{E}_{\mathrm{full}}}
\newcommand{\setEhalf}{\set{E}_{\mathrm{half}}}
\newcommand{\setF}{\set{F}}

\newcommand{\setG}{\set{G}}

\newcommand{\setI}{\set{I}}
\newcommand{\setJ}{\set{J}}

\newcommand{\setS}{\set{S}}

\newcommand{\rdiff}[1]{\frac{\mathrm{d}}{\mathrm{d}#1}}
\newcommand{\rddiff}[1]{\frac{\mathrm{d}^2}{\mathrm{d}{#1}^2}}

\newcommand{\vlambda}{\boldsymbol{\lambda}}

\newcommand{\vkappa}{\boldsymbol{\kappa}}

\newcommand{\vomega}{\boldsymbol{\omega}}

\newcommand{\va}{\vect{a}}

\newcommand{\vc}{\vect{c}}

\newcommand{\vm}{\vect{m}}

\newcommand{\vp}{\vect{p}}
\newcommand{\vu}{\vect{u}}

\newcommand{\vunit}{\vu}

\newcommand{\fch}[2]{\mathcal{#1}(\matr{#2})}

\newcommand{\lmpB}{\mathcal{B}}

\newcommand{\graphN}{\graph{N}}

\newcommand{\convhull}{\operatorname{conv}}

\newcommand{\FBethe}{F_{\mathrm{B}}}
\newcommand{\FBethesub}[1]{F_{\mathrm{B},#1}}
\newcommand{\UBethe}{U_{\mathrm{B}}}
\newcommand{\HBethe}{H_{\mathrm{B}}}
\newcommand{\UBethesub}[1]{U_{\mathrm{B},#1}}
\newcommand{\HBethesub}[1]{H_{\mathrm{B},#1}}

\newcommand{\ZBethe}{Z_{\mathrm{B}}}
\newcommand{\ZBetheM}[1]{Z_{\mathrm{B}, #1}}
\newcommand{\fracFBethe}[1]{F^{(#1)}_{\mathrm{B}}}

\newcommand{\fracHBethe}[1]{H^{(#1)}_{\mathrm{B}}}

\newcommand{\LBethe}{L_{\mathrm{Bethe}}}

\newcommand{\FpdualBethe}{F^{\#}_{\mathrm{Bethe}}}

\newcommand{\cover}[1]{\tilde{#1}}
\newcommand{\cset}[1]{\cover{\set{#1}}}
\newcommand{\cgraph}[1]{\cover{\graph{#1}}}

\newcommand{\card}[1]{\left\lvert #1 \right\rvert}
\newcommand{\cardbig}[1]{\big\lvert #1 \big\rvert}

\newcommand{\muleftijt}{\overleftarrow\mu_{\! i,j}^{(t)}}
\newcommand{\muleftijtmo}{\overleftarrow\mu_{\! i,j}^{(t-1)}}
\newcommand{\muleftijpt}{\overleftarrow\mu_{\! i,j'}^{(t)}}
\newcommand{\muleftijptmo}{\overleftarrow\mu_{\! i,j'}^{(t-1)}}
\newcommand{\muleftijpptmo}{\overleftarrow\mu_{\! i,j''}^{(t-1)}}

\newcommand{\murightijt}{\overrightarrow\mu_{\! i,j}^{(t)}}

\newcommand{\Lambdaleftijt}{\overleftarrow\Lambda_{\! i,j}^{(t)}}

\newcommand{\Lambdaleftijptmo}{\overleftarrow\Lambda_{\! i,j'}^{(t-1)}}
\newcommand{\Lambdaleftijextext}[2]{\overleftarrow\Lambda_{\! #1,j}^{(#2)}}

\newcommand{\Lambdaright}{\overrightarrow\Lambda}

\newcommand{\Lambdarightijt}{\overrightarrow\Lambda_{\! i,j}^{(t)}}

\newcommand{\Lambdarightijextext}[2]{\overrightarrow\Lambda_{\! i,#1}^{(#2)}}

\newcommand{\udLambdaleftij}{\overleftarrow{\mathrm{V}}_{\!\! i,j}}
\newcommand{\udLambdaleftijp}{\overleftarrow{\mathrm{V}}_{\!\! i,j'}}

\newcommand{\udLambdaleftijt}{\overleftarrow{\mathrm{V}}_{\!\! i,j}^{(t)}}

\newcommand{\udLambdaleftijtmo}{\overleftarrow{\mathrm{V}}_{\!\! i,j}^{(t-1)}}
\newcommand{\udLambdaleftijptmo}{\overleftarrow{\mathrm{V}}_{\!\! i,j'}^{(t-1)}}

\newcommand{\udLambdaleftijext}[1]{\overleftarrow{\mathrm{V}}_{\!\! i,j}^{(#1)}}

\newcommand{\udLambdaright}{\overrightarrow{\mathrm{V}}}
\newcommand{\udLambdarightij}{\overrightarrow{\mathrm{V}}_{\! i,j}}

\newcommand{\udLambdarightipj}{\overrightarrow{\mathrm{V}}_{\!\! i',j}}
\newcommand{\udLambdarightijt}{\overrightarrow{\mathrm{V}}_{\!\! i,j}^{(t)}}
\newcommand{\udLambdarightipjt}{\overrightarrow{\mathrm{V}}_{\!\! i',j}^{(t)}}

\newcommand{\udLambdarightijext}[1]{\overrightarrow{\mathrm{V}}_{\!\! i,j}^{(#1)}}
\newcommand{\udLambdarightipjtmo}{\overrightarrow{\mathrm{V}}_{\!\! i',j}^{(t-1)}}

\newcommand{\udLambdarightijextext}[2]{\overrightarrow{\mathrm{V}}_{\!\! i,#1}^{(#2)}}

\newcommand{\epsleftijt}{\overleftarrow\varepsilon_{\! i,j}^{(t)}}

\newcommand{\epsleftijtime}[1]{\overleftarrow\varepsilon_{\! i,j}^{(#1)}}
\newcommand{\epsleftijtmo}{\overleftarrow\varepsilon_{\! i,j}^{(t-1)}}
\newcommand{\epsleftijptmo}{\overleftarrow\varepsilon_{\! i,j'}^{(t-1)}}
\newcommand{\epsleftijtpo}{\overleftarrow\varepsilon_{\! i,j}^{(t+1)}}
\newcommand{\epsleftmaxt}{\overleftarrow\varepsilon_{\!\!\max}^{(t)}}
\newcommand{\epsleftmaxtime}[1]{\overleftarrow{\varepsilon}_{\max}^{(#1)}}
\newcommand{\epsleftmaxtmo}{\overleftarrow{\varepsilon}_{\!\!\max}^{(t-1)}}
\newcommand{\epsleftmaxtpo}{\overleftarrow{\varepsilon}_{\!\!\max}^{(t+1)}}

\newcommand{\epsrightijtime}[1]{\overrightarrow\varepsilon_{\! i,j}^{(#1)}}
\newcommand{\epsrightijt}{\overrightarrow\varepsilon_{\!\! i,j}^{(t)}}
\newcommand{\epsrightipjt}{\overrightarrow\varepsilon_{\! i',j}^{(t)}}
\newcommand{\epsrightijtmo}{\overrightarrow\varepsilon_{\! i,j}^{(t-1)}}

\newcommand{\deltaleftijt}{\overleftarrow\delta_{\!\! i,j}^{(t)}}
\newcommand{\deltarightijt}{\overrightarrow\delta_{\!\! i,j}^{(t)}}

\newcommand{\vlambdalefte}{\overleftarrow\vlambda_{\! e}}
\newcommand{\vlambdarighte}{\overrightarrow\vlambda_{\! e}}

\newcommand{\lambdalefteae}{\overleftarrow\lambda_{\! e, a_e}}
\newcommand{\lambdarighteae}{\overrightarrow\lambda_{\! e, a_e}}

\newcommand{\lambdaleftext}[1]{\overleftarrow\lambda_{\! #1}}
\newcommand{\lambdarightext}[1]{\overrightarrow\lambda_{\! #1}}

\newcommand{\vxi}{\boldsymbol{\xi}}

\newcommand{\lefteigvect}{u^{\mathrm{L}}}
\newcommand{\righteigvect}{u^{\mathrm{R}}}
\newcommand{\vlefteigvect}{\vu^{\mathrm{L}}}
\newcommand{\vrighteigvect}{\vu^{\mathrm{R}}}

\newcommand{\temp}{\tau}

\def\squarebox#1{\hbox to #1{\hfill\vbox to #1{\vfill}}}

\newcommand{\qedblack}{\hspace*{\fill}%
  $\blacksquare$\smallskip}
\newcommand{\qedwhite}{\hspace*{\fill}%
  $\square$\smallskip}

\newcommand{\defend}{\qedwhite} 
\newcommand{\exampleend}{\qedwhite} 
\newcommand{\remarkend}{\qedwhite} 
\newcommand{\assumptionend}{\qedwhite} 
\newcommand{\conjectureend}{\qedwhite}

\newcommand{\FGibbs}{F_{\mathrm{G}}}
\newcommand{\FGibbssub}[1]{F_{\mathrm{G}, #1}}
\newcommand{\UGibbs}{U_{\mathrm{G}}}
\newcommand{\HGibbs}{H_{\mathrm{G}}}
\newcommand{\ZGibbs}{Z_{\mathrm{G}}}

\newcommand{\del}{\partial}

\newcommand{\interior}{\operatorname{interior}}

\newcommand{\setGamma}[1]{\Gamma_{#1 \times #1}}
\newcommand{\setPMat}[1]{\set{P}_{#1 \times #1}}
\newcommand{\setPi}{\Pi}

\newcommand{\csetPsi}{\cover{\Psi}}
\newcommand{\cmatrP}{\cover{\matrP}}
\newcommand{\cdel}{\cover{\del}}
\newcommand{\csigma}{\cover{\sigma}}

\newcommand{\bsigma}{\bar{\sigma}}

\newtheorem{Lemma}{Lemma}
\newtheorem{Theorem}[Lemma]{Theorem}

\newtheorem{Corollary}[Lemma]{Corollary}
\newtheorem{Definition}[Lemma]{Definition}
\newtheorem{Remark}[Lemma]{Remark}
\newtheorem{Example}[Lemma]{Example}

\newtheorem{Conjecture}[Lemma]{Conjecture}
\newtheorem{Assumption}[Lemma]{Assumption}

\newenvironment{Proof}%
  {\noindent \emph{Proof:}}{\qedblack}

\newcommand{\Ftwo}{{\mathbb{F}}_{\!2}} 

\newcommand{\perm}{\operatorname{perm}}
\newcommand{\permBethe}{\operatorname{perm}_{\mathrm{B}}}
\newcommand{\permBetheM}[1]{\operatorname{perm}_{\mathrm{B}, #1}}
\newcommand{\fracpermBethe}[1]{\operatorname{perm}^{(#1)}_{\mathrm{B}}}

\newcommand{\ells}{\ell^{*}}

\newcommand{\hxi}{\hat \xi}

\newcommand{\hvxi}{\boldsymbol{\hat \xi}}

\title{The Bethe Permanent of a Non-Negative Matrix}

\author{Pascal O.~Vontobel
        \thanks{\statusstring \ 
          Some of the material in this paper was previously presented
          at the 48th Annual Allerton Conference on Communications, Control, 
          and Computing, Monticello, IL, USA, Sep.~29--Oct.~1, 2010, 
          and at the 2011 Information Theory and Applications Workshop,
          UC San Diego, La Jolla, CA, USA, Feb.~6--11, 2011.}%
        \thanks{P.~O.~Vontobel is with Hewlett--Packard Laboratories, 1501 Page
          Mill Road, Palo Alto, CA 94304, USA
          (e-mail: pascal.vontobel@ieee.org).}%
      }

\begin{document}

\maketitle

\begin{abstract}
  It has recently been observed that the permanent of a non-negative square
  matrix, \ie, of a square matrix containing only non-negative real entries,
  can very well be approximated by solving a certain Bethe free energy
  function minimization problem with the help of the sum-product algorithm. We
  call the resulting approximation of the permanent the Bethe permanent.

  In this paper we give reasons why this approach to approximating the
  permanent works well. Namely, we show that the Bethe free energy function is
  convex and that the sum-product algorithm finds its minimum efficiently. We
  then discuss the fact that the permanent is lower bounded by the Bethe
  permanent, and we comment on potential upper bounds on the permanent based
  on the Bethe permanent. We also present a combinatorial characterization of
  the Bethe permanent in terms of permanents of so-called lifted versions of
  the matrix under consideration.

  Moreover, we comment on possibilities to modify the Bethe permanent so that
  it approximates the permanent even better, and we conclude the paper with
  some observations and conjectures about permanent-based pseudo-codewords and
  permanent-based kernels.
\end{abstract}


\section{Introduction}
\label{sec:introduction:1}

Central to the topic of this paper is the definition of the permanent of a
square matrix (see, \eg, \cite{Minc:78}).

\begin{Definition}
  \label{def:permanent:1}

  Let $\matrtheta = (\theta_{i,j})_{i,j}$ be a real matrix of size $n \times
  n$. The permanent of $\matrtheta$ is defined to be the scalar
  \begin{align}
    \perm(\matrtheta)
      &= \sum_{\sigma}
           \prod_{i \in [n]}
           \theta_{i,\sigma(i)},
             \label{eq:def:permanent:1}
  \end{align}
  where the summation is over all $n!$ permutations of the set $[n] \defeq \{
  1, 2, \ldots, n \}$.
  \defend
\end{Definition}

Contrast this definition with the definition of the \emph{determinant} of
$\matrtheta$, \ie,
\begin{align*}
  \det(\matrtheta)
    &= \sum_{\sigma}
         \operatorname{sgn}(\sigma)
         \prod_{i \in [n]}
         \theta_{i,\sigma(i)},
\end{align*}
where $\operatorname{sgn}(\sigma)$ equals $+1$ if $\sigma$ is an even
permutation and equals $-1$ if $\sigma$ is an odd permutation.

\subsection{Complexity of Computing the Permanent}
\label{complexity:of:computing:the:permanent:1}

Because the definition of the permanent looks simpler than the definition of
the determinant, it is tempting to conclude that the permanent can be computed
at least as efficiently as the determinant. However, this does not seem to be
the case. Namely, whereas the arithmetic complexity (number of real additions
and multiplications) needed to compute the determinant is in $O(n^3)$, Ryser's
algorithm (one of the most efficient algorithms for computing the permanent)
requires $\Theta(n \cdot 2^n)$ arithmetic operations~\cite{Ryser:63:1}. This
clearly improves upon the brute-force complexity $O(n \cdot n!) = O\bigl(
n^{3/2} \cdot (n/e)^n \bigr)$ for computing the permanent, but is still
exponential in the matrix size.

In terms of complexity classes, the computation of the permanent is in the
complexity class \#P (``sharp P'' or ``number P'')~\cite{Valiant:79:1}, where
\#P is the set of the counting problems associated with the decision problems
in the class NP. Note that even the computation of the permanent of matrices
that contain only zeros and ones is \#P-complete. Therefore, the
above-mentioned complexity numbers for the computation of the permanent are
not surprising.

\subsection{Approximations to the Permanent}
\label{sec:approximations:to:the:permanent:1}

Given the difficulty of computing the permanent exactly, and given the fact
that in many applications it is good enough to compute an \emph{approximation}
to the permanent, this paper focuses on efficient methods to approximate the
permanent. This relaxation in requirements, from exact to approximate
evaluation of the permanent, allows one to devise algorithms that potentially
have much lower complexity.

Moreover, we will consider only the case where the matrix~$\matrtheta$
in~\eqref{eq:def:permanent:1} is non-negative, \ie, where all entries of
$\matrtheta$ are non-negative. It is to be expected that approximating the
permanent is simpler in this case because with this restriction the sum
in~\eqref{eq:def:permanent:1} contains only non-negative terms, \ie, the terms
in this sum ``interfere constructively.'' This is in contrast to the general
case where the sum in~\eqref{eq:def:permanent:1} contains positive and
negative terms, \ie, the terms in this sum ``interfere constructively and
destructively.''\footnote{Strictly speaking, there are also matrices
  $\matrtheta$ with positive and negative entries but where the product
  $\prod_{i \in [n]} \theta_{i,\sigma(i)}$ is non-negative for every
  $\sigma$.} Despite this restriction to non-negative matrices, many
interesting counting problems can be captured by this setup.

Earlier work on approximating the permanent of a non-nega\-tive matrix includes:
\begin{itemize}

\item Markov-chain-Monte-Carlo-based methods, which started with the work of
  Broder~\cite{Broder:86:1} and ultimately lead to a famous fully polynomial
  randomized approximation scheme (FPRAS) by Jerrum, Sinclair, and
  Vigoda~\cite{Jerrum:Sinclair:Vigoda:04:1} (for more details, in particular
  for complexity estimates of these and related methods, see for example the
  discussion in~\cite{Huber:Law:08:1});

\item Godsil-Gutman-estimator-based methods by Karmarkar, Karp, Lipton,
  Lov{\'a}sz, and Luby~\cite{Karmarkar:Karp:Lipton:Lovasz:Luby:93:1} and by
  Barvinok~\cite{Barvinok:99:1};

\item a divide-and-conquer approach by Jerrum and
  Vazirani~\cite{Jerrum:Vazirani:96:1};

\item a Sinkhorn-matrix-rescaling-based method by Linial, Samorodnitsky, and
  Wigderson~\cite{Linial:Samorodnitsky:Wigderson:00:1};

\item Bethe-approximation / sum-product-algorithm (SPA) based methods by
  Chertkov, Kroc, and Vergassola~\cite{Chertkov:Kroc:Vergassola:08:1} and by
  Huang and Jebara~\cite{Huang:Jebara:09:1}.

\end{itemize}
The study in this paper was very much motivated by these last two papers on
graphical-model-based methods, in particular because the resulting algorithms
are \emph{very efficient} and the obtained permanent estimates have an
\emph{accuracy that is good enough for many purposes}.

The main idea behind this graphical-model-based approach is to formulate a
factor graph whose partition function equals the permanent that we are looking
for. Consequently, the negative logarithm of the permanent equals the minimum
of the so-called Gibbs free energy function that is associated with this
factor graph. Although being an elegant reformulation of the permanent
computation problem, this does not yield any computational savings
yet. Nevertheless, it suggests to look for a function that is tractable and
whose minimum is close to the minimum of the Gibbs free energy function. One
such function is the so-called Bethe free energy
function~\cite{Yedidia:Freeman:Weiss:05:1}, and with this, paralleling the
above-mentioned relationship between the permanent and the minimum of the
Gibbs free energy function, the \emph{Bethe permanent} is defined such that
its negative logarithm equals the global minimum of the Bethe free energy
function. The Bethe free energy function is an interesting candidate because a
theorem by Yedidia, Freeman, and Weiss~\cite{Yedidia:Freeman:Weiss:05:1} says
that fixed points of the SPA correspond to stationary points of the Bethe free
energy function.

In general, this approach of replacing the Gibbs free energy function by the
Bethe free energy function comes with very few guarantees, though.
\begin{itemize}

\item The Bethe free energy function might have multiple local minima.

\item It is unclear how close the (global) minimum of the Bethe free energy
  function is to the minimum of the Gibbs free energy function.

\item It is unclear if the SPA converges, even to a local minimum of the Bethe
  free energy function. (As we will see, the factor graph that we use
  (see Fig.~\ref{fig:ffg:permanent:1}) is not sparse and has many short
  cycles, in particular many four-cycles. These facts might suggest that the
  application of the SPA to this factor graph is rather problematic.)

\end{itemize}
Luckily, in the case of the permanent approximation problem, one can formulate
a factor graph where the Bethe free energy function is very well behaved. In
particular, in this paper we discuss a factor graph that has the following
properties.
\begin{itemize}

\item We show that the Bethe free energy function is, when suitably
  parameterized, a \emph{convex} function; therefore it has no non-global
  local minima.

\item The minimum of the Bethe free energy function is \emph{quite close} to
  the minimum of the Gibbs free energy function. Namely, as was recently shown
  by Gurvits~\cite{Gurvits:11:1, Gurvits:11:2}, the permanent is lower bounded
  by the Bethe permanent. Moreover, we list conjectures on strict and
  probabilistic Bethe-permanent-based upper bounds on the permanent. In
  particular, for certain classes of square non-negative matrices, empirical
  evidence suggests that the permanent is upper bounded by some constant (that
  grows rather modestly with the matrix size) times the Bethe permanent.

\item We show that the SPA \emph{finds} the minimum of the Bethe free energy
  function under rather mild conditions. In fact, the error between the
  iteration-dependent estimate of the Bethe permanent and the Bethe permanent
  itself decays exponentially fast, with an exponent depending on the matrix
  $\matrtheta$. Interestingly enough, in the associated convergence analysis a
  key role is played by a certain Markov chain that maximizes the sum of its
  entropy rate plus some average state transition cost.

\end{itemize}
Besides leaving some questions open with respect to (w.r.t.) the Bethe free
energy function (see, \eg, the above-mentioned conjectures concerning
permanent upper bounds), these results by-and-large validate the empirical
success, as observed by Chertkov, Kroc, and
Vergassola~\cite{Chertkov:Kroc:Vergassola:08:1} and by Huang and
Jebara~\cite{Huang:Jebara:09:1}, of approximating the permanent by
graphical-model-based methods.

Let us remark that for many factor graphs with cycles the Bethe free energy
function is not as well behaved as the Bethe free energy function under
consideration in this paper. In particular, as discussed
in~\cite{Vontobel:10:7:subm}, every code picked from an ensemble of regular
low-density parity-check codes~\cite{Gallager:63}, where the ensemble is such
that the minimum Hamming distance grows (with high probability) linearly with
the block length, has a Bethe free energy function that is non-convex in
certain regions of its domain. Nevertheless, decoding such codes with
SPA-based decoders has been highly successful (see,
\eg~\cite{Richardson:Urbanke:08:1}).

\subsection{Related Work}
\label{sec:related:work:1}

The literature on permanents (and adjacent areas of counting perfect
matchings, counting zero/one matrices with specified row and column sums,
\etc) is vast. Therefore, we just mention works that are (to the best of our
knowledge) the most relevant to the present paper.

Besides the already cited papers~\cite{Chertkov:Kroc:Vergassola:08:1,
  Huang:Jebara:09:1} on Bethe-approximation-based methods to the permanent of
a non-negative matrix, some aspects of the Bethe free energy function were
analyzed by Watanabe and Chertkov in~\cite{Watanabe:Chertkov:10:1} and by
Chertkov, Kroc, Krzakala, Vergassola, and Zdeborov{\'a}
in~\cite{Chertkov:Kroc:Krzakala:Vergassola:Zdeborova:10:1}. (In particular,
the paper~\cite{Watanabe:Chertkov:10:1} applied the loop calculus technique by
Chertkov and Chernyak~\cite{Chertkov:Chernyak:06:1}.) Very recent work in that
line of research is presented in a paper by A.~B.~Yedidia and
Chertkov~\cite{Yedidia:Chertkov:11:1:subm} that studies so-called fractional
free energy functionals, and resulting lower and upper bounds on the permanent
of a non-negative matrix.

Because computing the permanent is related to counting perfect matchings, the
paper by Bayati and Nair~\cite{Bayati:Nair:06:1} on counting matchings in
graphs with the help of the SPA is very relevant. Note that their setup is
such that the perfect matching case can be seen as a limiting case (namely the
zero-temperature limit) of the matching setup. However, for the perfect
matching case (a case for which the authors of~\cite{Bayati:Nair:06:1} make no
claims) the convergence proof of the SPA in~\cite{Bayati:Nair:06:1} is
incomplete. Moreover, their matchings are weighted only inasmuch as the weight
of a matching depends on the size of the matching. Consequently, because all
perfect matchings have the same size, they all are assigned the same
weight. (See also the related paper by Bayati, Gamarnik, Katz, Nair, and
Tetali~\cite{Bayati:Gamarnik:Katz:Nair:Tetali:07:1}, and an extension to
counting perfect matchings in certain types of graph by Gamarnik and
Katz~\cite{Gamarnik:Katz:10:1}.) For an SPA convergence analysis of a slightly
generalized weighted matching setup, the interested reader is referred to a
recent paper by Williams and Lau~\cite{Williams:Lau:10:2}.

Very relevant to the present paper are also papers on max-product algorithm~/
min-sum algorithm based approaches to the maximum weight perfect matching
problem~\cite{Huang:Jebara:07:1, Bayati:Shah:Sharma:08:1,
  Bayati:Borgs:Chayes:Zecchina:11:1, Sanghavi:Malioutov:Willsky:11:1}. As
shown in these papers, these algorithms find the desired solution efficiently
for bipartite graphs, a fact which is strongly related to the observation that
the linear programming relaxation of the underlying integer linear program is
tight in this case. This tightness in relaxation, which is an immediate
consequence of a theorem by Birkhoff and von Neumann (see
Theorem~\ref{theorem:Birkhoff:von:Neumann:1}), goes also a long way towards
explaining why the Bethe free energy function under consideration in the
present paper is well behaved. Finally, let us remark that because the
difference between two perfect matchings corresponds to a union of disjoint
cycles, the max-product algorithm~/ min-sum algorithm convergence analysis
in~\cite{Huang:Jebara:07:1, Bayati:Shah:Sharma:08:1,
  Bayati:Borgs:Chayes:Zecchina:11:1, Sanghavi:Malioutov:Willsky:11:1} has some
resemblance with Wiberg's max-product algorithm / min-sum algorithm
convergence analysis for so-called cycle codes~\cite{Wiberg:96}.

Linial, Samorodnitsky, and
Wigderson~\cite{Linial:Samorodnitsky:Wigderson:00:1} published a deterministic
strongly polynomial algorithm to compute the permanent of an $n \times n$
non-negative matrix within a multiplicative factor of $e^n$. This is related
to the present paper because their approach is based on Sinkhorn's matrix
rescaling method, which can be seen as finding the minimum of a certain free
energy type function.

The present paper has some similarities with recent papers by Barvinok on
counting zero/one matrices with prescribed row and column
sums~\cite{Barvinok:10:1} and by Barvinok and Samorodnitsky on computing the
partition function for perfect matchings in
hypergraphs~\cite{Barvinok:Samorodnitsky:11:1}. However, these papers pursue
what would be called a mean-field theory approach in the physics
literature~\cite{Yedidia:01:1}. An exception to the previous statement is
Section~3.2 in~\cite{Barvinok:10:1}, which contains Bethe-approximation-type
computations. (See the references in that section for further papers that
investigate similar approaches.)

As mentioned in the abstract, the present paper discusses a combinatorial
characterization of the Bethe permanent in terms of permanents of so-called
lifted versions of the matrix under consideration. For this we use results
from~\cite{Vontobel:10:7:subm} that give a combinatorial characterization of
the Bethe partition function of a factor graph in terms of the partition
function of graph covers of this factor graph. Interestingly, very similar
objects were considered by Greenhill, Janson, and
Ruci{\'n}ski~\cite{Greenhill:Janson:Rucinski:10:1}; we will comment on this
connection in
Section~\ref{sec:connections:to:results:by:greenhill:janson:rucinsky:1}.

Finally, as already mentioned in the previous subsection, Gurvits's recent
papers~\cite{Gurvits:11:1, Gurvits:11:2} contain important observations
w.r.t.\ the relationship between the permanent and the Bethe permanent of a
non-negative matrix, and puts them into the context of Schrijver's permanental
inequality.

\subsection{Overview of the Paper}

This paper is structured as follows. We conclude this introductory section
with a discussion of some of the notation that is used. In
Section~\ref{sec:factor:graph:representation:1} we then introduce the main
normal factor graph (NFG) for this paper, in Section~\ref{sec:bethe:entropy:1}
we formally define the Bethe permanent, in
Section~\ref{sec:properties:Bethe:entropy:1} we discuss properties of the
Bethe entropy function and the Bethe free energy function, in
Section~\ref{sec:factor:graph:spa:1} we analyze the SPA, in
Section~\ref{sec:finite:graph:cover:interpretation:Bethe:permanent:1} we give
a ``combinatorial characterization'' of the Bethe permanent in terms of graph
covers of the above-mentioned NFG, in
Section~\ref{sec:permanent:Bethe:permanent:relationship:1} we discuss
Bethe-permanent-based bounds on the permanent, in
Section~\ref{sec:fractional:Bethe:entropy:1} we list some thoughts on using
the concept of the ``fractional Bethe entropy function,'' in
Section~\ref{sec:comments:and:conjectures:1} we list some observations and
conjectures, and we conclude the paper
in~Section~\ref{sec:conclusions:1}. Finally, the appendix contains some of the
proofs.

\subsection{Basic Notations and Definitions}
\label{sec:notation:1}

This subsection discusses the most important notations that will be used in
this paper. More notational definitions will be given in later sections.

We let $\R$ be the field of real numbers, $\Rp$ be the set of non-negative
real numbers, $\Rpp$ be the set of positive real numbers, $\Z$ be the ring of
integers, $\Zp$ be the set of non-negative integers, $\Zpp$ be the set of
positive integers, and for any positive integer $L$ we define $[L] \defeq \{
1, \ldots, L \}$. Scalars are denoted by non-boldface characters, whereas
vectors and matrices by boldface characters. For any positive integer $L$, the
matrix $\matrallone{L}$ is the all-one matrix of size $L\times L$.

\begin{Assumption}
  \label{assumption:non:negative:matrix:theta:1}

  Throughout this paper, if not mentioned otherwise, $n$ is a positive integer
  and $\matrtheta = (\theta_{i,j})_{i,j}$ is a non-negative matrix of size $n
  \times n$. Moreover, we assume that $\matrtheta$ is such that
  $\perm(\matrtheta) > 0$, \ie, there is at least one permutation $\sigma$ of
  $[n]$ such that $\prod_{i \in [n]} \theta_{i,\sigma(i)} > 0$.
  \assumptionend
\end{Assumption}

We use calligraphic letters for sets, and the size of a set $\set{S}$ is
denoted by $\card{\set{S}}$. For a finite set $\setS$, we let $\setPi_{\setS}$
be the set of probability mass functions over $\setS$, \ie,
\begin{align*}
  \setPi_{\setS}
    &\defeq 
       \left\{
           \vect{p} = \big( p_s \big)_{s \in \setS}
         \ \middle| \ 
           p_s \geq 0 \text{ for all $s \in \setS$}, \ 
           \sum_{s \in \setS} p_s = 1
         \right\}.
\end{align*}
Moreover, for any positive integer $L$, we define $\setPMat{L}$ to be the set
of all $L \times L$ permutation matrices, \ie,
\begin{align*}
  \setPMat{L}
    &\defeq
       \left\{
           \matr{P}
       \ \middle|
           \begin{array}{l}
             \text{$\matr{P}$ is a matrix of size $L \times L$} \\
             \text{$\matr{P}$ contains exactly one $1$ per row} \\
             \text{$\matr{P}$ contains exactly one $1$ per column} \\
             \text{$\matr{P}$ contains $0$s otherwise}
           \end{array}\!\!
       \right\}.
\end{align*}
Clearly, there is a bijection between $\setPMat{L}$ and the set of all
permutations of $[L]$. 
Finally, for any positive integer $L$, we let $\setGamma{L}$ be the set of
doubly stochastic matrices of size $L \times L$, \ie,
\begin{align*}
  \setGamma{L}
    &\defeq 
       \left\{
           \matrgamma = \big( \gamma_{i,j} \big)
         \middle| \!
           \begin{array}{l}
             \text{$\gamma_{i,j} \geq 0$ for all $(i,j) \in [L] \times [L]$} \\
             \text{$\sum_{j \in [L]} \gamma_{i,j} = 1$ for all $i \in [L]$} \\
             \text{$\sum_{i \in [L]} \gamma_{i,j} = 1$ for all $j \in [L]$} \\
           \end{array}\!\!
         \right\}.
\end{align*}

The convex hull~\cite{Boyd:Vandenberghe:04:1} of some subset $\set{S}$ of some
multi-dimensional real space is denoted by $\convhull(\set{S})$. In the
following, when talking about the interior of a polytope, we will mean the
relative interior~\cite{Boyd:Vandenberghe:04:1} of that polytope.

When appropriate, we will identify the set of $L \times L$ real matrices with
the $L^2$-dimensional real space. In that sense, $\setGamma{L}$ can be seen as
a polytope in the $L^2$-dimensional real space. Clearly, $\setGamma{L}$ is a
convex set, and every permutation matrix of size $L \times L$ is a doubly
stochastic matrix of size $L \times L$. Most interestingly, every doubly
stochastic matrix of size $L \times L$ can be written as a convex combination
of permutation matrices of size $L \times L$; this observation is a
consequence of the important Birkhoff--von~Neumann Theorem.

\begin{Theorem}[Birkhoff--von Neumann Theorem]
  \label{theorem:Birkhoff:von:Neumann:1}

  For any positive integer $L$, the set of doubly stochastic matrices of size
  $L \times L$ is a polytope whose vertex set equals the set of permutation
  matrices of size $L \times L$, \ie,
  \begin{align*}
    \operatorname{vertex-set}(\setGamma{L})
      &= \setPMat{L}.
  \end{align*}
  As a consequence, the set of doubly stochastic matrices of size $L \times L$
  is the convex hull of the set of all permutation matrices of size $L \times
  L$, \ie,
  \begin{align*}
    \setGamma{L}
      &= \convhull(\setPMat{L}).
  \end{align*}
\end{Theorem}

\begin{Proof}
  See, \eg, \cite[Section 8.7]{Horn:Johnson:90}.
\end{Proof}

Finally, all logarithms will be natural logarithms and the value of $0 \cdot
\log(0)$ is defined to be equal to $0$.

\section{Normal Factor Graph Representation}
\label{sec:factor:graph:representation:1}

Factor graphs are a convenient way to represent multivariate
functions~\cite{Kschischang:Frey:Loeliger:01}. In this paper we use a variant
called ``normal factor graphs (NFGs)''~\cite{Forney:01:1} (also called
``Forney-style factor graphs''~\cite{Loeliger:04:1}), where variables are
associated with edges.

As already mentioned in the introduction, the main idea behind the
graphical-model-based approach to estimating the permanent is to formulate an
NFG such that its partition function equals the permanent. There are of course
different ways to do this and typically different formulations will yield
different results when estimating the permanent with sub-optimal algorithms
like the SPA. It is well known that when the NFG has no cycles, then the SPA
computes the partition function exactly, however, for the given problem any
NFG \emph{without} cycles yields highly inefficient SPA update rules for
reasonably large $n$ (otherwise there would be a contradiction to the
considerations in Section~\ref{complexity:of:computing:the:permanent:1}), and
so we will focus on NFGs \emph{with} cycles. The NFG that is introduced in the
following definition and that is based on a complete bipartite graph with two
times $n$ vertices, is a rather natural candidate, and, as we will see, has
very interesting and useful properties.

\begin{figure}
  \begin{center}
    \epsfig{file=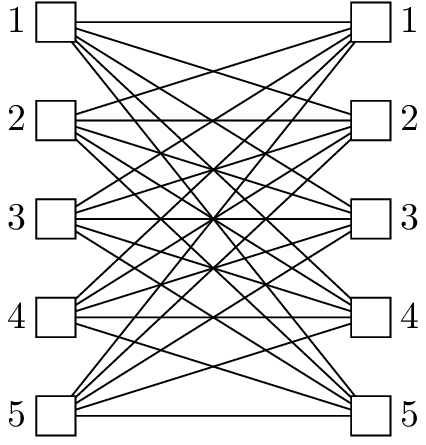, width=0.30\linewidth}
  \end{center}
  \caption{The NFG $\graphN(\matrtheta)$ is based on a complete bipartite
    graph with two times $n$ vertices (here $n = 5$). The function nodes on
    the left-hand side represent the local functions $\{ g_i \}_{i \in
      \setI}$, the function nodes on the right-hand side represent the
    functions $\{ g_j \}_{j \in \setJ}$, and with the edge $e = (i,j)$ we
    associate the variable $A_e = A_{i,j}$. (See
    Definition~\ref{def:permanent:normal:factor:graph:1} for more details.)}
  \label{fig:ffg:permanent:1}
\end{figure}

\begin{Definition}
  \label{def:permanent:normal:factor:graph:1}

  We define the NFG $\graphN(\matrtheta) \defeq
  \graphN(\setF,\setE, \setA, \setG)$ as follows (see also
  Fig.~\ref{fig:ffg:permanent:1}).
  \begin{itemize}
   
  \item The set of vertices (henceforth also called function nodes) is $\setF
    \defeq \setI \ {\dot \cup} \ \setJ$, where $\setI \defeq [n]$ will be
    called the set of left vertices and $\setJ \defeq [n]$ will be called the
    set of right vertices.\footnote{Here, $\setF \defeq \setI \ {\dot \cup} \
      \setJ$ stands for the more cumbersome $\setF \defeq \bigl( \{
      \mathrm{left} \} \times \setI \bigr) \cup \bigl( \{ \mathrm{right} \}
      \times \setJ \bigr)$. In the following, $i$ (and variations thereof)
      will refer to a left vertex and $j$ (and variations thereof) will refer
      to a right vertex. In that spirit, variables like $\eta_i$ and $\eta_j$
      are different variables, also if $i = j$.}

  \item The set of full-edges is $\setEfull \defeq \setI \times \setJ =
    \bigl\{ (i,j) \bigm| i \in \setI, j \in \setJ \bigr\}$ and the set of
    half-edges is $\setEhalf = \emptyset$, \ie, the empty set. (A full-edge is
    an edge connecting two vertices, whereas a half-edge is an edge that is
    connected to only one vertex.) The set of edges is $\setE \defeq \setEfull
    \cup \setEhalf = \setEfull$.

  \item With every edge $e = (i,j) \in \setE$ we associate the variable $A_e =
    A_{i,j}$ with alphabet $\setA_e = \setA_{i,j} \defeq \{ 0, 1 \}$; a
    realization of $A_e = A_{i,j}$ will be denoted by $a_e = a_{i,j}$.
 
  \item The set $\setA \defeq \prod_e \setA_e = \prod_{i,j} \setA_{i,j}$ will
    be called the configuration set, and so
    \begin{align*}
      \va 
        &\ \defeq \ 
           (a_e)_{e \in \setE}
         = (a_{i,j})_{(i,j) \in \setI \times \setJ}
             \ \in \ 
             \setA
    \end{align*}
    will be called a configuration. For a given vector $\va$, we also define
    the sub-vectors
    \begin{align*}
      \va_i
        &\defeq
           (a_{i,j})_{j \in \setJ}
      \quad \text{and} \quad
      \va_j
         \defeq
           (a_{i,j})_{i \in \setI}.
    \end{align*}
    When convenient, the vector $\va$ will be considered to be an $n \times n$
    matrix. Then $\va_i$ corresponds to the $i$th row of $\va$, and $\va_j$
    corresponds to the $j$th column of $\va$. (Note that we will also use the
    notations $\va_i \defeq (a_{i,j})_{j \in \setJ}$ and $\va_j \defeq
    (a_{i,j})_{i \in \setI}$ when there is not necessarily an underlying
    configuration $\va$ of the whole NFG.)

  \item For every $i \in \setI$ we define the local functions\footnote{Here
      and in the following, $\vunit_j$, $j \in \setJ$, stands for the
      length-$n$ vector where all entries are zero except for the $j$th entry
      that equals $1$. The vector $\vunit_i$, $i \in \setI$, is defined
      similarly.}\,\footnote{Here and in the following, we will use the
      short-hands $\sum_i$, $\sum_j$, $\sum_{i'}$, $\sum_{j'}$, $\sum_e$,
      $\sum_{e'}$ for $\sum_{i \in \setI}$, $\sum_{j \in \setJ}$, $\sum_{i'
        \in \setI}$, $\sum_{j' \in \setJ}$, $\sum_{e \in \setE}$, $\sum_{e'
        \in \setE}$, respectively, with similar conventions for products.}
    \begin{align*}
      g_i: \ 
        &\prod_{j'} \setA_{i,j'}
           \to
           \R, \quad
         \va_i
           \mapsto 
           \begin{cases}
             \sqrt{\theta_{i,j}} & \text{(if $\va_i = \vunit_j$)} \\
             0                  & \text{(otherwise)}
           \end{cases}
    \end{align*}
    Similarly, for every $j \in \setJ$ we define the local functions
    \begin{align*}
      g_j: \ 
        &\prod_{i'} \setA_{i',j}
           \to
           \R, \quad
         \va_j
           \mapsto 
           \begin{cases}
             \sqrt{\theta_{i,j}} & \text{(if $\va_j = \vunit_i$)} \\
             0                  & \text{(otherwise)}
           \end{cases}
    \end{align*}

  \item For every $i \in \setI$ we define the function node alphabet $\setA_i$
    to be the set
    \begin{align*}
      \setA_i
        &\defeq
            \left\{
              \va_i \in \prod_{j'} \setA_{i,j'}
            \
            \middle|
            \
              g_i(\va_i) \neq 0
            \right\}
         = \left\{
             \vunit_j \ \middle| \ j \in \setJ
           \right\}.
    \end{align*}
    Similarly, for every $j \in \setJ$ we define the function node alphabet
    $\setA_j$ to be the set
    \begin{align*}
      \setA_j
        &\defeq
            \left\{
              \va_j \in \prod_{i'} \setA_{i',j}
            \
            \middle|
            \
              g_j(\va_j) \neq 0
            \right\}
         = \left\{
             \vunit_i \ \middle| \ i \in \setI
           \right\}.
    \end{align*}
    (The sets $\setA_i$ and $\setA_j$ are also known as the local constraint
    codes of the function nodes $i$ and $j$, respectively.)

  \item The global function $g$ is defined to be
    \begin{align*}
      g: \
         \setA
           \to \R,
         \quad
         \va
           \mapsto
           \left(
             \prod_i
               g_i(\va_i)
           \right)
           \cdot
           \left(
             \prod_j
               g_j(\va_j)
           \right).
    \end{align*}

  \item A configuration $\vc$ with $g(\vc) \neq 0$ will be called a valid
    configuration. The set of all valid configurations, \ie,
    \begin{align*}
      \codeC
        &\defeq
           \bigl\{
             \vc \in \setA
             \bigm|
             g(\vc) \neq 0
           \bigr\} \\
        &= \left\{
             (c_{i,j})_{i,j \in \setI \times \setJ}
           \, \middle| \!\!
             \begin{array}{rl}
               \text{$c_{i,j} \in \setA_{i,j}$,} & \!\!\!\!
               \text{$(i,j) \in \setI \times \setJ$} \\
               \text{$\vc_i \in \setA_i$,} & \!\!\!\!
               \text{$i \in \setI$} \\
               \text{$\vc_j \in \setA_j$,} & \!\!\!\!
               \text{$j \in \setJ$}
             \end{array} \!\!
           \right\},
    \end{align*}
    will be called the global behavior of $\graphN(\matrtheta)$. Considering
    the elements of $\codeC$ as $n \times n$ matrices, it can easily
    be verified that $\codeC = \setPMat{n}$. This allows us to
    associate with $\vc \in \codeC$ the permutation $\sigma_{\vc}: \,
    [n] \to [n]$ that maps $i \in \setI$ to $j \in \setJ$ if $c_{i,j} = 1$.
    \defend

  \end{itemize}

\end{Definition}

\begin{Lemma}
  \label{lemma:global:local:function:rewritten:1}

  Consider the NFG $\graphN(\matrtheta)$ and let $\vc \in
  \codeC$ be a valid configuration of it. Then
  \begin{align*}
    g_i(\vc_i)
      &= \sqrt{\theta_{i,\sigma_{\vc}(i)}},
           \quad i \in \setI, \\
    g_j(\vc_j)
      &= \sqrt{\theta_{\sigma_{\vc}^{-1}(j),j}},
           \quad j \in \setJ, \\
    g(\vc)
      &= \prod_i
           \theta_{i,\sigma_{\vc}(i)}
       = \prod_j
           \theta_{\sigma^{-1}_{\vc}(j),j}.
  \end{align*}
\end{Lemma}

\begin{Proof}
  The first two expressions follow easily from the definitions of $g_i$ and
  $g_j$ in Definition~\ref{def:permanent:normal:factor:graph:1}. The third
  expression is a consequence of
  \begin{align*}
    g(\vc)
       &= \left(
            \prod_i
              g_i(\vc_i)
          \right)
          \cdot
          \left(
            \prod_j
              g_j(\vc_j)
          \right) \\
       &= \left(
            \prod_i
              \sqrt{\theta_{i,\sigma_{\vc}(i)}}
          \right)
          \cdot
          \left(
            \prod_j
              \sqrt{\theta_{\sigma_{\vc}^{-1}(j),j}}
          \right) \\
       &= \left(
            \prod_i
              \sqrt{\theta_{i,\sigma_{\vc}(i)}}
          \right)
          \cdot
          \left(
            \prod_{i'}
              \sqrt{\theta_{i',\sigma_{\vc}(i')}}
          \right) \\
       &= \prod_i
            \theta_{i,\sigma_{\vc}(i)}.
  \end{align*}\vskip-0.45cm
\end{Proof}

\begin{Definition}
  The (Gibbs) partition function of the NFG $\graphN(\matrtheta)$ is defined
  to be the sum of the global function over all configurations, or,
  equivalently, the sum of the global function over all valid configurations,
  \ie,
  \begin{align}
    \ZGibbs
      &\defeq
         \sum_{\va \in \setA} g(\va)
       = \sum_{\vc \in \codeC} g(\vc).
           \label{eq:Z:Gibbs:1}
  \end{align}
  In the following, when confusion can arise what NFG a certain Gibbs
  partition function is referring to, we will use $\ZGibbs\bigl(
  \graphN(\matrtheta) \bigr)$, etc., instead of $\ZGibbs$.\footnote{Note that
    ``function'' in ``partition function'' refers to the fact that the
    expression in~\eqref{eq:Z:Gibbs:1} typically is a function of some
    parameters like the temperature $T$ (see the discussion below). A better
    word for ``partition function'' would possibly be ``partition sum'' or
    ``state sum,'' which would more closely follow the German
    ``Zustandssumme'' whose first letter is used to denote the partition
    function.}  \defend
\end{Definition}

\begin{Definition}
  \label{def:F:Gibbs:1}

  The Gibbs free energy function associated with the NFG $\graphN(\matrtheta)$
  is defined to be
  \begin{align*}
    \FGibbs: \ 
      &\setPi_{\codeC}
         \to 
         \R, \quad
       \vp
         \mapsto 
           \UGibbs(\vp)
           -
           \HGibbs(\vp),
  \end{align*}
  where
  \begin{alignat*}{2}
    \UGibbs: \ 
       &
       \setPi_{\codeC}
         \to 
         \R, \quad
       \vp
         &
         \, \mapsto\, 
         & 
         -
         \sum_{\vc \in \codeC}
           p_{\vc}
           \cdot
           \log
             \big(
               g(\vc)
             \big), \\
    \HGibbs: \ 
       &
       \setPi_{\codeC}
         \to 
         \R, \quad
       \vp
         &
         \, \mapsto\,  
         &
         -
         \sum_{\vc \in \codeC}
           p_{\vc}
           \cdot
           \log
             \big(
               p_{\vc}
             \big).
  \end{alignat*}
  Here, $\UGibbs$ is called the Gibbs average energy function and $\HGibbs$ is
  called the Gibbs entropy function. In the following, when confusion can
  arise what NFG a certain Gibbs free energy function is referring to, we will
  use $\FGibbssub{\graphN(\matrtheta)}$, etc., instead of $\FGibbs$. Similar
  comments apply to $\UGibbs$ and $\HGibbs$.
  \defend
\end{Definition}

For more details on these functions we refer to, \eg,
\cite{Yedidia:Freeman:Weiss:05:1}. For a discussion of these functions in the
context of NFGs we refer to, \eg, \cite{Vontobel:10:7:subm}. Note that
$\HGibbs$ is a concave function of $\vp$, that $\UGibbs$ is a linear function
of $\vp$, and that, consequently, $\FGibbs$ is a convex function of $\vp$.

\begin{Lemma}
  \label{lemma:permanent:Gibbs:free:energy:reformulation:1}

  The permanent of $\matrtheta$ can be expressed in terms of the partition
  function or in terms of the minimum of the Gibbs free energy function of
  $\graphN(\matrtheta)$. Namely,
  \begin{align}
    \perm(\matrtheta)
      &= \ZGibbs
       = \exp
           \left(
             -
             \min_{\vp}
               \FGibbs(\vp)
           \right),
             \label{eq:permanent:vs:partition:function:vs:FGibbs:min:1}
  \end{align}
  where the minimization is over $\vp \in \setPi_{\codeC}$.
\end{Lemma}

\begin{Proof}
  The first equality is a straightforward consequence of
  Definitions~\ref{def:permanent:1}
  and~\ref{def:permanent:normal:factor:graph:1}, along with
  Lemma~\ref{lemma:global:local:function:rewritten:1}. For the second equality
  we refer to, \eg, \cite{Yedidia:Freeman:Weiss:05:1, Vontobel:10:7:subm}.
\end{Proof}

The partition function $\ZGibbs$ and the Gibbs free energy function $\FGibbs$
were specified for temperature $T = 1$ in the above definitions. For a general
temperature parameter $T \in \Rpp$, these functions have to be replaced by
$\ZGibbs \defeq \sum_{\vc \in \codeC} g(\vc)^{1/T}$ and by $\FGibbs(\vp)
\defeq \UGibbs(\vp) - T \cdot \HGibbs(\vp)$, respectively, and
Lemma~\ref{lemma:permanent:Gibbs:free:energy:reformulation:1} has to be
replaced by $\ZGibbs = \exp \left( - \frac{1}{T} \min_{\vp} \FGibbs(\vp)
\right)$. Of course, $\ZGibbs = \perm(\matrtheta)$ does not hold anymore,
unless a suitable $T$-dependence is built into the definition of
$\perm(\matrtheta)$.

\section{The Bethe Permanent}
\label{sec:bethe:entropy:1}

Although the reformulation of the permanent in
Lemma~\ref{lemma:permanent:Gibbs:free:energy:reformulation:1} in terms of a
convex minimization problem is elegant, from a computational perspective it
does not represent much progress. However, it suggests to look for a
minimization problem that can be solved efficiently and whose minimum value is
related to the desired quantity. This is the approach that is taken in this
section and will be based on the Bethe approximation of the Gibbs free energy
function: the resulting approximation of the permanent of a non-negative
square matrix will be called the Bethe permanent. (Note that in this section
we give the technical details only; for a general discussion w.r.t.\ the
motivations behind the Bethe approximation we refer
to~\cite{Yedidia:Freeman:Weiss:05:1}, and for a discussion of the Bethe
approximation in the context of NFGs we refer to~\cite{Vontobel:10:7:subm}.)

\begin{Definition}
  \label{def:permanent:normal:factor:graph:lmp:1}

  Consider the NFG $\graphN(\matrtheta)$. We let
  \begin{align*}
    \vbel
      &\defeq
         \big( 
           (\vbel_i)_{i \in \setI},
           (\vbel_j)_{j \in \setJ}, 
           (\vbel_e)_{e \in \setE}
         \big)
  \end{align*}
  be a collection of vectors based on the real vectors
  \begin{align*}
    \vbel_i
      &\defeq
         (\bel_{i,\va_i})_{\va_i \in \setA_i}, \\
    \vbel_j
      &\defeq
         (\bel_{j,\va_j})_{\va_j \in \setA_j}, \\
    \vbel_e
      &\defeq
         (\bel_{e,a_e})_{a_e \in \setA_e}.
  \end{align*}
  Moreover, we define the sets
  \begin{align*}
    \lmpB_i
      &\defeq
         \setPi_{\setA_i},
           \quad i \in \setI, \\
    \lmpB_j
      &\defeq
         \setPi_{\setA_j},
           \quad j \in \setJ, \\
    \lmpB_e
      &\defeq
         \setPi_{\setA_e},
           \quad e \in \setE,
  \end{align*}
  and call $\lmpB_i$, $\lmpB_j$, and $\lmpB_e$, the $i$th local marginal
  polytope, the $j$th local marginal polytope, and the $e$th local marginal
  polytope, respectively. (Sometimes $\lmpB_i$ is also called the $i$th belief
  polytope, etc.)

  With this, the local marginal polytope (or belief polytope) $\lmpB$ is
  defined to be the set
  \begin{align*}
    \lmpB
      &= \left\{
           \ \vbel \ 
         \ 
         \middle|
         \ 
           \begin{array}{c}
             \vbel_i \in \lmpB_i \text{ for all $i \in \setI$} \\
             \vbel_j \in \lmpB_j \text{ for all $j \in \setJ$} \\
             \vbel_e \in \lmpB_e \text{ for all $e \in \setE$} \\[0.25cm]
             \sum\limits_{\va'_i \in \setA_i: \, a'_{i,j} = a_e} 
               \bel_{i, \va'_i} = \bel_{e, a_e} \\
             \text{for all $e = (i,j) \in \setE$,
                   $a_e \in \setA_e$} \\[0.25cm]
             \sum\limits_{\va'_j \in \setA_j: \, a'_{i,j} = a_e} 
               \bel_{j, \va'_j} = \bel_{e, a_e} \\
             \text{for all $e = (i,j) \in \setE$,
                   $a_e \in \setA_e$}
           \end{array}
         \right\},
  \end{align*}
  where $\vbel \in \lmpB$ is called a pseudo-marginal vector. (The two
  constraints that were listed last in the definition of $\lmpB$ will be
  called ``edge consistency constraints.'')
  \defend
\end{Definition}

\begin{Definition}
  \label{def:Bethe:free:energy:1}

  The Bethe free energy function associated with the NFG $\graphN(\matrtheta)$
  is defined to be the function
  \begin{align*}
    \FBethe: \ 
      &\lmpB
         \to \R, \quad
       \vbel
       \mapsto
         \UBethe(\vbel)
         -
         \HBethe(\vbel),
  \end{align*}
  where
  \begin{align*}
    \UBethe: \ 
      &\lmpB
         \to \R, 
       \quad
       \vbel
       \mapsto
         \sum_i
           \UBethesub{i}(\vbel_i)
         +
         \sum_j
           \UBethesub{j}(\vbel_j) \\
    \HBethe: \ 
      &\lmpB
         \to \R,
       \quad
       \vbel
       \mapsto
         \sum_i
           \HBethesub{i}(\vbel_i)
         +
         \sum_j
           \HBethesub{j}(\vbel_j) \\
      &\quad\quad\quad\quad\quad\quad\quad\quad
         -
         \sum_{e}
           \HBethesub{e}(\vbel_e),
  \end{align*}
  with\footnote{Here and in the following, we use the short-hand
    $\sum_{\va_i}$ for $\sum_{\va_i \in \setA_i}$, \etc.}
  \begin{alignat*}{2}
    \UBethesub{i}: \ 
      &\lmpB_i
         \to \R, \quad
       \vbel_i
       &
       \, \mapsto\,
       &
         -
         \sum_{\va_i}
           \bel_{i,\va_i}
           \cdot
           \log
             \big(
               g_i(\va_i)
             \big), \\
    \UBethesub{j}: \ 
      &\lmpB_j
         \to \R, \quad
       \vbel_j
       &
       \, \mapsto\,
       &
         -
         \sum_{\va_j}
           \bel_{j,\va_j}
           \cdot
           \log
             \big(
               g_j(\va_j)
             \big), \\
    \HBethesub{i}: \ 
      &\lmpB_i
         \to \R, \quad
       \vbel_i
       &
       \, \mapsto\,
       &
         -
         \sum_{\va_i}
           \bel_{i,\va_i}
           \cdot
           \log
             (
               \bel_{i,\va_i}
             ), \\
    \HBethesub{j}: \ 
      &\lmpB_j
         \to \R, \quad
       \vbel_j
       &
       \, \mapsto\,
       &
         -
         \sum_{\va_j}
           \bel_{j,\va_j}
           \cdot
           \log
             (
               \bel_{j,\va_j}
             ), \\
    \HBethesub{e}: \ 
      &\lmpB_e
         \to \R, \quad
       \vbel_e
       &
       \, \mapsto\,
       &
         -
         \sum_{a_e}
           \bel_{e,a_e}
           \cdot
           \log
             (
               \bel_{e,a_e}
             ).
  \end{alignat*}
  Here, $\UBethe$ is the Bethe average energy function and $\HBethe$ is the
  Bethe entropy function. In the following, when confusion can arise what NFG
  a certain Bethe free energy function is referring to, we will use
  $\FBethesub{\graphN(\matrtheta)}$, etc., instead of $\FBethe$. Similar
  comments apply to $\UBethe$ and $\HBethe$.
  \defend
\end{Definition}

With this, the Bethe partition function of an NFG is \emph{defined} such that
an equality analogous to the second equality
in~\eqref{eq:permanent:vs:partition:function:vs:FGibbs:min:1} holds.

\begin{Definition}
  \label{def:Bethe:partition:function:1}

  The Bethe partition function of the NFG $\graphN(\matrtheta)$ is defined
  to be
  \begin{align*} 
    \ZBethe
      &\defeq
         \exp
           \left(
             -
             \min_{\vbel \in \lmpB}
               \FBethe(\vbel)
           \right).
  \end{align*}
  In the following, when confusion can arise what NFG a
  certain Bethe partition function is referring to, we will use
  $\ZBethe(\graphN)$, etc., instead of $\ZBethe$.
  \defend
\end{Definition}

The next definition is the main definition of this paper and was motivated by
the work of Chertkov, Kroc, and
Vergasso\-la~\cite{Chertkov:Kroc:Vergassola:08:1} and by the work of Huang and
Jebara~\cite{Huang:Jebara:09:1}.

\begin{Definition}
  \label{def:Bethe:permanent:1}
  
  Consider the NFG $\graphN(\matrtheta)$. The Bethe permanent
  of $\matrtheta$, which will be denoted by $\permBethe(\matrtheta)$, is
  defined to be
  \begin{align*} 
    \permBethe(\matrtheta)
      &\defeq
         \ZBethe\big( \graphN(\matrtheta) \big).
  \end{align*}
  \vskip-0.5cm
  \defend
\end{Definition}

A similar comment w.r.t.\ a temperature parameter $T \in \Rpp$ as at the end
of Section~\ref{sec:factor:graph:representation:1} applies also to the
definition of the Bethe partition function and the Bethe free energy
function. In the following, however, we will only consider the case $T =
1$. An exception is Section~\ref{sec:fractional:Bethe:entropy:1} on the
fractional Bethe approximation: this approximation can be viewed as
introducing multiple temperature parameters, namely one temperature parameter
for every term of $\HBethe$, and therefore includes the single temperature
parameter case as a special case.

\section{Properties of the Bethe Entropy Function \\
               and the Bethe Free Energy Function}
\label{sec:properties:Bethe:entropy:1}

There are relatively few general statements about the shape of the Bethe
entropy function. In this section we show that Bethe entropy function
associated with $\graphN(\matrtheta)$ has many special properties.
\begin{itemize}

\item In general, the Bethe entropy function is not a concave
  function. However, here we show that the Bethe entropy function associated
  with $\graphN(\matrtheta)$ is, when suitably parameterized, a concave
  function.

  Similarly, the Bethe free energy function is in general not a convex
  function. However, because the Bethe free energy function is the difference
  of the Bethe average energy function and the Bethe entropy function, because
  the Bethe average energy function is linear in its arguments, and because
  the Bethe entropy function is concave, the Bethe free energy function
  associated with $\graphN(\matrtheta)$ is convex and does \emph{not} have
  non-global local minima.\footnote{The fact that convexity / non-convexity of
    a function depends on its parameterization might explain the non-convexity
    observations in~\cite[Section~3.3]{Huang:Jebara:09:1} w.r.t.\ the Bethe
    free energy function.}

\item In general, the Bethe entropy function can take on positive, zero, and
  negative values. However, here we show that the Bethe entropy function
  associated with $\graphN(\matrtheta)$ is non-negative.

\item Very often, the directional derivative of the Bethe entropy function
  away from a vertex of its domain is $+\infty$ or $-\infty$. For the Bethe
  entropy function of $\graphN(\matrtheta)$ we show that the directional
  derivative away from any vertex of its domain has a (non-negative) finite
  value. (As we will see in Section~\ref{sec:factor:graph:spa:1}, this
  observation will have important consequences for the SPA convergence
  analysis.)

\end{itemize}

\subsection{Reformulation of the Bethe Entropy Function \\
                          and the Bethe Free Energy Function}

As mentioned in Section~\ref{sec:related:work:1}, the successes of max-product
algorithm~/ min-sum algorithm based approaches to the bipartite graph maximum
weight perfect matching problem in the papers~\cite{Huang:Jebara:07:1,
  Bayati:Borgs:Chayes:Zecchina:11:1, Bayati:Shah:Sharma:08:1,
  Sanghavi:Malioutov:Willsky:11:1} was heavily based on a theorem by Birkhoff
and von Neumann (see Theorem~\ref{theorem:Birkhoff:von:Neumann:1}). This
theorem is equally central to the results of the present paper. Namely, in the
next lemma we introduce a parameterization of the belief polytope $\lmpB$
based on $\setGamma{n}$ that will be used for the rest of the paper.

\begin{Lemma}
  \label{lemma:lmpB:parametrization:1}

  Consider the NFG $\graphN(\matrtheta)$. Its belief polytope $\lmpB$ can be
  parameterized by $\setGamma{n}$, the set of doubly stochastic matrices of
  size $n \times n$. In particular, we define the parameterization such that
  the matrix $\matrgamma = (\gamma_{i,j})_{(i,j) \in \setI \times \setJ} \in
  \setGamma{n}$ indexes the pseudo-marginal vector $\vbel \in \lmpB$ with
  \begin{align*}
    \Big.
      \bel_{i,\va_i}
    \Big|_{\va_i = \vunit_j}
      &= \Big.
           \bel_{j,\va_j}
         \Big|_{\va_j = \vunit_i}
       = \gamma_{i,j},
  \end{align*}
  and
  \begin{align*}
    \Big.
      \bel_{e,a_e}
    \Big|_{a_e = 0}
      &= 1
         -
         \gamma_{i,j},
    \quad
    \Big.
      \bel_{e,a_e}
    \Big|_{a_e = 1}
       = \gamma_{i,j},
  \end{align*}
  for every $i \in \setI$, $j \in \setJ$, and $e = (i,j) \in \setE$.
\end{Lemma}

\begin{Proof}
  It is straightforward to verify that the pseudo-marginal vector $\vbel$
  which is specified in the lemma statement is indeed in $\lmpB$. Moreover,
  one can verify that for every pseudo-marginal vector $\vbel \in \lmpB$ there
  is a $\matrgamma \in \setGamma{n}$ such that $\matrgamma$ indexes $\vbel$.
\end{Proof}

In the following, for a given matrix $\matrgamma = (\gamma_{i,j})_{(i,j) \in
  \setI \times \setJ}$, the $i$th row of $\matrgamma$ will be denoted by
$\vgamma_i = (\gamma_{i,j})_{j \in \setJ}$ and the $j$th column of
$\matrgamma$ will be denoted by $\vgamma_j = (\gamma_{i,j})_{i \in \setI}$.

The above observations allow us to express the Bethe free energy function and
related functions in terms of $\matrgamma \in \setGamma{n}$.

\begin{Lemma}
  \label{lemma:Bethe:free:energy:as:function:of:gamma:1}

  Consider the NFG $\graphN(\matrtheta)$. Then
  \begin{align*}
    \FBethe: \ 
      &\setGamma{n}
         \to \R, \quad
       \matrgamma
       \mapsto
         \UBethe(\matrgamma)
         -
         \HBethe(\matrgamma),
  \end{align*}
  where
  \begin{align*}
    \UBethe: \ 
      &\setGamma{n}
         \to \R,
       \quad
       \matrgamma
       \mapsto
         \sum_i
           \UBethesub{i}(\matrgamma_i)
         +
         \sum_j
           \UBethesub{j}(\matrgamma_j), \\
    \HBethe: \ 
      &\setGamma{n}
         \to \R,
       \quad
       \matrgamma
       \mapsto
         \sum_i
           \HBethesub{i}(\matrgamma_i)
         +
         \sum_j
           \HBethesub{j}(\matrgamma_i) \\
      &\quad\quad\quad\quad\quad\quad\quad\quad\quad\quad
         -
         \sum_{i,j}
           \HBethesub{(i,j)}(\gamma_{i,j}),
  \end{align*}
  with
  \begin{align*}
    \UBethesub{i}: \ 
       \setPi_{[n]}
         &\to \R, \quad
       \matrgamma_i
       \mapsto
         -
         \frac{1}{2}
         \sum_{j}
           \gamma_{i,j}
           \cdot
           \log(\theta_{i,j}), \\
    \UBethesub{j}: \ 
       \setPi_{[n]}
         &\to \R, \quad
       \matrgamma_j
       \mapsto
         -
         \frac{1}{2}
         \sum_{i}
           \gamma_{i,j}
           \cdot
           \log(\theta_{i,j}), \\
    \HBethesub{i}: \ 
      \setPi_{[n]}
         &\to \R, \quad
       \matrgamma_i
       \mapsto
         -
         \sum_j
           \gamma_{i,j}
           \cdot
           \log(\gamma_{i,j}), \\
    \HBethesub{j}: \ 
       \setPi_{[n]}
         &\to \R, \quad
       \matrgamma_j
       \mapsto
         -
         \sum_i
           \gamma_{i,j}
           \cdot
           \log(\gamma_{i,j}), \\
    \HBethesub{(i,j)}: \ 
      [0,1]
         &\to \R \\
      \gamma_{i,j}
         &\mapsto
         - \,
         \gamma_{i,j}
         \log(\gamma_{i,j})
         -
         (1 \! - \! \gamma_{i,j})
         \log(1 \! - \! \gamma_{i,j}),
  \end{align*}
\end{Lemma}

\begin{Proof}
  This follows straightforwardly from Definition~\ref{def:Bethe:free:energy:1}
  and Lemma~\ref{lemma:lmpB:parametrization:1}.
\end{Proof}

\begin{Corollary}
  \label{cor:FBethe:as:function:of:gamma:1}

  It holds that
  \begin{align*} 
    \permBethe(\matrtheta)
      &= \exp
           \left(
             -
             \min_{\matrgamma \in \setGamma{n}}
               \FBethe(\matrgamma)
           \right),  
  \end{align*}
  where
  \begin{align*}
    \FBethe(\matrgamma)
      &= \UBethe(\matrgamma)
         -
         \HBethe(\matrgamma), \\
    \UBethe(\matrgamma)
      &= -
         \sum_{i,j}
           \gamma_{i,j}
           \log(\theta_{i,j}), \\
    \HBethe(\matrgamma)
      &= -
         \sum_{i,j}
           \gamma_{i,j} \log(\gamma_{i,j})
         +
         \sum_{i,j}
           (1-\gamma_{i,j}) \log(1-\gamma_{i,j}).
  \end{align*}
\end{Corollary}

\begin{Proof}
  This follows from Definitions~\ref{def:Bethe:partition:function:1}
  and~\ref{def:Bethe:permanent:1} and from
  Lemma~\ref{lemma:Bethe:free:energy:as:function:of:gamma:1}.
\end{Proof}

If the sign in front of the second half of the expression for
$\HBethe(\matrgamma)$ in Corollary~\ref{cor:FBethe:as:function:of:gamma:1}
were a minus sign, then $\HBethe(\matrgamma)$ could be expressed as a sum of
binary entropy functions, and therefore the concavity of $\HBethe(\matrgamma)$
would be immediate. However, the presence of the plus sign means that a more
careful look at $\HBethe(\matrgamma)$ is required to determine if it is
concave or not.

\begin{Assumption}
  \label{assumption:positive:matrix:theta:1}

  For the rest of this section we assume that $n \geq 2$ and that $\matrtheta$
  is a \emph{positive} matrix of size $n \times n$. This simplifies the
  wording of most results without hurting their generality too much. In
  practice, two possible ways to deal with the issue of zero entries in
  $\matrtheta$ are the following.
  \begin{itemize}
  
  \item One can change the matrix $\matrtheta$ so that zero entries become
    tiny positive entries.

  \item One can redefine $\graphN(\matrtheta)$ by removing the edge $e =
    (i,j)$, along with redefining the local functions $g_i$ and $g_j$, if
    $\theta_{i,j} = 0$
  \assumptionend

  \end{itemize}
\end{Assumption}

\subsection{Concavity of the Bethe Entropy Function \\
                    and Convexity of the Bethe Free Energy Function}

Towards showing that $\HBethe(\matrgamma)$ is a concave function of
$\matrgamma$, and subsequently that $\FBethe(\matrgamma)$ is a convex function
of $\matrgamma$, we first study two useful functions. Namely, in
Definition~\ref{ess:function:1} and
Lemma~\ref{lemma:ess:function:properties:1} we look at a function called $s$,
and in Definition~\ref{def:md:ess:function:1} and
Theorem~\ref{theorem:md:ess:function:properties:1} we look at a function
called $S$. Note that in this section we use the short-hands $\sum_{\ell}$ and
$\sum_{\ell \neq \ells}$ for $\sum_{\ell \in [n]}$ and $\sum_{\ell \in [n]: \,
  \ell \neq \ells}$, respectively.

\begin{Definition}
  \label{ess:function:1}

  Let $s$ be the function
  \begin{align*}
    s: \ [0,1] &\to \R,
         \quad
         \xi    \mapsto -\xi \log(\xi) + (1-\xi) \log(1-\xi).
  \end{align*}
  Note that in contrast to the binary entropy function, there is a plus sign
  (not a minus sign) in front of the second term.
  \defend
\end{Definition}

\begin{figure}
  \psfrag{xlabel}{\hskip0.15cm$\xi$}
  \psfrag{ylabel}{$s(\xi)$}
  \epsfig{file=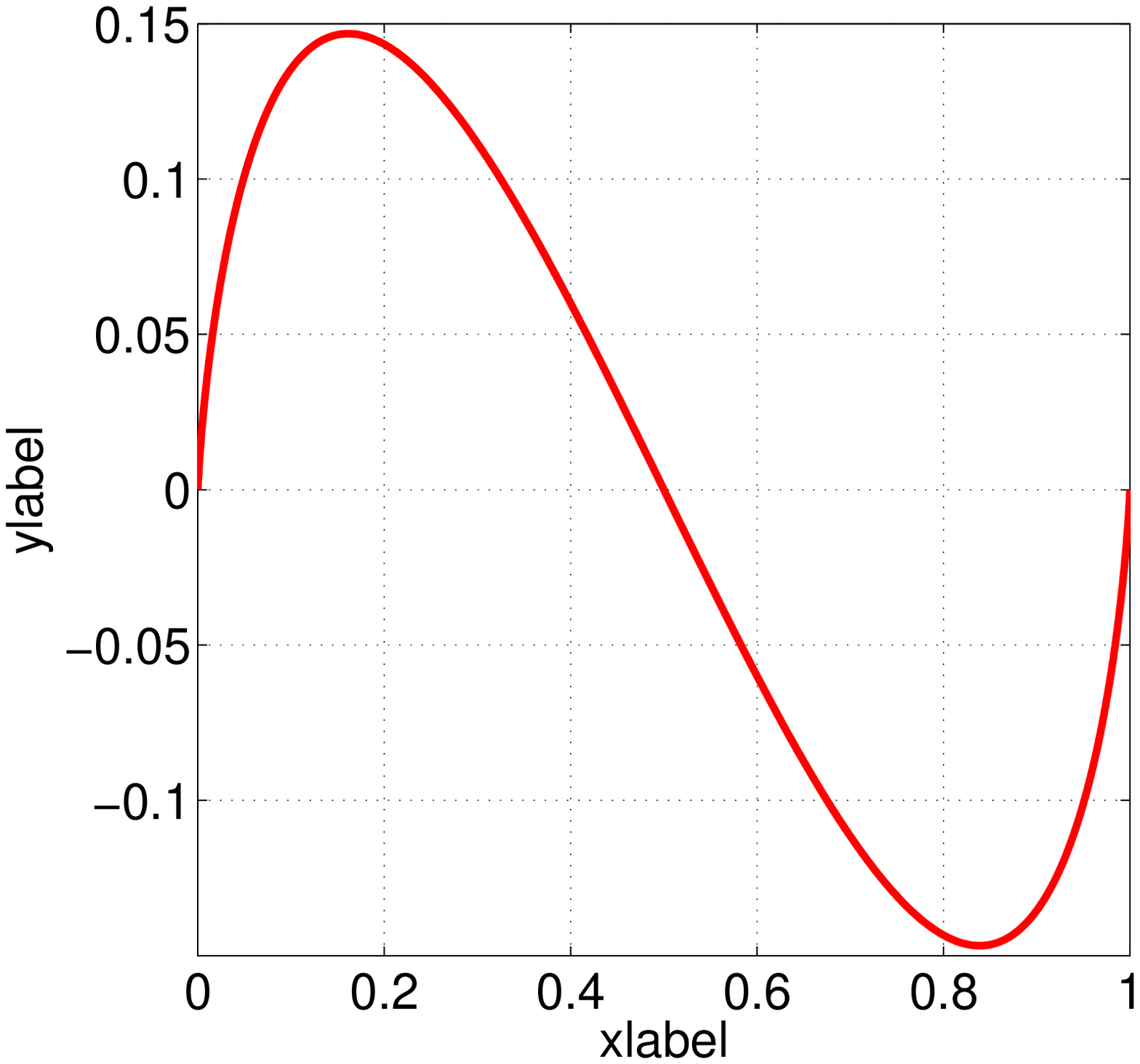, height=0.45\linewidth}
  \hfill
  \psfrag{xlabel}{\hskip0.1cm$\xi_1$}
  \psfrag{ylabel}{\hskip0.05cm$\xi_2$}
  \epsfig{file=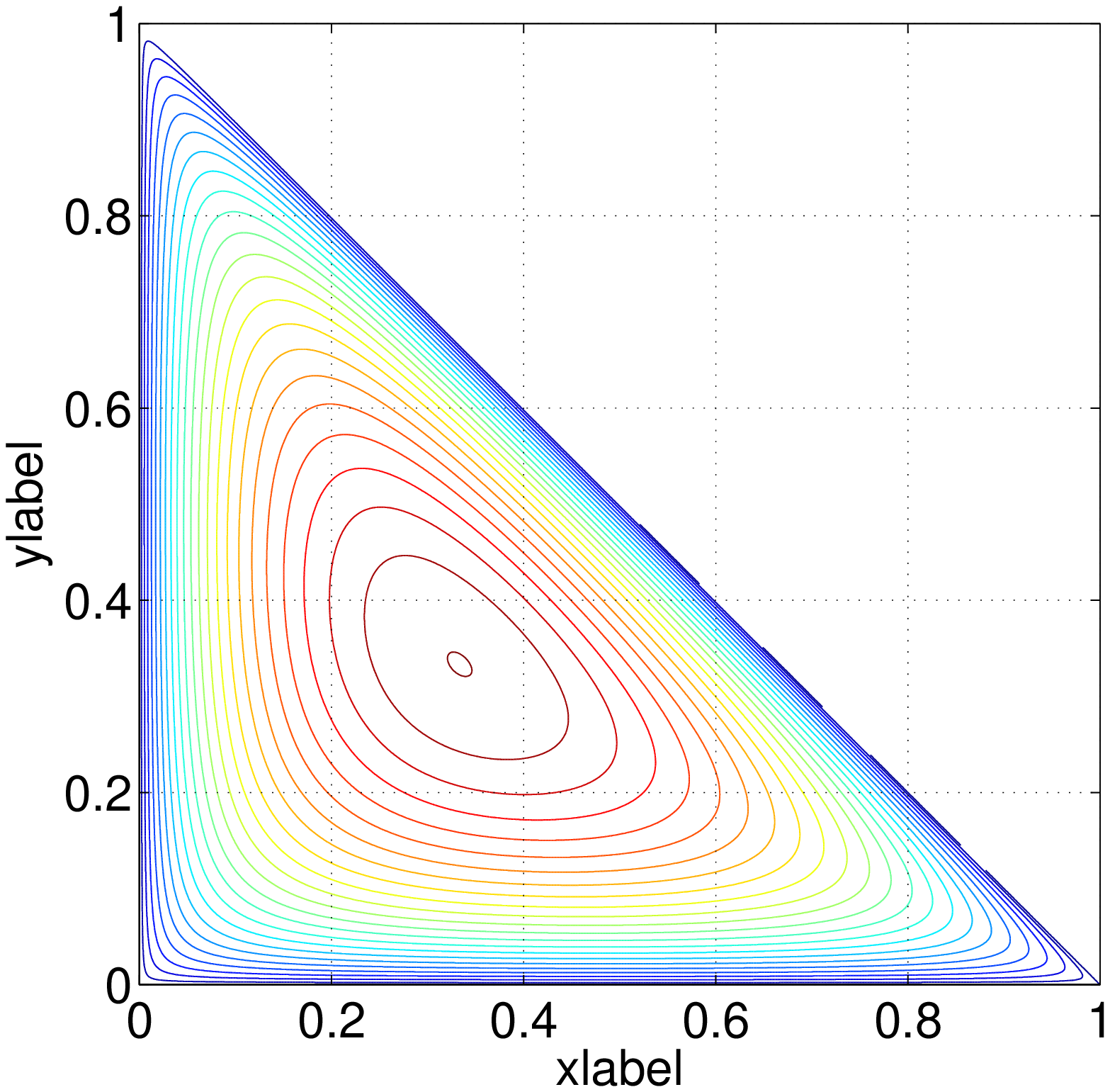, height=0.45\linewidth}
  \caption{Left: plot of the function $s$, see
    Definition~\ref{ess:function:1}. Right: contour plot of the function
    $(\xi_1, \xi_2) \mapsto S(\xi_1, \xi_2, 1 \! - \! \xi_1 \! - \! \xi_2)$,
    see Definition~\ref{def:md:ess:function:1}.}
  \label{fig:ess:function:1}
\end{figure}

\begin{Lemma}
  \label{lemma:ess:function:properties:1}

  The function $s$ that is specified in Definition~\ref{ess:function:1} has
  the following properties.
  \begin{itemize}

  \item As can be seen from Fig.~\ref{fig:ess:function:1}~(left), the graph of
    the function $s$ is s-shaped.
    
  \item The first-order derivative of $s$ is
  \begin{align*}
    \rdiff{\xi}
      s(\xi)
      &= -
         2
         -
         \log
           \big(
             \xi (1-\xi)
           \big).
  \end{align*}

  \item The second-order derivative of $s$ is
  \begin{align*}
    \rddiff{\xi}
      s(\xi)
      &= -
         \frac{1}{\xi}
         +
         \frac{1}{1-\xi}
       = -
         \frac{1-2\xi}{\xi (1-\xi)}.
  \end{align*}
  Clearly, the function $s(\xi)$ is strictly concave in the interval $0 \leq
  \xi < 1/2$ and strictly convex in the interval $1/2 < \xi \leq 1$.

  \item The graph of $s$ has a point-symmetry at $(1/2,0)$.

  \end{itemize}
\end{Lemma}

\begin{Proof}
  The proof of this lemma is based on straightforward calculus and is
  therefore omitted.
\end{Proof}

\begin{Definition}
  \label{def:md:ess:function:1} 

  Let $S$ be the function
  \begin{align*}
    S: \ 
       \setPi_{[n]} \to \R,
       \ 
       \vxi
         \mapsto 
         \sum_{\ell}
           s(\xi_{\ell})
       = 
       &  -
          \sum_{\ell}
            \xi_{\ell} \log(\xi_{\ell}) \\
       &  +
          \sum_{\ell}
            (1-\xi_{\ell}) \log(1-\xi_{\ell}).
  \end{align*}
  \vskip-0.25cm
  \defend
\end{Definition}

Fig.~\ref{fig:ess:function:1}~(right) shows the function $S$ for $n = 3$. More
precisely, that plot shows the contour plot of the function $(\xi_1, \xi_2)
\mapsto S(\xi_1, \xi_2, 1-\xi_1-\xi_2)$.

Clearly, if the domain of the function $S$ were the set $[0,1]^n$, then $S$
would not be concave everywhere because $s$ is not concave
everywhere. Therefore, the observation that is made in the following theorem,
namely that $S$ is concave, is non-trivial. (Note that because the function
$s$ is concave in $[0, 1/2]$, the function $S$ is concave in $\setPi_{[n]}
\cap [0,1/2]^n$. Therefore, as we will see, most of the work in the proof of
the following theorem will be devoted to proving the concavity of the function
$S$ in $\setPi_{[n]} \setminus [0,1/2]^n$.)

\begin{Theorem}
  \label{theorem:md:ess:function:properties:1}

  The function $S$ from Definition~\ref{def:md:ess:function:1} is concave and
  satisfies $S(\vxi) \geq 0$ for all $\vxi \in \setPi_{[n]}$. Moreover,
  \begin{itemize}

  \item For $n = 2$, it holds that $S(\vxi) = 0$ for all $\vxi \in
    \setPi_{[n]}$.

  \item For $n \geq 3$, the function $S$ is at almost all points in its domain
    a strictly concave function. However there are points in its domain and
    corresponding directions in which the function $S$ is linear.

  \end{itemize}
\end{Theorem}

\begin{Proof}
  See Appendix~\ref{sec:proof:theorem:md:ess:function:properties:1}.
\end{Proof}

After the original submission of the present paper, an alternative proof of
the concavity of the function $S$ has been given by Gurvits,
see~\cite[Section~5.1]{Gurvits:11:2}.

Interestingly, the functions $s$ and $S$ have recently appeared also in
another context~\cite{Lad:Sanfilippo:Agro:11:1}. (We refer
to~\cite{Lad:Sanfilippo:Agro:11:1} for details.) In particular, that paper
gives a direct proof of $S(\vxi) \geq 0$ for all $\vxi \in \setPi_{[n]}$; this
is in contrast to the proof of that statement in
Theorem~\ref{theorem:md:ess:function:properties:1} which was mainly based on
the concavity of $S$.

\begin{Lemma}
  \label{lemma:Bethe:entropy:function:in:terms:of:S:1}

  The Bethe entropy function can be expressed in terms of the function $S$ as
  follows
  \begin{alignat*}{2}
    \HBethe: \ 
      &\setGamma{n}
      &\ \to\ 
      &\R \\
      &\matrgamma
      &\ \mapsto \ 
      &
           \frac{1}{2}
           \sum_i
             S(\matrgamma_i)
           +
           \frac{1}{2}
           \sum_j
             S(\matrgamma_j).
  \end{alignat*}
\end{Lemma}

\begin{Proof}
  This result follows from
  \begin{align*}
    &
    \HBethe(\matrgamma) \\
      &\onestareq
         -
         \sum_{i,j}
           \gamma_{i,j} \log(\gamma_{i,j})
         +
         \sum_{i,j}
           (1-\gamma_{i,j}) \log(1-\gamma_{i,j}) \\
      &= \frac{1}{2}
           \sum_{i}
           \left(
             \! - \!
             \sum_{j}
               \gamma_{i,j} \log(\gamma_{i,j})
             +
             \sum_{j}
               (1\!-\!\gamma_{i,j}) \log(1\!-\!\gamma_{i,j})
           \right) 
           + \\
      &\quad\ 
         \frac{1}{2}
           \sum_{j}
           \left(
             \! - \!
             \sum_{i}
               \gamma_{i,j} \log(\gamma_{i,j})
             +
             \sum_{i}
               (1\!-\!\gamma_{i,j}) \log(1\!-\!\gamma_{i,j})
           \right) \\
      &\twostarseq
         \frac{1}{2}
         \sum_i
           S(\matrgamma_i)
         +
         \frac{1}{2}
         \sum_j
           S(\matrgamma_j),
  \end{align*}
  where at step~$\onestar$ we have used
  Corollary~\ref{cor:FBethe:as:function:of:gamma:1} and where at
  step~$\twostars$ we have used Definition~\ref{def:md:ess:function:1}.
\end{Proof}

\begin{Theorem}
  \label{theorem:concavity:Bethe:entropy:1:details}

  The Bethe entropy function $\HBethe(\matrgamma)$ is a concave function of
  $\matrgamma \in \setGamma{n}$. Moreover, for all $\matrgamma \in
  \setGamma{n}$ it holds that $\HBethe(\matrgamma) \geq 0$.
\end{Theorem}

\begin{Proof}
  Lemma~\ref{lemma:Bethe:entropy:function:in:terms:of:S:1} showed that
  $\HBethe(\matrgamma)$ can be written as a sum of $S$-functions. The
  concavity of $\HBethe(\matrgamma)$ then follows from
  Theorem~\ref{theorem:md:ess:function:properties:1} and the fact that the sum
  of concave functions is a concave function. Similarly, the non-negativity of
  $\HBethe(\matrgamma)$ follows from
  Theorem~\ref{theorem:md:ess:function:properties:1} and the fact that the sum
  of non-negative functions is a non-negative function.
\end{Proof}

\begin{Corollary}
  \label{cor:convexity:Bethe:free:energy:function:1}

  The Bethe free energy function $\FBethe(\matrgamma)$ is a convex function of
  $\matrgamma \in \setGamma{n}$.
\end{Corollary}

\begin{Proof}
  This follows from $\FBethe(\matrgamma) = \UBethe(\matrgamma) -
  \HBethe(\matrgamma)$
  (see Corollary~\ref{cor:FBethe:as:function:of:gamma:1}), from the fact
  that $\UBethe(\matrgamma)$ is a linear function of $\matrgamma$
  (see Corollary~\ref{cor:FBethe:as:function:of:gamma:1}), and from the
  fact that $\HBethe(\matrgamma)$ is a concave function of $\matrgamma$
  (see Theorem~\ref{theorem:concavity:Bethe:entropy:1:details}).
\end{Proof}

\subsection{Behavior of the Bethe Entropy Function 
                    and the Bethe Free Energy Function
                    at a Vertex of their Domain}

In this section we study the Bethe entropy function and the Bethe free energy
function near a vertex of their domain. Because both functions can be
expressed in terms of the function $S$, we first study the behavior of $S$
near a vertex of its domain.

\begin{Lemma}
  \label{lemma:linear:behavior:md:ess:1}

  Let
  \begin{align*}
    \vxi(\temp)
      &\defeq \vxi + \temp \cdot \hvxi,
  \end{align*}
  where the vector $\vxi \in \setPi_{[n]}$ is a vertex of $\setPi_{[n]}$ and
  where $\hvxi \neq \vect{0}$ is such that $\vxi(\temp) \in \setPi_{[n]}$ for
  small non-negative $\temp$. This means that there is an $\ells \in [n]$ such
  that $\vxi$ satisfies $\xi_{\ells} = 1$ and $\xi_{\ell} = 0$, $\ell \neq
  \ells$, and such that $\hvxi$ satisfies $\hxi_{\ells} < 0$, $\hxi_{\ell}
  \geq 0$, $\ell \neq \ells$, and $\sum_{\ell} \hxi_{\ell} = 0$. Then, for $0
  < \temp \ll 1$, we have
  \begin{align}
    S\big( \vxi(\temp) \big)
      &= \temp
         \cdot |\hxi_{\ells}|
         \cdot 
         \left(
           -
           \sum_{\ell \neq \ell'}
             \frac{|\hxi_{\ell}|}
                  {|\hxi_{\ells}|}
             \log
               \left(
                 \frac{|\hxi_{\ell}|}
                      {|\hxi_{\ells}|}
               \right)
         \right)
         +
         O(\temp^2),
           \label{eq:S:value:near:vertex:1}
  \end{align}
  \ie, the function $S\bigl( \vxi(\temp) \bigr)$ can very well be approximated
  by a linear function for $0 < \temp \ll 1$. Note that the coefficient of
  $\temp$ in~\eqref{eq:S:value:near:vertex:1} is non-negative.
\end{Lemma}

\begin{Proof}
  See Appendix~\ref{sec:proof:lemma:linear:behavior:md:ess:1}.
\end{Proof}

\mbox{}

A word of caution: the behavior of the function $S$ is somewhat special around
a vertex $\vxi$ of $\setPi_{[n]}$: namely, \emph{in general there is no
  gradient vector $\vect{G}$} such that $S(\vxi + \temp \cdot \hvxi) = S(\vxi)
+ \temp \cdot \sum_{\ell} G_{\ell} \hxi_{\ell} + O(\temp^2) = \temp \cdot
\sum_{\ell} G_{\ell} \hxi_{\ell} + O(\temp^2)$ for $0 < \temp \ll 1$ and for all
possible direction vectors $\hvxi$.

Lemma~\ref{lemma:linear:behavior:md:ess:1} has the following consequences for
the behavior of the Bethe entropy function at a vertex of its domain.

\begin{Lemma}
  \label{lemma:linear:behavior:HBethe:1}

  Let
  \begin{align*}
    \matrgamma(\temp)
      &\defeq \matrgamma + \temp \cdot \hmatrgamma,
  \end{align*}
  where $\matrgamma \in \codeC$ is a vertex of $\setGamma{n}$ and where
  $\hmatrgamma \neq \vect{0}$ is such that $\matrgamma(\temp) \in
  \setGamma{n}$ for small non-negative $\temp$. This means that $\matrgamma$
  corresponds to the permutation $\sigma_{\matrgamma}$. (In the following
  statement we will use the short-hands $\sigma \defeq \sigma_{\matrgamma}$
  and $\bsigma \defeq \sigma_{\matrgamma}^{-1}$.) Then, for $0 < \temp \ll 1$,
  we have
  \begin{align*}
    &
    \HBethe\big( \matrgamma(\temp) \big) \\
      &= \temp
         \sum_i
           |\hgamma_{i,\sigma(i)}|
           \cdot 
           \!
           \left(\!
             - \!\!\!
             \sum_{j \neq \sigma(i)}
             \frac{|\hgamma_{i,j}|}
             {|\hgamma_{i,\sigma(i)}|}
             \log
               \left(
                 \frac{|\hgamma_{i,j}|}
                      {|\hgamma_{i,\sigma(i)}|}
               \right)
           \!\!
           \right) \!
         +
         O(\temp^2) \\
      &= \temp
         \sum_j
           |\hgamma_{\bsigma(j),j}|
           \cdot 
           \!
           \left(\!
             - \!\!\!
             \sum_{i \neq \bsigma(j)}
             \frac{|\hgamma_{i,j}|}
             {|\hgamma_{\bsigma(j),j}|}
             \log
               \left(
                 \frac{|\hgamma_{i,j}|}
                      {|\hgamma_{\bsigma(j),j}|}
               \right)
           \!\!
           \right) \!
         +
         O(\temp^2),
  \end{align*}
  \ie, the function $\HBethe\big( \matrgamma(\temp) \big)$ can very well be
  approximated by a linear function for $0 < \temp \ll 1$. Note that the
  coefficient of $\temp$ is non-negative.
\end{Lemma}

\begin{Proof}
  See Appendix~\ref{sec:proof:lemma:linear:behavior:HBethe:1}.
\end{Proof}

\mbox{}

Assume that $\hmatrgamma$ in Lemma~\ref{lemma:linear:behavior:HBethe:1} is
chosen such that $\sum_i |\hgamma_{i,\sigma(i)}| = 1$. (If this is not the
case, then $\hmatrgamma$ can be rescaled by a positive real number such that
this condition is satisfied.) The coefficient of $\temp$ in the first display
equation of Lemma~\ref{lemma:linear:behavior:HBethe:1} can be given the
following meaning. It is the entropy rate of the time-invariant Markov chain
corresponding to the (backtrackless) random walk on the NFG
$\graphN(\matrtheta)$ (see Fig.~\ref{fig:ffg:permanent:1}) with the following
properties:\footnote{For a discussion of the entropy rate of a time-invariant
  Markov chain, see, \eg, \cite[Section~4.2]{Cover:thomas:06:1}.}
\begin{itemize}

\item The probability of being at vertex $i \in \setI$ is
  $|\hgamma_{i,\sigma(i)}|$.

\item The probability of going to vertex $j \in \setJ \setminus \{ \sigma(i)
  \}$, conditioned on being at vertex $i \in \setI$, is $|\hgamma_{i,j}| /
  |\hgamma_{i,\sigma(i)}|$. \\
  The probability of going to vertex $\sigma(i) \in \setJ$, conditioned on
  being at vertex $i \in \setI$, is $0$.

\item The probability of being at vertex $j \in \setJ$ is
  $|\hgamma_{\bsigma(j),j}|$.

\item The probability of going to vertex $\bsigma(j) \in \setI$, conditioned
  on being at vertex $j \in \setJ$, is $1$. \\
  The probability of going to vertex $i' \in \setI \setminus \{ \bsigma(j)
  \}$, conditioned on being at vertex $j \in \setJ$, is $0$.

\end{itemize}
The above two half-steps of the random walk can be combined into one step:
\begin{itemize}

\item The probability of being at vertex $i \in \setI$ is
  $|\hgamma_{i,\sigma(i)}|$.

\item For $i, i' \in \setI$ with $i \neq i'$, the probability of going to
  vertex $\sigma(i')$ and then to vertex $i'$, conditioned on being at vertex
  $i$, is $|\hgamma_{i,\sigma(i')}| / |\hgamma_{i,\sigma(i)}|$.

\end{itemize}

An analogous interpretation can be given to the coefficient of $\temp$ in the
second display equation of Lemma~\ref{lemma:linear:behavior:HBethe:1}. Observe
that the condition $\sum_i |\hgamma_{i,\sigma(i)}| = 1$ is equivalent to the
condition $\sum_j |\hgamma_{\bsigma(j),j}| = 1$.

Note that similar random walks appeared in the analysis of the Bethe entropy
function for so-called cycle codes~(\confer~\cite{Vontobel:10:2}) and in the
analysis of linear programming decoding of low-density parity-check
codes~(\confer~\cite{Vontobel:10:3}, which gives a random walk interpretation
of a result by Arora, Daskalakis, Steurer~\cite{Arora:Daskalakis:Steurer:09:1}
and its extensions by Halabi and Even~\cite{Halabi:Even:11:1}). Actually,
given the fact that the symmetric difference of two perfect matchings
corresponds to a union of cycles in $\graphN(\matrtheta)$, the similarity of
the random walks here and of the random walks in the above-mentioned context
of cycle codes is not totally surprising.

We come now to the main result of this subsection. Although this result is
interesting in its own right, it will be especially important for the
convergence analysis of the SPA in Section~\ref{sec:factor:graph:spa:1}.

\begin{Theorem}
  \label{theorem:FBethe:extremal:behavior:1}
  
  Let
  \begin{align*}
    \matrgamma(\temp)
      &\defeq \matrgamma + \temp \cdot \hmatrgamma,
  \end{align*}
  where $\matrgamma \in \codeC$ is a vertex of $\setGamma{n}$ and where
  $\hmatrgamma \neq \vect{0}$ is such that $\matrgamma(\temp) \in
  \setGamma{n}$ for small non-negative $\temp$. This means that $\matrgamma$
  corresponds to the permutation $\sigma_{\matrgamma}$. (In the following
  statement we will use the short-hands $\sigma \defeq \sigma_{\matrgamma}$
  and $\bsigma \defeq \sigma_{\matrgamma}^{-1}$.) We also assume that
  $\hmatrgamma$ is normalized as follows
  \begin{align}
    \sum_i
      |\hgamma_{i,\sigma(i)}|
      &= \sum_j
           |\hgamma_{\bsigma(j),j}|
       = 1.
           \label{eq:hgamma:normalization:constraint:1}
  \end{align}
  Then, for $0 < \temp \ll 1$, we have
  \begin{align}
    \FBethe\big( \matrgamma(\temp) \big)
      &\geq
         -
         \sum_{i} \log( \theta_{i,\sigma(i)} )
         -
         \temp
           \cdot
           \log(\rho)
         +
         O(\temp^2),
           \label{eq:FBethe:lower:bound:1}
  \end{align}
  where $\rho$ is the maximal (real) eigenvalue of the $n \times n$ matrix
  $\matrA$ with entries
  \begin{align*}
    A_{i,i'}
      &\defeq
         \begin{cases}
           \frac{\theta_{i,\sigma(i')}}
                {\theta_{i,\sigma(i)}}
             & \text{(if $i \neq i'$)} \\
           0
             & \text{(otherwise)}
         \end{cases}.
  \end{align*}
  Note that equality holds in~\eqref{eq:FBethe:lower:bound:1} for the matrix
  $\hmatrgamma$ with entries
  \begin{align*}
    \hgamma_{i,\sigma(i')}
      &\defeq
         \begin{cases}
           +
           \kappa
           \cdot
           \frac{
             \lefteigvect_i 
             \cdot
             A_{i,i'}
             \cdot
             \righteigvect_{i'}
           }
           {\rho}
             & \text{(if $i \neq i'$)} \\
           -
           \kappa
           \cdot
           \lefteigvect_i 
           \cdot
           \righteigvect_{i}
             & \text{(otherwise)}
         \end{cases},
  \end{align*}
  where $\vlefteigvect$ and $\vrighteigvect$ are, respectively, the left and
  right eigenvectors of $\matrA$ with eigenvalue $\rho$, and where $\kappa$ is
  a suitable normalization constant such
  that~\eqref{eq:hgamma:normalization:constraint:1} is satisfied.
\end{Theorem}

\begin{Proof}
  See Appendix~\ref{sec:proof:theorem:FBethe:extremal:behavior:1}.
\end{Proof}

\begin{Corollary}
  \label{cor:FBethe:minimum:at:vertex:1}

  Consider a vertex $\matrgamma$ of $\setGamma{n}$ and define $\rho$ for
  $\matrgamma$ as in Theorem~\ref{theorem:FBethe:extremal:behavior:1}.
  \begin{itemize}

  \item If $\rho < 1$ then $\FBethe$ has its unique minimum at
    $\matrgamma$.

  \item If $\rho > 1$ then $\FBethe$ is not minimal at $\matrgamma$.

  \end{itemize}
\end{Corollary}

\begin{Proof}
  Consider the setup of Theorem~\ref{theorem:FBethe:extremal:behavior:1}. From
  that theorem we know that
  \begin{align*}
    \FBethe\big( \matrgamma(\temp) \big)
      &\geq -
         \sum_{i} \log( \theta_{i,\sigma(i)} )
         -
         \temp
           \cdot
           \log(\rho)
         +
         O(\temp^2),
  \end{align*}
  with equality for the direction matrix $\hmatrgamma$ that was specified
  there. Moreover, from
  Corollary~\ref{cor:convexity:Bethe:free:energy:function:1} we know that
  $\FBethe$ is convex over $\setGamma{n}$. Therefore, if $\log(\rho) < 0$
  (\ie, $\rho < 1$) then $\FBethe$ has a unique minimum at $\matrgamma$. On
  the other hand, if $\log(\rho) > 0$ (\ie, $\rho > 1$) then $\FBethe$ cannot
  be minimal at $\matrgamma$.

  Note that for $\log(\rho) = 0$ (\ie, $\rho = 1$), the minimality~/
  non-minimality of $\FBethe$ at $\matrgamma$ is determined by the $O(\temp^2)$
  term.
\end{Proof}

Typically, the Bethe entropy function and the Bethe free energy function have
a positive or negative infinite directional derivative away from a vertex of
their domain because of the appearance of terms like $c \cdot \temp \cdot
\log(\temp)$. However, because for the function $S$ all these $c \cdot \temp \cdot
\log(\temp)$ terms cancel in the vicinity of a vertex of its domain (see the proof
of Theorem~\ref{theorem:md:ess:function:properties:1}, in particular
Eq.~\eqref{eq:S:first:orderderivative:1} in
Appendix~\ref{sec:vxi:at:vertex:1}), the directional derivatives of the Bethe
entropy function and the Bethe free energy function are finite away from a
vertex of their domain.

Let us conclude this section by pointing out that the observations that were
made in this subsection give an alternative viewpoint of some of the results
that were presented in~\cite[Section~3]{Watanabe:Chertkov:10:1}.

\section{Sum-Product-Algorithm-Based Search of the 
               Minimum of the Bethe Free Energy Function}
\label{sec:factor:graph:spa:1}

\medskip

\begin{Assumption}
  \label{sec:positive:and:perturbed:matrix:theta:1}

  In this section we make the following two assumptions, both with the goal of
  simplifying the wording of most results without hurting their generality too
  much.\footnote{The purpose of these assumptions is, in particular, to avoid
    dealing with matrices $\matrtheta$ which have the following
    property. Namely, consider the subgraph induced by the edge subset
    $\bigl\{ (i,j) \in \setE \bigm| \theta_{i,j} > 0 \}$. Assume that one of
    the connected components of this subgraph is a cycle (necessarily of even
    length), and consider the partition of the edge set of this cycle into two
    sets $\setE'$ and $\setE''$ such that the edges of this cycle are
    alternatingly placed into $\setE'$ and $\setE''$, respectively. If
    $\prod_{(i,j) \in \setE'} \theta_{i,j} = \prod_{(i,j) \in \setE''}
    \theta_{i,j}$ holds, then the SPA exhibits a periodic behavior unless the
    initial messages correspond to SPA fixed point messages. A matrix having
    this property is, \eg, the matrix $\matrtheta = \bigl( \begin{smallmatrix}
      1 & 1 \\ 1 & 1 \end{smallmatrix} \bigr)$. Here, the relevant cycle
    $(1,1)-(1,2)-(2,2)-(2,1)-(1,1)$ has length four and one verifies that
    $\theta_{1,1} \cdot \theta_{2,2} = \theta_{1,2} \cdot \theta_{2,1}$.}
  \begin{itemize}

  \item We assume that $n \geq 3$ and that $\matrtheta$ is a \emph{positive}
    matrix of size $n \times n$.

  \item We assume that the minimum of the Bethe free energy function $\FBethe$
    is either in the interior of $\setGamma{n}$ or at a vertex of
    $\setGamma{n}$, but not at a non-vertex boundary point of
    $\setGamma{n}$. A possibility to guarantee this with probability~$1$ is to
    apply tiny random perturbations to the entries of~$\matrtheta$.
    \assumptionend

  \end{itemize}
\end{Assumption}

In Definition~\ref{def:Bethe:permanent:1} we have defined the Bethe permanent
of a square matrix $\matrtheta$ via the minimum of the Bethe free energy
function of the NFG $\graphN(\matrtheta)$. In
Corollary~\ref{cor:convexity:Bethe:free:energy:function:1} we have seen that
the Bethe free energy function is a convex function, \ie, it behaves very
favorably. This means that we could use any generic optimization algorithm
(see, \eg,~\cite{Bertsekas:99:1, Boyd:Vandenberghe:04:1}) to find the minimum
of the Bethe free energy function, and with that the Bethe permanent of
$\matrtheta$. However, given the special structure of the optimization
problem, there is the hope that there are more efficient approaches.

A natural candidate for searching this minimum is the
SPA~\cite{Kschischang:Frey:Loeliger:01, Forney:01:1, Loeliger:04:1}. The
reason for this is that a theorem by Yedidia, Freeman, and
Weiss~\cite{Yedidia:Freeman:Weiss:05:1} says that fixed points of the SPA
correspond to stationary points of the Bethe free energy
function.\footnote{Strictly speaking, for NFGs with hard constraints, \ie,
  NFGs that contain local functions that can assume the value zero for certain
  points in their domain (which is the case for $\graphN(\matrtheta)$), this
  statement has only been proven for \emph{interior} stationary points of the
  Bethe free energy function
  (see~\cite[Theorem~2]{Yedidia:Freeman:Weiss:05:1}). For SPA fixed points
  with some beliefs equal to zero it is only conjectured that they correspond
  to edge-stationary points of the Bethe free energy function
  (\confer~discussion in~\cite[Section VI.D]{Yedidia:Freeman:Weiss:05:1}).}
Given the convexity of the Bethe free energy function, the following two
questions must therefore be answered:
\begin{itemize}

\item If the minimum of $\FBethe$ is in the interior of $\setGamma{n}$, does
  the SPA always converge to a fixed point?

\item If the minimum of $\FBethe$ is at a vertex of $\setGamma{n}$, does the
  SPA find that vertex?

\end{itemize}
In this section we answer both questions affirmatively, independently of the
matrix $\matrtheta$, and (nearly) independently of the chosen initial
messages.

The rest of this section is structured as follows. First we discuss the
details of the SPA message update rules in
Section~\ref{sec:factor:graph:spa:update:rules:1}. Afterwards, we state the
SPA convergence result in Section~\ref{sec:spa:convergence:1}.

\subsection{Sum-Product Algorithm Message Update Rules}
\label{sec:factor:graph:spa:update:rules:1}

In this subsection we derive the SPA message update rules for the NFG
$\graphN(\matrtheta)$ in Fig.~\ref{fig:ffg:permanent:1}. Here we only give the
technical details; for a general discussion w.r.t.\ the motivations behind the
SPA we refer to~\cite{Kschischang:Frey:Loeliger:01, Forney:01:1,
  Loeliger:04:1}. Note that analogous SPA message update rules were already
stated in~\cite{Huang:Jebara:09:1,
  Chertkov:Kroc:Krzakala:Vergassola:Zdeborova:10:1}. (In contrast
to~\cite{Huang:Jebara:09:1}, we use an undampened version of the SPA.)

On a high level, the SPA works as follows. With every edge in
Fig.~\ref{fig:ffg:permanent:1} we associate a right-going message and a
left-going message. Every iteration of the SPA consists then of two
half-iterations, in the first half-iteration the right-going messages are
updated based on the left-going messages and in the second half-iteration the
left-going messages are updated based on the right-going messages. Finally,
once some suitable convergence criterion is met or a fixed number of
iterations has been reached, the pseudo-marginal vector (belief vector) is
computed based on the messages at the last iteration.

Mathematically, we define for every $t \geq 0$ and every edge $(i,j) \in \setI
\times \setJ$ a left-going message $\muleftijt: \setA_{i,j} \to \R$, and for
every $t \geq 1$ and every edge $(i,j) \in \setI \times \setJ$ a right-going
message $\murightijt: \setA_{i,j} \to \R$.

For every left-going and for every right-going message it turns out to be
sufficient to keep track of the likelihood ratios
\begin{align*}
  \Lambdarightijt
    &\defeq
       \frac{\murightijt(0)}
            {\murightijt(1)},
  \quad\quad
  \Lambdaleftijt
     \defeq
       \frac{\muleftijt(0)}
            {\muleftijt(1)},
\end{align*}
respectively. Actually, for the NFG under consideration it is
more convenient to deal with the inverses of these quantities, and so we
define the inverse likelihood ratios as follows
\begin{align*}
  \udLambdarightijt
    &\defeq
       \left(
         \Lambdarightijt
       \right)^{-1},
  \quad\quad
  \udLambdaleftijt
     \defeq
       \left(
         \Lambdaleftijt
       \right)^{-1}.
\end{align*}

\begin{Lemma}
  \label{lemma:sum:product:algorithm:LR:update:rules:1}

  Consider the NFG $\graphN(\matrtheta)$. The inverse
  likelihood ratio update rules for the left-hand side and right-hand side
  function nodes of $\graphN(\matrtheta)$ are given by, respectively,
  \begin{alignat*}{2}
    \udLambdarightijt
      &= \frac{\sqrt{\theta_{i,j}}}
              {\sum_{j' \neq j}
                 \sqrt{\theta_{i,j'}}
                 \cdot
                 \udLambdaleftijptmo},
         &&\quad t \geq 1, \ (i,j) \in \setI \times \setJ, \\
    \udLambdaleftijt
      &= \frac{\sqrt{\theta_{i,j}}}
              {\sum_{i' \neq i}
                 \sqrt{\theta_{i',j}}
                 \cdot
                 \udLambdarightipjt
              },
         &&\quad t \geq 1, \ (i,j) \in \setI \times \setJ.
  \end{alignat*}
  The beliefs at the left-hand side and right-hand side function nodes of
  $\graphN(\matrtheta)$ are given by, respectively,
  \begin{alignat*}{2}
    \left.
      \belivait
    \right|_{\va_i = \vu_j}
      &\propto
         \sqrt{\theta_{i,j}}
         \cdot
         \udLambdaleftijt,
           &&\quad t \geq 0, \ (i,j) \in \setI \times \setJ, \\
    \left.
      \beljvajt
    \right|_{\va_j = \vu_i}
      &\propto
         \sqrt{\theta_{i,j}}
         \cdot
         \udLambdarightijt,
           &&\quad t \geq 1, \ (i,j) \in \setI \times \setJ.
  \end{alignat*}
  Here the proportionality constants are defined such that for every function
  node the beliefs sum to $1$. At a fixed point of the SPA, the beliefs
  satisfy the edge consistency constraints, \ie, for every $e = (i,j) \in
  \setE$ and every $a_e \in \setA_e$, it holds that $\sum_{\va'_i \in \setA_i:
    \, a'_{i,j} = a_e} \bel^{(t)}_{i, \va'_i} = \sum_{\va'_j \in \setA_j: \,
    a'_{i,j} = a_e} \bel^{(t)}_{j, \va'_j}$.
\end{Lemma}

\begin{Proof}
  See Appendix~\ref{sec:proof:lemma:sum:product:algorithm:LR:update:rules:1}.
\end{Proof}

Let us remark on the side that the above update equations can be reformulated
such that we only multiply by factors like $\theta_{i,j}$ instead of by
factors like $\sqrt{\theta_{i,j}}$. We leave the details to the reader.

\begin{Remark}
  \label{remark:gauge:invariance:1}

  The SPA messages for the NFG $\graphN(\matrtheta)$ exhibit
  the following property, a property that we will henceforth call ``message
  gauge invariance.'' Namely, consider the messages
  \begin{align*}
    \left\{
      \udLambdaleftijt
    \right\}_{i,j,t}
    \quad
    \text{and}
    \quad
    \left\{
      \udLambdarightijt
    \right\}_{i,j,t}
  \end{align*}
  that are connected by the update equations in
  Lemma~\ref{lemma:sum:product:algorithm:LR:update:rules:1}. It is then easy
  to show that for any $C \in \Rpp$ the messages
  \begin{align*}
    \left\{
      C
      \cdot
      \udLambdaleftijt
    \right\}_{i,j,t}
    \quad
    \text{and}
    \quad
    \left\{
      \frac{1}{C}
      \cdot
      \udLambdarightijt
    \right\}_{i,j,t}
  \end{align*}
  also satisfy the update equations in
  Lemma~\ref{lemma:sum:product:algorithm:LR:update:rules:1}. Moreover, the
  beliefs $\bigl\{ \belivait \bigr\}_{i,\va_i,t}$ and $\bigl\{ \beljt(\va_j)
  \bigr\}_{j,\va_j,t}$ are left unchanged by this rescaling of the inverse
  likelihood ratios. This is because the normalization that appears in the
  definition of $\bigl\{ \belivait \bigr\}_{i,\va_i,t}$ and $\bigl\{
  \beljt(\va_j) \bigr\}_{j,\va_j,t}$ removes the influence of this message
  rescaling.  \remarkend
\end{Remark}

Strictly speaking, the Bethe free energy function can only be evaluated at
fixed points of the SPA. However, very often it is desirable to track the
progress towards the minimum Bethe free energy function value. This can be
done via the so-called pseudo-dual function of the Bethe free energy
function~\cite{Regalia:Walsh:07:1, Mezard:Montanari:09:1}. This function has
the following two properties: it can be evaluated at any point during the SPA
computations, and at a fixed point of the SPA its value equals the value of
the Bethe free energy function. However, in general it is \emph{not} a
non-increasing or a non-decreasing function of the iteration number.

\begin{Lemma}
  \label{lemma:pseudo:dual:Bethe:free:energy:1}

  Consider the NFG $\graphN(\matrtheta)$. For any set of
  left-going messages $\bigl\{ \udLambdaleftij \bigr\}_{i,j}$ and any set of
  right-going messages $\bigl\{ \udLambdarightij \bigr\}_{i,j}$\,, the
  pseudo-dual function of the Bethe free energy function is
  \begin{align*}
    \FpdualBethe
      \left(
        \bigl\{ \udLambdaleftij \bigr\}, 
        \bigl\{ \udLambdarightij \bigr\}
      \right)
      &= -
         \sum_i
           \log
             \left(
               \sum_j
                 \sqrt{\theta_{i,j}}
                 \cdot
                 \udLambdaleftij
             \right) \\
       &\quad\ 
         -
         \sum_j
           \log
             \left(
               \sum_i
                 \sqrt{\theta_{i,j}}
                 \cdot
                 \udLambdarightij
             \right) \\
       &\quad\ 
         +
         \sum_{i,j}
           \log
             \left(
               1
               +
               \udLambdaleftij
               \cdot
               \udLambdarightij
             \right)
  \end{align*}
\end{Lemma}

\begin{Proof}
  See Appendix~\ref{sec:proof:lemma:pseudo:dual:Bethe:free:energy:1}.
\end{Proof}

In particular, if desired, we can evaluate $\FpdualBethe$ after every
half-iteration of the SPA, \ie, we can compute $\FpdualBethe \bigl( \bigl\{
\udLambdaleftijtmo \bigr\}, \bigl\{ \udLambdarightijt \bigr\} \bigr)$ and
$\FpdualBethe \bigl( \bigl\{ \udLambdaleftijt \bigr\}, \bigl\{
\udLambdarightijt \bigr\} \bigr)$ for every $t \geq 1$.

\subsection{Convergence of the Sum-Product Algorithm}
\label{sec:spa:convergence:1}

Note that there are rather few general results concerning the behavior of
message-passing type algorithms for NFGs with cycles. For certain classes of
graphical models and message-passing type algorithms, early results showed
that under the assumption that the algorithm converges then the obtained
estimates are correct (see, \eg, the results in~\cite{Weiss:Freeman:01:1,
  Weiss:Freeman:01:2}). Later, conditions for convergence were established for
a variety of graphical models and message-passing type algorithms (see, \eg,
\cite{Rusmevichientong:VanRoy:01:1, Mooij:Kappen:07:1,
  Malioutov:Johnson:Willsky:06:1, Ruozzi:Thaler:Tatikonda:09:1} and references
therein). However, these results do not seem to be applicable to the NFG under
consideration in this paper.

The SPA convergence proof that is the most relevant for the present paper is
the one in the paper by Bayati and Nair~\cite{Bayati:Nair:06:1} (see also the
comments that we made about this paper in
Section~\ref{sec:related:work:1}). However, the fact that the graphical model
in~\cite{Bayati:Nair:06:1} counts matchings (and not only perfect matchings
like here), implies a different behavior of the Bethe free energy function
near the boundary of its domain, and so no separate analysis of interior and
boundary minima of the Bethe free energy is required in the convergence proof
in~\cite{Bayati:Nair:06:1}. The SPA convergence analysis for a slightly
generalized weighted matching setup was recently presented by Williams and
Lau~\cite{Williams:Lau:10:2}.

Note that, interestingly enough, establishing convergence for the SPA on
$\graphN(\matrtheta)$ is independent of the choice of $\matrtheta$, which is
in contrast to, say, Gaussian graphical models where the convergence behavior
not only depends on the connectivity of the underlying graph but also on the
values of the non-zero entries of the information matrix describing the
Gaussian graphical model. (Of course, the convergence \emph{speed} of the SPA
on $\graphN(\matrtheta)$ does depend on the choice of $\matrtheta$.)

\begin{Theorem}
  \label{theorem:convergence:spa:1}

  Consider the SPA for NFG $\graphN(\matrtheta)$, for which the message update
  rules were established in
  Lemma~\ref{lemma:sum:product:algorithm:LR:update:rules:1}. For any initial
  set of inverse likelihood ratios $\bigl\{ \udLambdaleftijext{0}
  \bigr\}_{i,j}$ that satisfies $0 < \udLambdaleftijext{0} < \infty$, $(i,j)
  \in \setI \times \setJ$, the pseudo-marginals computed by the SPA converge
  to the pseudo-marginals that minimize the Bethe free energy function of
  $\graphN(\matrtheta)$. More precisely, we can make the following
  statements. (We remind the reader of the assumptions that were made in
  Assumption~\ref{sec:positive:and:perturbed:matrix:theta:1}.)
  \begin{itemize}

  \item If the minimum of $\FBethe$ is in the interior of $\setGamma{n}$, then
    the inverse likelihood ratios
    \begin{align*}
      \Bigl.
        \bigl\{
          \udLambdaleftijext{t}
        \bigr\}_{i,j,t}
      \Bigr|_{t \to \infty}
      \text{ and }
      \Bigl.
        \bigl\{
          \udLambdarightijext{t}
        \bigr\}_{i,j,t}
      \Bigr|_{t \to \infty}
    \end{align*}
    stay bounded and converge (modulo the message gauge invariance mentioned
    in Remark~\ref{remark:gauge:invariance:1}) to the fixed point inverse
    likelihood ratios corresponding to the minimum of $\FBethe$.

  \item If the minimum of $\FBethe$ is at the vertex $\matrgamma$ of
    $\setGamma{n}$, then the inverse likelihood ratios satisfy
    \begin{alignat*}{2}
      \Bigl.
        \udLambdaleftijext{t}
      \Bigr|_{j = \sigma_{\matrgamma}(i)}
        &\ \xrightarrow{t \to \infty} \ \infty,
      \quad\quad
      &
      \Bigl.
          \udLambdarightijext{t}
      \Bigr|_{j = \sigma_{\matrgamma}(i)}
        &\ \xrightarrow{t \to \infty} \ \infty, \\
      \Bigl.
        \udLambdaleftijext{t}
      \Bigr|_{j \neq \sigma_{\matrgamma}(i)}
        &\ \xrightarrow{t \to \infty} \ 0,
      \quad\quad
      &
      \Bigl.
        \udLambdarightijext{t}
      \Bigr|_{j \neq \sigma_{\matrgamma}(i)}
        &\ \xrightarrow{t \to \infty} \ 0.
    \end{alignat*}

  \end{itemize}
  Finally, 
  \begin{align*}
    \left|\,
      \exp
      \bigg(\!\!\!
        -
        \FpdualBethe
          \Big(
            \bigl\{ \udLambdaleftijt \bigr\}, 
            \bigl\{ \udLambdarightijt \bigr\}
          \Big) \!\!
      \bigg)
      -
      \permBethe(\matrtheta)
    \right|
      &\leq
         C
         \cdot
         \e^{-\nu \cdot t}
  \end{align*}
  for some constants $C, \nu \in \Rpp$ that depend on the matrix $\matrtheta$
  and the initial messages.
\end{Theorem}

\begin{Proof}
  See Appendix~\ref{sec:proof:theorem:convergence:spa:1}.
\end{Proof}

Explicit convergence speed estimates (in particular, values for $C$ and $\nu$)
can be extracted from the proof of
Theorem~\ref{theorem:convergence:spa:1}. However, we think that a more
sophisticated analysis might yield tighter convergence speed estimates; we
leave this as an open problem for future research.

\section{Finite-Graph-Cover Interpretation \\
               of the Bethe Permanent}
\label{sec:finite:graph:cover:interpretation:Bethe:permanent:1}

Note that the definition of the permanent of $\matrtheta$ in
Definition~\ref{def:permanent:1} has a ``combinatorial flavor.'' In
particular, it can be seen as a sum over all weighted perfect matchings of a
complete bipartite graph. This is in contrast to the definition of the Bethe
permanent of $\matrtheta$ (see
Definitions~\ref{def:Bethe:partition:function:1}
and~\ref{def:Bethe:permanent:1}) that has an ``analytical flavor.'' In this
section we show that it is possible to represent the Bethe permanent by an
expression that has a ``combinatorial flavor.'' We do this by applying the
results from~\cite{Vontobel:10:7:subm}, that hold for general NFGs, to the NFG
$\graphN(\matrtheta)$. The key concept in that respect are so-called finite
graph covers. (We keep the discussion here somewhat brief and we refer
to~\cite{Vontobel:10:7:subm} for all the details. See
also~\cite{Vontobel:11:1}.)

Besides being of interest in its own right, we think that the combinatorial
interpretation of the Bethe permanent discussed in this section can lead to
alternative proofs of known results or to proofs of new results for the Bethe
permanent. See, \eg,
Appendix~\ref{sec:proof:cor:Bethe:permanent:all:one:matrix:bound:1} that gives
an alternative proof of a special case of
Theorem~\ref{theorem:Bethe:permanent:upper:bound:1} in the next section.

This section is structured as follows. In
Section~\ref{sec:degree:M:Bethe:permanent:1} we define the degree-$M$ Bethe
permanent of a non-negative square matrix with the help of finite graph covers
and show that in the limit $M \to \infty$ the degree-$M$ Bethe permanent
converges to the Bethe permanent. Towards obtaining a better understanding of
the degree-$M$ Bethe permanent, we then study various examples of $2 \times 2$
matrices in
Sections~\ref{sec:Bethe:permanent:n:2:1:initial}--\ref{sec:degree:M:Bethe:permanent:n:2:1:general:matrix}. Because
the Bethe permanent can be computed with the help of the SPA, and because the
SPA is a locally operating algorithm on the relevant NFG, it is not surprising
that finite graph covers play a central role in the above-mentioned
combinatorial interpretation of the Bethe permanent; this aspect will be
discussed in Section~\ref{sec:relevance:finite:graph:covers:1}.

\subsection{The Degree-$M$ Bethe Permanent of a Non-Negative Matrix}
\label{sec:degree:M:Bethe:permanent:1}

\begin{Definition}[see, \eg, \cite{Massey:77:1, Stark:Terras:96:1}]
  \label{def:graph:cover:1}

  A {\em cover} of a graph $\graph{G}$ with vertex set $\set{V}$ and edge set
  $\set{E}$ is a graph $\graph{G}$ with vertex set $\set{\cover{V}}$ and edge
  set $\set{\cover{E}}$, along with a surjection $\pi: \set{\cover{V}} \to
  \set{V}$ which is a graph homomorphism (\ie, $\pi$ takes adjacent vertices
  of $\graph{G}$ to adjacent vertices of $\graph{G}$) such that for each
  vertex $v \in \set{V}$ and each $\cover{v} \in \pi^{-1}(v)$, the
  neighborhood $\del(\cover{v})$ of $\cover{v}$ is mapped bijectively to
  $\del(v)$. A cover is called an {\em $M$-cover}, where $M \in \Zpp$, if
  $\bigl| \pi^{-1}(v) \bigr| = M$ for every vertex $v$ in
  $\set{V}$.\footnote{The number $M$ is also known as the degree of the
    cover. (Not to be confused with the degree of a vertex.)}  \defend
\end{Definition}

Because NFGs are graphs, it is straightforward to extend this definition to
NFGs. (Of course, the variables that are associated with the $M$ copies of an
edge are allowed to take on different values.) For an $M$-cover, the left-hand
side function nodes will be labeled by elements of $\setI \times [M]$, the
right-hand side function nodes will be labeled by elements of $\setJ \times
[M]$, and the edges will be labeled by elements of a cover-dependent subset of
$\set I \times [M] \times \setJ \times [M]$. We will denote the set of all
$M$-covers $\cgraph{N}$ of $\graphN(\matrtheta)$ by
$\cset{N}_M(\matrtheta)$. (Note that we distinguish two $M$-covers with
different function node labels, even if the underlying graphs are isomorphic;
see also the comments on labeled graph covers
after~\cite[Definition~19]{Vontobel:10:7:subm}.)

\begin{figure}
  \begin{center}
    \begin{tabular}{ccc}
    \begin{minipage}[c]{0.25\linewidth}
      \begin{center}
        \subfigure[]
                  {\epsfig{file=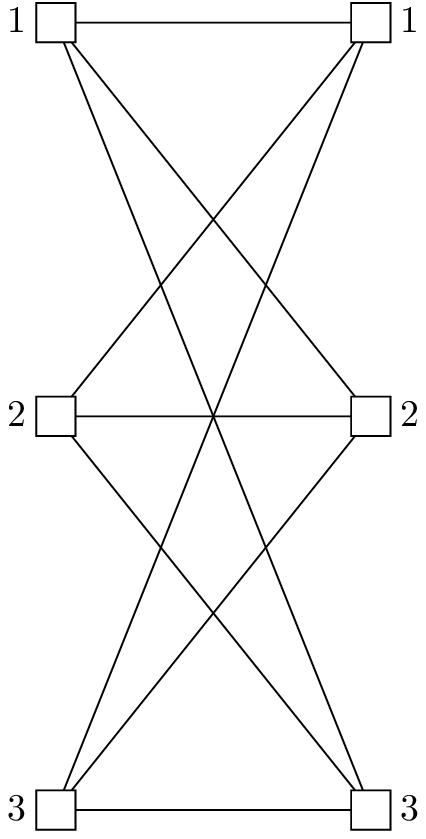,
                       scale=0.5}}
    \label{fig:gc:graph:cover:interpretation:3:by:3:base}
      \end{center}
    \end{minipage}
    &
    \ 
    \begin{minipage}[c]{0.30\linewidth}
      \begin{center}
        \subfigure[]
                  {\epsfig{file=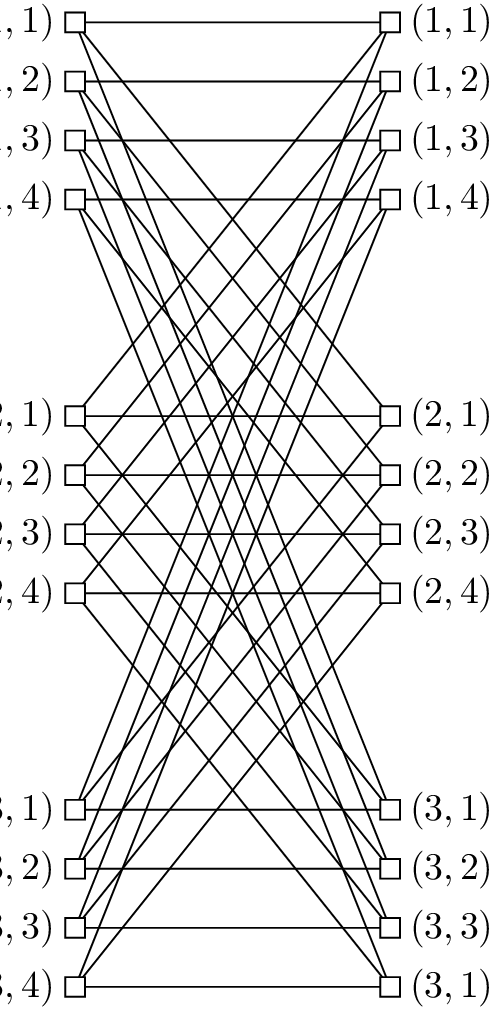,
                       scale=0.5}}
    \label{fig:gc:graph:cover:interpretation:3:by:3:fourfold:cover:1}
      \end{center}
    \end{minipage}
    &
    \ 
    \begin{minipage}[c]{0.30\linewidth}
      \begin{center}
        \subfigure[]
              {\epsfig{file=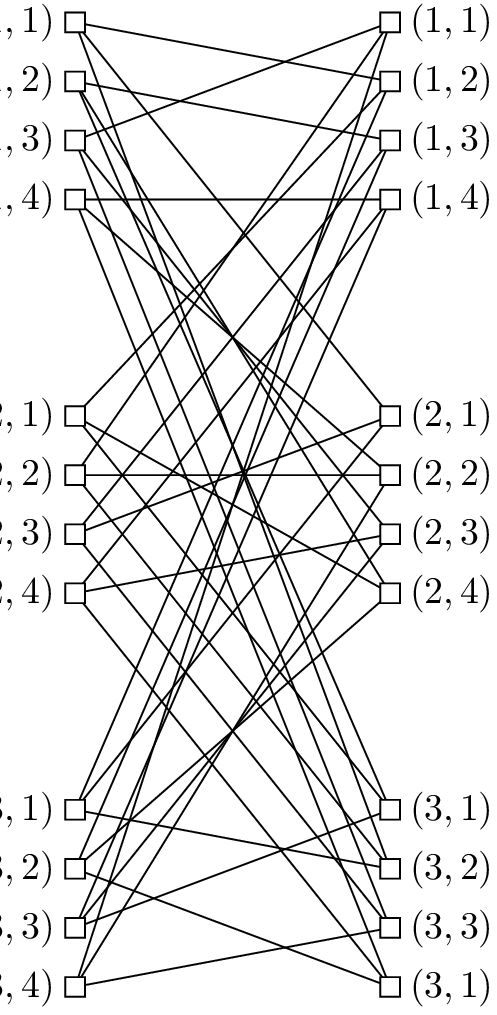,
                       scale=0.5}}
      \end{center}
    \end{minipage}
    \end{tabular}
  \end{center}
  \caption{(a) NFG $\graphN(\matrtheta)$ for $n = 3$. (b) ``Trivial''
    $4$-cover of $\graphN(\matrtheta)$ (c) A possible $4$-cover of
    $\graphN(\matrtheta)$.}
  \label{fig:gc:graph:cover:interpretation:3:by:3:1}
\end{figure}

\begin{Example}
  Let $n = 3$. The NFG $\graphN(\matrtheta)$ is shown in
  Fig.~\ref{fig:gc:graph:cover:interpretation:3:by:3:1}(a). There is only
  one $1$-cover of $\graphN(\matrtheta)$, namely $\graphN(\matrtheta)$
  itself. Two possible $4$-covers of $\graphN(\matrtheta)$ are shown in
  Figs.~\ref{fig:gc:graph:cover:interpretation:3:by:3:1}(b)--(c). The
  $4$-cover in Fig.~\ref{fig:gc:graph:cover:interpretation:3:by:3:1}(b) is
  ``trivial'' in the sense that it consists of $4$ disconnected copies of
  $\graphN(\matrtheta)$. On the other hand, the $4$-cover in
  Fig.~\ref{fig:gc:graph:cover:interpretation:3:by:3:1}(c) is ``nontrivial''
  in the sense that it consists of $4$ copies of $\graphN(\matrtheta)$ that
  are intertwined.
  \exampleend
\end{Example}

\begin{Lemma}
  \label{lemma:set:of:graph:covers:1}

  It holds that
  \begin{align}
    \cardbig{\cset{N}_M(\matrtheta)}
      &= (M!)^{(n^2)}.
           \label{eq:number:M:covers:1}
  \end{align}
\end{Lemma}

\begin{Proof}
  This follows from~\cite[Lemma~20]{Vontobel:10:7:subm} and the fact that the
  NFG $\graphN(\matrtheta)$ has $n^2$ full-edges.
\end{Proof}

\mbox{}

The following definition is the main definition of this section.

\begin{Definition}
  \label{def:Bethe:permanent:degree:M:partition:function:1}

  For any $M \in \Zpp$ we define the degree-$M$ Bethe permanent of
  $\matrtheta$ to be
  \begin{align*}
    \permBetheM{M}(\matrtheta)
      &\defeq
         \sqrt[M]{\Big\langle \!
                    \ZGibbs(\cgraph{N})
                  \! \Big\rangle_{\cgraph{N} \in \cset{N}_{M}}},
  \end{align*}
  where the angular brackets represent the arithmetic average of
  $\ZGibbs(\cgraph{N})$ over all $\cgraph{N} \in \cset{N}_{M}$. (Note that the
  right-hand side is based on the Gibbs partition function, not the Bethe
  partition function.)  \defend
\end{Definition}

As we will now show, one can express $\ZGibbs(\cgraph{N})$ for any $M$-cover
$\cgraph{N}$ of $\graphN(\matrtheta)$ as the permanent of some matrix that is
derived from $\matrtheta$.

\begin{Definition}
  \label{def:maththeta:lifting:1}

  For any $M \in \Zpp$ we define $\csetPsi_M$ to be the set
  \begin{align*}
    \csetPsi_M
      &\defeq
         \left\{
            \cmatrP = \big\{ \cmatrP^{(i,j)} \big\}_{i \in \setI, j \in \setJ}
            \ \middle| \ 
            \cmatrP^{(i,j)} \in \setPMat{M}
         \right\}.
  \end{align*}
  Moreover, for $\cmatrP \in \csetPsi_M$ we define the $\cmatrP$-lifting of
  $\matrtheta$ to be the following $(nM) \times (nM)$ matrix
  \begin{align*}
    \matrtheta^{\uparrow\cmatrP}
      &\defeq
         \begin{pmatrix}
           \theta_{1,1} \cmatrP^{(1,1)}
             & \cdots
             & \theta_{1,n} \cmatrP^{(1,n)} \\
           \vdots                   
             &
             &  \vdots \\
           \theta_{n,1} \cmatrP^{(n,1)} 
             & \cdots
             & \theta_{n,n} \cmatrP^{(n,n)}
         \end{pmatrix}.
  \end{align*}
  \defend
\end{Definition}

For any positive integer $M$ it is straightforward to see that there is a
bijection between the set $\cset{N}_M(\matrtheta)$ of all $M$-covers of
$\graphN(\matrtheta)$ and the set $\{ \matrtheta^{\uparrow\cmatrP} \}_{\cmatrP
  \in \csetPsi_M}$. In particular, because of
Lemma~\ref{lemma:permanent:Gibbs:free:energy:reformulation:1}, for an
$M$-cover $\cgraph{N}$ and its corresponding matrix
$\matrtheta^{\uparrow\cmatrP}$ it holds that $\ZGibbs(\cgraph{N}) =
\perm(\matrtheta^{\uparrow\cmatrP})$. Therefore, we have the following
reformulation of
Definition~\ref{def:Bethe:permanent:degree:M:partition:function:1}.

\begin{Definition}[Reformulation of
  Definition~\ref{def:Bethe:permanent:degree:M:partition:function:1}]
  \label{def:Bethe:permanent:degree:M:partition:function:2}

  For any $M \in \Zpp$ we define the degree-$M$ Bethe permanent of
  $\matrtheta$ to be
  \begin{align}
    \permBetheM{M}(\matrtheta)
      &\defeq
         \sqrt[M]{\Big\langle \!
                    \perm
                      \left(
                        \matrtheta^{\uparrow\cmatrP}
                      \right) 
                  \! \Big\rangle_{\cmatrP \in \csetPsi_M}},
                    \label{eq:degree:M:Bethe:permanent:1}
  \end{align}
  where the angular brackets represent the arithmetic average of $\perm\bigl(
  \matrtheta^{\uparrow\cmatrP} \bigr)$ over all $\cmatrP \in \csetPsi_M$. (Note
  that the permanent, not the Bethe permanent, appears on the right-hand
  side of the above expression.)
  \defend
\end{Definition}

In order to better appreciate the right-hand side of the above expression, it
is worthwhile to make the following two observations.
\begin{itemize}

\item For $M = 1$, the averaging is trivial because $\csetPsi_M$ contains only
  one element. Moreover, letting $\cmatrP$ be this single element, it holds
  that $\matrtheta^{\uparrow\cmatrP} = \matrtheta$. Therefore
  \begin{align*}
    \permBetheM{1}(\matrtheta)
      &= \perm(\matrtheta).
  \end{align*}

\item For any $M \in \Zpp$, the ``trivial'' $M$-cover of $\graphN(\matrtheta)$
  is given by the choice $\cmatrP = \bigl\{ \cmatrP^{(i,j)} \bigr\}_{i \in
    \setI, j \in \setJ}$ with $\cmatrP^{(i,j)} = \cmatrI$, $(i,j) \in \setI
  \times \setJ$, where $\cmatrI$ is the identity matrix of size $M
  \times M$. For this $M$-cover we obtain
  \begin{align*}
    \perm(\matrtheta^{\uparrow\cmatrP})
      &= \perm(\matrtheta)^M,
  \end{align*}
  \ie
  \begin{align*}
    \sqrt[M]{\perm(\matrtheta^{\uparrow\cmatrP})}
      &= \perm(\matrtheta).
  \end{align*}

\end{itemize}

\mbox{}

\noindent
With this, we are ready for the main result of this section.

\begin{Theorem}
  \label{theorem:Behte:permanent:degree:M:partition:function:1}
 
  It holds that
  \begin{align*}
    \limsup_{M \to \infty} \ 
      \permBetheM{M}(\matrtheta)
      &= \permBethe(\matrtheta).
  \end{align*}
\end{Theorem}

\begin{Proof}
  This follows from Definitions~\ref{def:Bethe:permanent:1}
  and~\ref{def:Bethe:permanent:degree:M:partition:function:2}, along with the
  application of~\cite[Theorem~33]{Vontobel:10:7:subm} to $\graphN =
  \graphN(\matrtheta)$.
\end{Proof}

Theorem~\ref{theorem:Behte:permanent:degree:M:partition:function:1}, together
with the relation $\permBetheM{1}(\matrtheta) = \perm(\matrtheta)$, are
visualized in Fig.~\ref{fig:degree:M:Bethe:permanent:1}. Because the
permanents that appear on the right-hand side
of~\eqref{eq:degree:M:Bethe:permanent:1} are combinatorial objects,
Definition~\ref{def:Bethe:permanent:degree:M:partition:function:2} and
Theorem~\ref{theorem:Behte:permanent:degree:M:partition:function:1} give the
promised ``combinatorial characterization'' of the Bethe permanent.

\subsection{The Bethe Permanent for Matrices of 
                    Size $2 \times 2$}
\label{sec:Bethe:permanent:n:2:1:initial}

\begin{figure*}
  \begin{center}
    \begin{tabular}{c||c|c}
    \begin{minipage}[c]{0.23\linewidth}
      \begin{center}
        \subfigure[]{\epsfig{file=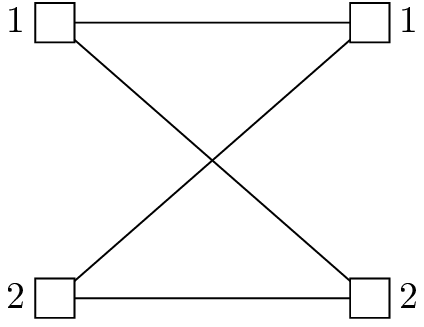, scale=0.50}}

        \subfigure[]{\epsfig{file=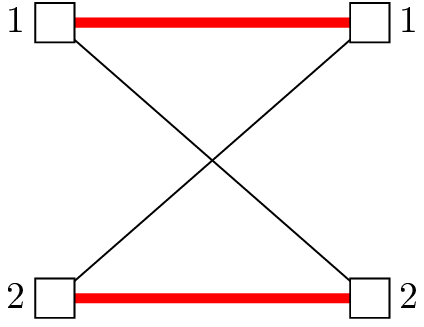,
            scale=0.4}}
        \ 
        \subfigure[]{\epsfig{file=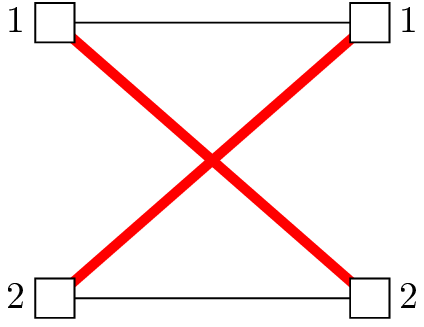,
          scale=0.4}}
      \end{center}
    \end{minipage}
    &
    \begin{minipage}[c]{0.46\linewidth}
      \begin{center}
    \subfigure[]{\epsfig{file=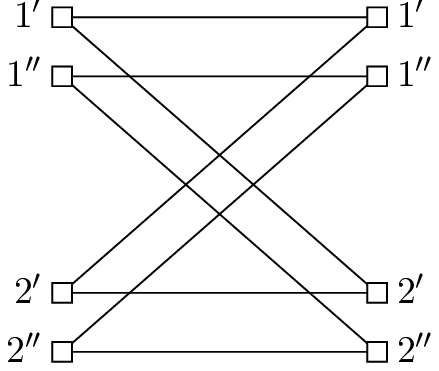, 
        scale=0.50}}

    \subfigure[]{\epsfig{file=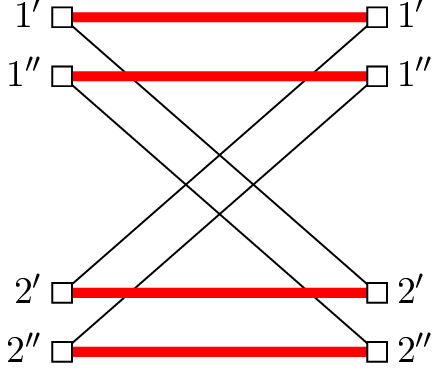, 
        scale=0.4}}
    \
    \subfigure[]{\epsfig{file=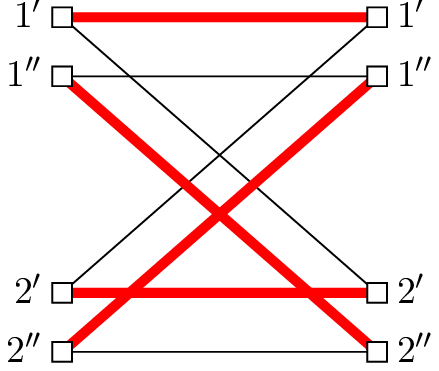,
        scale=0.4}}
    \
    \subfigure[]{\epsfig{file=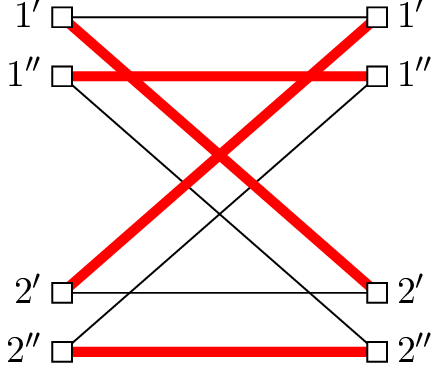,
        scale=0.4}}
    \
    \subfigure[]{\epsfig{file=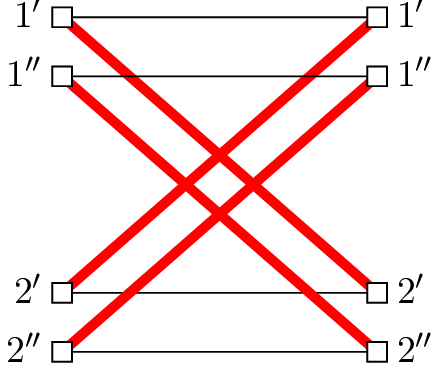,
        scale=0.4}}
      \end{center}
    \end{minipage}
    &
    \begin{minipage}[c]{0.23\linewidth}
      \begin{center}
    \subfigure[]{\epsfig{file=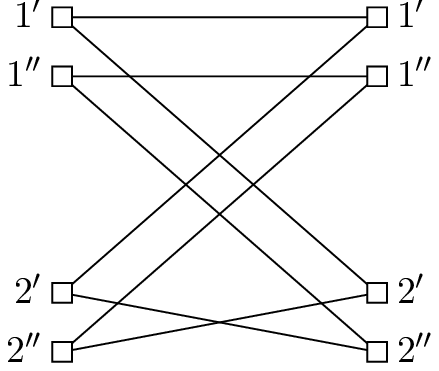,
        scale=0.50}}

    \subfigure[]{\epsfig{file=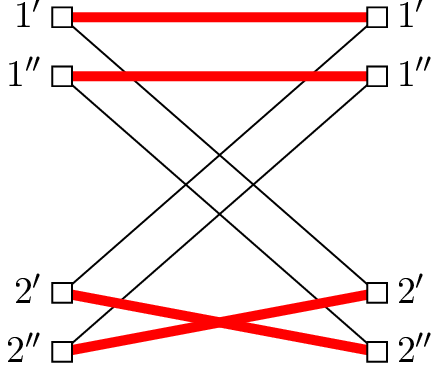,
        scale=0.4}}
    \
    \subfigure[]{\epsfig{file=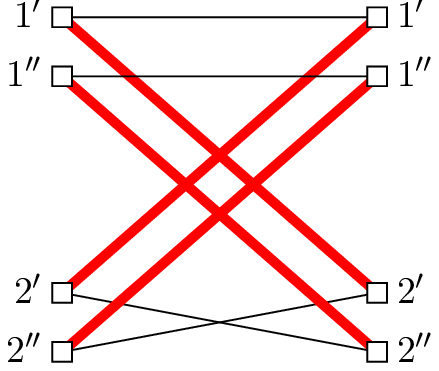,
        scale=0.4}}
      \end{center}
    \end{minipage}
    \end{tabular}
  \end{center}
  \caption{Graphs (NFGs) that are discussed in
    Sections~\ref{sec:degree:M:Bethe:permanent:n:2:1:initial}--\ref{sec:degree:M:Bethe:permanent:n:2:1:general:matrix}.
    (a) Base graph. (b)--(c) Perfect matchings of the graph in (a). (d) A
    possible double cover of the graph in (a). (e)--(h) Perfect matchings of
    the graph in (d). (i) A possible double cover of the graph in
    (a). (j)--(k) Perfect matchings of the graph in (i).}
  \label{fig:gc:2:by:2:case:degree:2:1}
\end{figure*}

In this and the following subsections we illustrate the concepts and results
that have been presented so far in this section by having a detailed look at
the case $n = 2$, \ie, we study the permanent, the Bethe permanent, and the
degree-$M$ Bethe permanent for the matrix
\begin{align*}
  \matrtheta
    &= \begin{pmatrix}
         \theta_{1,1} & \theta_{1,2} \\
         \theta_{2,1} & \theta_{2,2}
       \end{pmatrix}.
\end{align*}
The corresponding NFG $\graphN(\matrtheta)$ is shown in
Fig.~\ref{fig:gc:2:by:2:case:degree:2:1}(a). Of course, nobody would use the
Bethe permanent to approximate the permanent of a $2 \times 2$ matrix,
however, it gives some good insights into the strengths and the weaknesses of
the Bethe approximation to the permanent.

\begin{Lemma}
  \label{lemma:permanent:Bethe:permanent:n:2:1}

  For $n = 2$ it holds that
  \begin{align*}
    \perm(\matrtheta)
      &= \theta_{1,1} \theta_{2,2}
         +
         \theta_{2,1} \theta_{1,2}, \\
    \permBethe(\matrtheta)
      &= \max(
              \theta_{1,1} \theta_{2,2}, \ 
              \theta_{2,1} \theta_{1,2}
             ).
  \end{align*}
\end{Lemma}

\begin{Proof}
  The result for $\perm(\matrtheta)$ follows from
  Definition~\ref{def:permanent:1}. On the other hand, in order to obtain
  $\permBethe(\matrtheta)$, we apply
  Corollary~\ref{cor:FBethe:as:function:of:gamma:1}. The crucial step in
  Corollary~\ref{cor:FBethe:as:function:of:gamma:1} is to minimize
  $\FBethe(\matrgamma)$ over $\matrgamma \in \setGamma{2}$. Because
  $\HBethe(\matrgamma) = 0$, $\matrgamma \in \setGamma{2}$, minimizing
  $\FBethe(\matrgamma)$ is equivalent to minimizing $\UBethe(\matrgamma) = -
  \sum_{i,j} \gamma_{i,j} \log(\theta_{i,j})$.
  \begin{itemize}

  \item For $\theta_{1,1} \theta_{2,2} = \theta_{1,2} \theta_{2,1}$ the
    minimum is achieved at every $\matrgamma \in \setGamma{2}$.

  \item For $\theta_{1,1} \theta_{2,2} > \theta_{1,2} \theta_{2,1}$
    the minimum is achieved at $\matrgamma = \bigl(\begin{smallmatrix} 1 & 0 \\
      0 & 1 \end{smallmatrix}\bigr)$.

  \item For $\theta_{1,1} \theta_{2,2} < \theta_{1,2} \theta_{2,1}$
    the minimum is achieved at $\matrgamma = \bigl(\begin{smallmatrix} 0 & 1 \\
      1 & 0 \end{smallmatrix}\bigr)$.

  \end{itemize}
\end{Proof}

\begin{figure}
  \begin{alignat*}{2}
    &\Big.
       \permBetheM{M}(\matrtheta)
     \Big|_{M \to \infty}
        &&= \permBethe(\matrtheta) \\
    &\hskip1cm \Big\vert \\
    &\Big.
       \permBetheM{M}(\matrtheta)
     \Big. && \\
    &\hskip1cm \Big\vert \\
    &\Big.
       \permBetheM{M}(\matrtheta)
     \Big|_{M = 1}
       &&= \perm(\matrtheta)
  \end{alignat*}
  \caption{The degree-$M$ Bethe permanent of the non-negative matrix
    $\matrtheta$ for different values of $M$.}
  \label{fig:degree:M:Bethe:permanent:1}
\end{figure}

\begin{Example}
  \label{example:permanent:Bethe:permanent:all:one:matrix:2:1}

  For $n = 2$ and $\theta_{i,j} = 1$, $(i,j) \in \setI \times \setJ$, we have
  \begin{align*}
    \perm(\matrtheta)
      &= 2, \\
    \permBethe(\matrtheta)
      &= 1.
  \end{align*}
  Recall that $\perm(\matrtheta)$ represents the sum of all the weighted
  perfect matchings of the complete bipartite graph $\graphN(\matrtheta)$, and
  so, for the special choice $\theta_{i,j} = 1$, $(i,j) \in \setI \times
  \setJ$, the quantity $\perm(\matrtheta)$ represents the number of perfect
  matchings of $\graphN(\matrtheta)$. As is illustrated in
  Figs.~\ref{fig:gc:2:by:2:case:degree:2:1}(b)--(c), the graph
  $\graphN(\matrtheta)$ has two perfect matchings, thereby combinatorially
  verifying $\perm(\matrtheta) = 2$.
  \exampleend
\end{Example}

\subsection{The Degree-$M$ Bethe Permanent for Matrices of 
                    Size $2 \times 2$ --- Initial Considerations}
\label{sec:degree:M:Bethe:permanent:n:2:1:initial}

One of the goals of this and the next subsections is to obtain a better
combinatorial understanding of the result $\permBethe(\matrtheta) = 1$ for $n
= 2$, in particular, why it is different from $\perm(\matrtheta)$, yet not too
different.

Towards this goal, let us study the degree-$M$ Bethe permanent of $\matrtheta$
as specified in
Definition~\ref{def:Bethe:permanent:degree:M:partition:function:2}.  Therein,
the average is taken over $\card{\csetPsi_M} = (M!)^4$ matrices
\begin{align*}
  \matrtheta^{\uparrow\cmatrP}
    &= \begin{pmatrix}
         \theta_{1,1} \cmatrP_{1,1} & \theta_{1,2} \cmatrP_{1,2} \\
         \theta_{2,1} \cmatrP_{2,1} & \theta_{2,2} \cmatrP_{2,2}
       \end{pmatrix},
       \quad \matrtheta^{\uparrow\cmatrP} \in \csetPsi_M.
\end{align*}
We can simplify the analysis by realizing that the permanent of
$\matrtheta^{\uparrow\cmatrP}$ equals the permanent of a modified matrix
$\matrtheta^{\uparrow\cmatrP}$, where the first block row is multiplied from
the left by $\cmatrP_{1,1}^{-1}$, where the second block row is multiplied
from the left by $\cmatrP_{2,1}^{-1}$, and where the second block column is
multiplied from the right by $\cmatrP_{1,2}^{-1} \cdot \cmatrP_{1,1}$, \ie,
\begin{align*}
  \perm
    \big(
      \matrtheta^{\uparrow\cmatrP}
    \big)
    &= \perm
         \begin{pmatrix}
           \theta_{1,1} \cmatrI & \theta_{1,2} \cmatrI \\
           \theta_{2,1} \cmatrI & \theta_{2,2} \cmatrP_{2,1}^{-1} \cmatrP_{2,2}
                                                 \cmatrP_{1,2}^{-1} \cmatrP_{1,1}
         \end{pmatrix},
\end{align*}
where $\cmatrI$ is the identity matrix of size $M \times M$. Therefore, we
can rewrite $\permBetheM{M}(\matrtheta)$ as follows
\begin{align}
  \permBetheM{M}(\matrtheta)
    &\defeq
       \sqrt[M]{\Big\langle \!
                  \perm
                    \begin{pmatrix}
                      \theta_{1,1} \cmatrI & \theta_{2,1} \cmatrI \\
                      \theta_{2,1} \cmatrI & \theta_{2,2} \cmatrP'_{2,2}
                    \end{pmatrix}
                \! \Big\rangle_{\cmatrP'_{2,2} \in \setPMat{M}}},
           \label{eq:degree:M:Bethe:permanent:simplified:average:1}
\end{align}
\ie, an average over the $M!$ permutation matrices of size $M \times M$.

\subsection{The Degree-$M$ Bethe Permanent for Matrices of 
                    Size $2 \times 2$ --- All-One Matrix}
\label{sec:degree:M:Bethe:permanent:n:2:1:all:one:matrix}

In this subsection we consider the cases $M = 2$, $M = 3$, and general $M$ for
the special choice
\begin{align*}
  \matrtheta
    &= \begin{pmatrix}
         1 & 1 \\
         1 & 1
       \end{pmatrix}.
\end{align*}

\begin{figure*}
  \begin{center}

    \begin{minipage}[c]{0.13\linewidth}
      \begin{center}
        \subfigure[]
              {\epsfig{file=ffg_permanent_2by2_0_1.fig.eps, scale=0.45}}
        \label{fig:gc:graph:cover:interpretation:2:by:2:case:1:case:base}
      \end{center}
    \end{minipage}
    \
    \begin{minipage}[c]{0.13\linewidth}
      \begin{center}
        \subfigure[$8$ pms.]
              {\epsfig{file=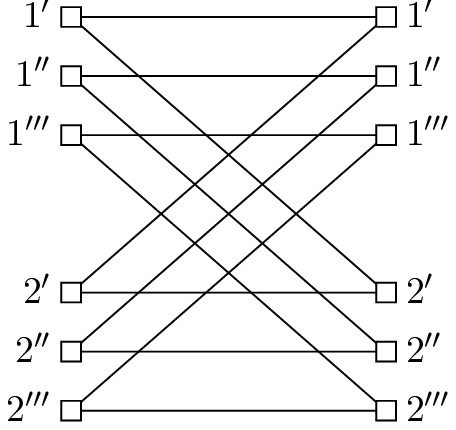, scale=0.45}}        
      \end{center}
    \end{minipage}
    \
    \begin{minipage}[c]{0.13\linewidth}
      \begin{center}
        \subfigure[$4$ pms.]
                  {\epsfig{file=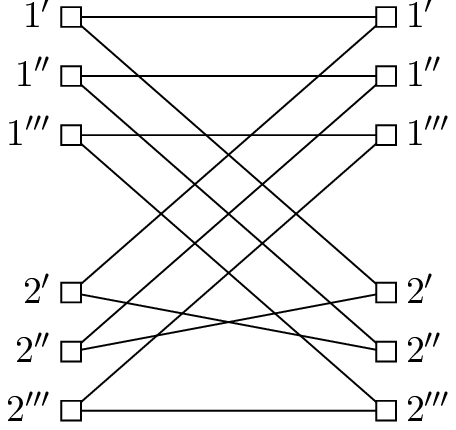, scale=0.45}}
        \label{fig:gc:graph:cover:interpretation:2:by:2:case:1:case:2}
      \end{center}
    \end{minipage}
    \
    \begin{minipage}[c]{0.13\linewidth}
      \begin{center}
        \subfigure[$4$ pms.]
                  {\epsfig{file=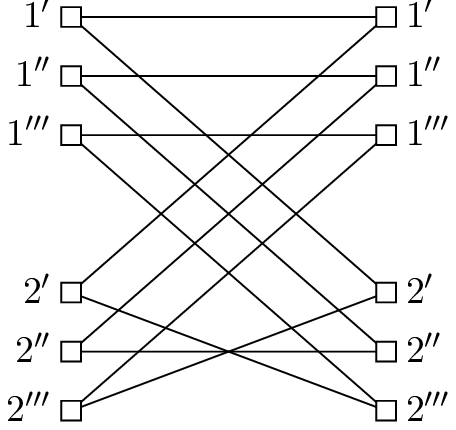, scale=0.45}}
      \end{center}
    \end{minipage}
    \
    \begin{minipage}[c]{0.13\linewidth}
      \begin{center}
        \subfigure[$4$ pms.]
                  {\epsfig{file=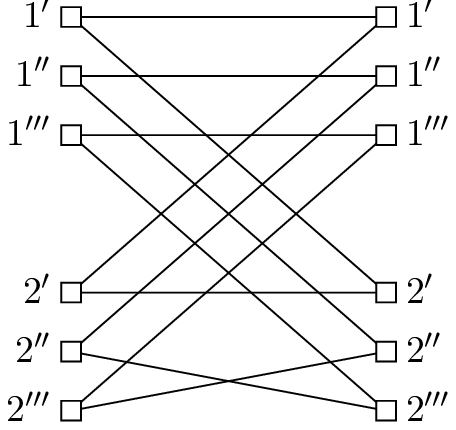, scale=0.45}}
      \end{center}
    \end{minipage}
    \
    \begin{minipage}[c]{0.13\linewidth}
      \begin{center}
        \subfigure[$2$ pms.]
                  {\epsfig{file=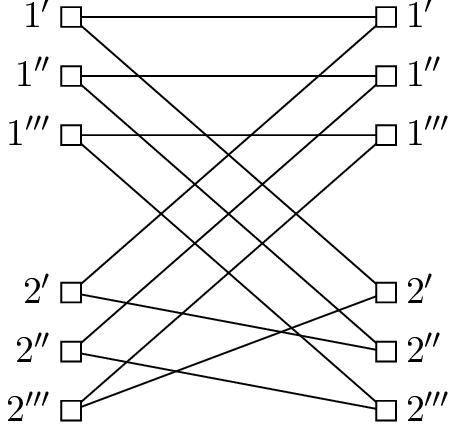, scale=0.45}}
      \end{center}
    \end{minipage}
    \
    \begin{minipage}[c]{0.13\linewidth}
      \begin{center}
        \subfigure[$2$ pms.]
                  {\epsfig{file=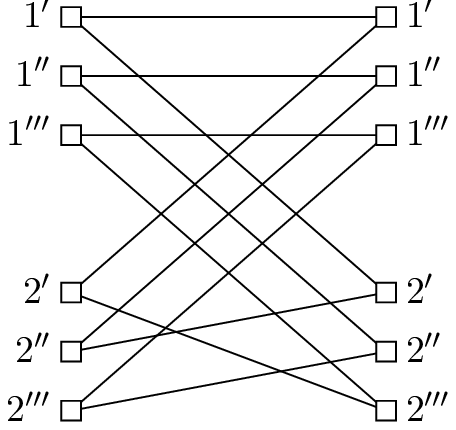, scale=0.45}}
      \end{center}
    \end{minipage}
  \end{center}
  \caption{Graphs (NFGs) that are discussed in
    Sections~\ref{sec:degree:M:Bethe:permanent:n:2:1:initial}--\ref{sec:degree:M:Bethe:permanent:n:2:1:general:matrix}.
    (a) Base graph. (b)--(g) Possible triple covers of the graph in
    (a). (``pms.'' stands for ``perfect matchings''.)}
  \label{fig:gc:graph:cover:interpretation:2:by:2:case:1}
\end{figure*}

\begin{figure*}
  \begin{center}
    \subfigure[]
              {\epsfig{file=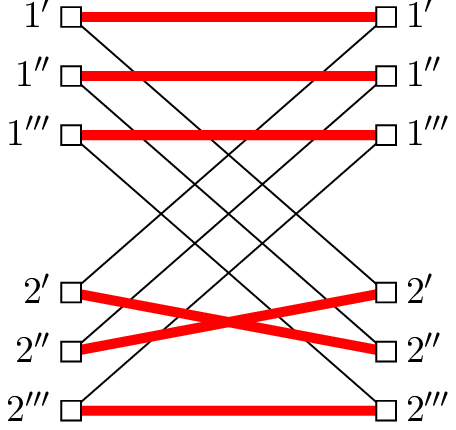, scale=0.45}}
    \quad\quad
    \subfigure[]
              {\epsfig{file=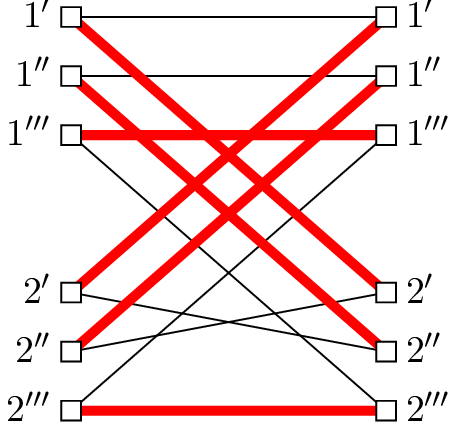, scale=0.45}}
    \quad\quad
    \subfigure[]
              {\epsfig{file=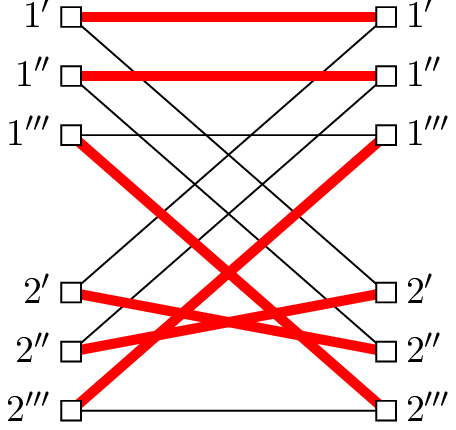, scale=0.45}}
    \quad\quad
    \subfigure[]
              {\epsfig{file=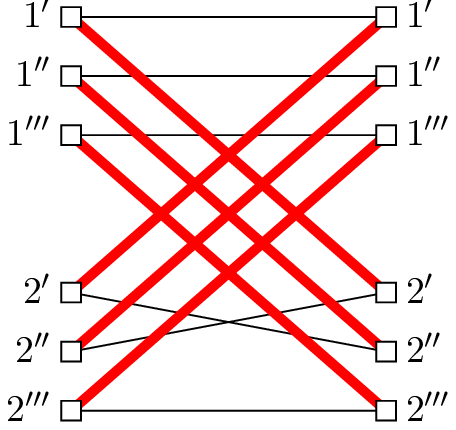, scale=0.45}}
  \end{center}
  \caption{The four perfect matchings of the triple cover in
    Fig.~\ref{fig:gc:graph:cover:interpretation:2:by:2:case:1}(c).}
  \label{fig:gc:graph:cover:interpretation:2:by:2:case:1:2}
\end{figure*}

\begin{Example}
  \label{example:degree:2:Bethe:permanent:all:one:matrix:1}

  Let $n = 2$, $M = 2$, and $\theta_{i,j} = 1$, $(i,j) \in \setI \times
  \setJ$. We make the following observations.
  \begin{itemize}
    
  \item The average
    in~\eqref{eq:degree:M:Bethe:permanent:simplified:average:1} is over $2! =
    2$ matrices, namely over
    \begin{align*}
      \hskip-0.25cm
      \matrtheta^{\uparrow (1)}
        & \defeq
            \left(
              \begin{array}{cc|cc}
                1 \! & 0 \! & 1 \! & 0 \\
                0 \! & 1 \! & 0 \! & 1 \\
                \hline
                1 \! & 0 \! & 1 \! & 0 \\
                0 \! & 1 \! & 0 \! & 1
              \end{array}
            \right), \quad
      \matrtheta^{\uparrow (2)}
         \defeq
           \left(
             \begin{array}{cc|cc}
               1 \! & 0 \! & 1 \! & 0 \\
               0 \! & 1 \! & 0 \! & 1 \\
               \hline
               1 \! & 0 \! & 0 \! & 1 \\
               0 \! & 1 \! & 1 \! & 0
             \end{array}
           \right).
    \end{align*}

  \item The matrix $\matrtheta^{\uparrow (1)}$ corresponds to the double cover
    of $\graphN(\matrtheta)$ shown in
    Fig.~\ref{fig:gc:2:by:2:case:degree:2:1}(d). Because that graph has $4$
    perfect matchings,
    see Figs.~\ref{fig:gc:2:by:2:case:degree:2:1}(e)--(h), we have
    \begin{align*}
      \perm(\matrtheta^{\uparrow (1)}) = 4.
    \end{align*}

  \item The matrix $\matrtheta^{\uparrow (2)}$ corresponds to the double cover
    of $\graphN(\matrtheta)$ shown in
    Fig.~\ref{fig:gc:2:by:2:case:degree:2:1}(i). Because that graph has $2$
    perfect matchings,
    see Figs.~\ref{fig:gc:2:by:2:case:degree:2:1}(j)--(k), we have
    \begin{align*}
      \perm(\matrtheta^{\uparrow (1)}) = 2.
    \end{align*}

  \end{itemize}
  Putting everything together, we obtain the degree-$2$ Bethe permanent of
  $\matrtheta$, \ie,
  \begin{align*}
    \permBetheM{2}(\matrtheta)
      &= \sqrt[2]{\frac{1}{2!}
                  \cdot
                  \left(
                    4
                    +
                    2
                  \right)}
       = \sqrt[2]{\frac{1}{2!}
                  \cdot
                  6
                 }
       = \sqrt[2]{3}
       \approx
         1.732.
  \end{align*}
  We note that the graph in Fig.~\ref{fig:gc:2:by:2:case:degree:2:1}(d)
  consists of $M$ independent copies of the graph in
  Fig.~\ref{fig:gc:2:by:2:case:degree:2:1}(a), therefore it is not
  surprising that $\perm(\matrtheta^{\uparrow (1)}) = \perm(\matrtheta)^M =
  2^2 = 4$. On the other hand, the graph in
  Fig.~\ref{fig:gc:2:by:2:case:degree:2:1}(i) consists of $M$
  \textbf{coupled} copies of the graph in
  Fig.~\ref{fig:gc:2:by:2:case:degree:2:1}(a), which implies that we cannot
  choose the perfect matchings independently. Therefore, it is not surprising
  that we have $\perm(\matrtheta^{\uparrow (2)}) \neq \perm(\matrtheta)^M =
  2^2 = 4$, which finally results in $\permBetheM{2}(\matrtheta) \neq
  \perm(\matrtheta)$. Nevertheless, these considerations also show why
  $\permBetheM{2}(\matrtheta)$ is not too different from $\perm(\matrtheta)$.
  \exampleend
\end{Example}

\begin{Example}
  \label{example:degree:3:Bethe:permanent:all:one:matrix:1}

  Let $n = 2$, $M = 3$, and $\theta_{i,j} = 1$, $(i,j) \in \setI \times
  \setJ$. The average
  in~\eqref{eq:degree:M:Bethe:permanent:simplified:average:1} is over $3! = 6$
  matrices. These matrices correspond to the triple covers of
  $\graphN(\matrtheta)$ shown in
  Fig.~\ref{fig:gc:graph:cover:interpretation:2:by:2:case:1}(b)--(g).
  Computing the number of perfect matchings for each of these cases, we obtain
  \begin{align*}
    \permBetheM{3}(\matrtheta)
      &= \sqrt[3]{\frac{1}{3!}
                  \cdot
                  \left(
                    8
                    +
                    4
                    +
                    4
                    +
                    4
                    +
                    2
                    +
                    2
                  \right)} \\
      &= \sqrt[3]{\frac{1}{3!}
                  \cdot
                  24
                 }
       = \sqrt[3]{4}
       \approx
         1.587.
  \end{align*}
  In particular, for the triple cover in
  Fig.~\ref{fig:gc:graph:cover:interpretation:2:by:2:case:1}(c) we show its
  $4$ perfect matchings explicitly in
  Fig.~\ref{fig:gc:graph:cover:interpretation:2:by:2:case:1:2}.

  Overall, we can make similar observations as at the end of
  Example~\ref{example:degree:2:Bethe:permanent:all:one:matrix:1} concerning
  the \textbf{coupling} of the $M$ copies of $\graphN(\matrtheta)$ that make
  up a degree-$M$ cover and its influence on the number of perfect matchings.
  \exampleend
\end{Example}

\begin{Example}
  \label{example:degree:M:Bethe:permanent:all:one:matrix:1}

  Let $n = 2$, $M \in \Zpp$, and $\theta_{i,j} = 1$, $(i,j) \in \setI \times
  \setJ$. The average
  in~\eqref{eq:degree:M:Bethe:permanent:simplified:average:1} is over $M!$
  matrices that correspond to the $M$-covers of $\graphN(\matrtheta)$. For
  each of these matrices, their permanent equals the number of perfect
  matchings in the corresponding $M$-cover. We make the following observations
  (see
  Figs.~\ref{fig:gc:2:by:2:case:degree:2:1}--\ref{fig:gc:graph:cover:interpretation:2:by:2:case:1:2}
  for illustrations for the cases $M = 2$ and $M = 3$).
  \begin{itemize}

  \item Every $M$-cover consists of up to $M$ cycles.

  \item Every cycle supports two perfect matchings (independently of the cycle
    length and independently of the perfect matchings chosen on the rest of the
    graph).

  \end{itemize}
  Therefore, if an $M$-cover has $c$ cycles then it has $2^c$ perfect
  matchings. The average
  in~\eqref{eq:degree:M:Bethe:permanent:simplified:average:1} can then be
  evaluated with suitable combinatorial tools, for example by using the
  so-called cycle index of the symmetric group over $M$ elements (see, \eg,
  \cite{Biggs:89:1}), and we obtain
  \begin{align*}
    \permBetheM{M}(\matrtheta)
      &= \sqrt[M]{M+1}.
  \end{align*}
  Therefore, in the limit $M \to \infty$, we get
  \begin{align*}
    \permBethe(\matrtheta)
      &= \limsup_{M \to \infty} \
           \permBetheM{M}(\matrtheta)
       = 1.
  \end{align*}
  This confirms the result for $\permBethe(\matrtheta)$ in
  Example~\ref{example:permanent:Bethe:permanent:all:one:matrix:2:1}, which
  was obtained by analytical means.
  \exampleend
\end{Example}

\subsection{The Degree-$M$ Bethe Permanent for Matrices of 
                    Size $2 \times 2$ --- General Non-Negative Matrix}
\label{sec:degree:M:Bethe:permanent:n:2:1:general:matrix}

In this subsection we consider the cases $M = 2$, $M = 3$, and general $M$ for
the general non-negative matrix
\begin{align*}
  \matrtheta
    &= \begin{pmatrix}
         \theta_{1,1} & \theta_{1,2} \\
         \theta_{2,1} & \theta_{2,2}
       \end{pmatrix}.
\end{align*}
A particular goal of this subsection is to compare the degree-$M$ Bethe
permanent of $\matrtheta$ with the permanent of $\matrtheta$. In fact, as we
will see, for every considered case in this subsection we have
$\permBetheM{M}(\matrtheta) \leq \perm(\matrtheta)$.

\begin{Example}
  Let $n = 2$ and $M = 2$. We perform similar computations as in
  Example~\ref{example:degree:2:Bethe:permanent:all:one:matrix:1}, but for a
  general non-negative matrix $\matrtheta$. Towards computing
  $\permBetheM{2}(\matrtheta)$ as given
  in~\eqref{eq:degree:M:Bethe:permanent:simplified:average:1}, we make the
  following observations.
  \begin{itemize}

    \item The average
    in~\eqref{eq:degree:M:Bethe:permanent:simplified:average:1} is over $2! =
    2$ matrices, namely over
    \begin{align*}
      \hskip-0.25cm
      \matrtheta^{\uparrow (1)}
        & \defeq
            \left(
              \begin{array}{cc|cc}
                \theta_{1,1} \! & 0 \! & \theta_{1,2} \! & 0 \\
                0 \! & \theta_{1,1} \! & 0 \! & \theta_{1,2} \\
                \hline
                \theta_{2,1} \! & 0 \! & \theta_{2,2} \! & 0 \\
                0 \! & \theta_{2,1} \! & 0 \! & \theta_{2,2}
              \end{array}
            \right), \\
      \matrtheta^{\uparrow (2)}
        &\defeq
           \left(
             \begin{array}{cc|cc}
                \theta_{1,1} \! & 0 \! & \theta_{1,2} \! & 0 \\
                0 \! & \theta_{1,1} \! & 0 \! & \theta_{1,2} \\
                \hline
                \theta_{2,1} \! & 0 \! & 0 \! & \theta_{2,2} \\
                0 \! & \theta_{2,1} \! & \theta_{2,2} \! & 0
              \end{array}
           \right).
    \end{align*}

  \item We obtain
    \begin{align*}
      \perm\big( \matrtheta^{\uparrow (1)} \big)
        &= (\theta_{1,1} \theta_{2,2} + \theta_{1,2} \theta_{2,1})^2 \\
        &= \theta_{1,1}^2 \theta_{2,2}^2 
           + 
           2\theta_{1,1} \theta_{1,2} \theta_{2,1} \theta_{2,2}
           +
           \theta_{1,2}^2 \theta_{2,1}^2.
    \end{align*}
    Note that the coefficients add up to $4$ because $\matrtheta^{\uparrow
      (1)}$ corresponds to the double cover of $\graphN(\matrtheta)$ shown in
    Fig.~\ref{fig:gc:2:by:2:case:degree:2:1}(d), which admits $4$ (weighted)
    perfect matchings.

  \item We obtain
    \begin{align*}
      \perm\big( \matrtheta^{\uparrow (2)} \big)
        &= \theta_{1,1}^2 \theta_{2,2}^2 
           +
           \theta_{1,2}^2 \theta_{2,1}^2.
    \end{align*}
    Note that the coefficients add up to $2$ because $\matrtheta^{\uparrow
      (2)}$ corresponds to the double cover of $\graphN(\matrtheta)$ shown in
    Fig.~\ref{fig:gc:2:by:2:case:degree:2:1}(i), which admits $2$ (weighted)
    perfect matchings.

  \end{itemize}
  Putting everything together, we obtain for the square of the degree-$2$
  Bethe partition function of $\matrtheta$
  \begin{align*}
    \big( \permBetheM{2}(\matrtheta) \big)^2
      &= \frac{1}{2}
         \cdot
           \big(
             \perm(\matrtheta^{\uparrow (1)})
             +
             \perm(\matrtheta^{\uparrow (2)})
           \big) \\
      &= \theta_{1,1}^2 \theta_{2,2}^2 
         + 
         \theta_{1,1} \theta_{1,2} \theta_{2,1} \theta_{2,2}
         +
         \theta_{1,2}^2 \theta_{2,1}^2.
  \end{align*}
  Given the observations that
  \begin{align*}
    \perm\big( \matrtheta^{\uparrow (1)} \big)
      &\leq \big( \perm(\matrtheta) \big)^2, \\
    \perm\big( \matrtheta^{\uparrow (2)} \big)
      &\leq \big( \perm(\matrtheta) \big)^2,
  \end{align*}
  it is not surprising that we also have the inequality
  \begin{align*}
    \big( \permBetheM{2}(\matrtheta) \big)^2
      &\leq
         \big( \perm(\matrtheta) \big)^2,
  \end{align*}
  \ie,
  \begin{align*}
    \permBetheM{2}(\matrtheta)
      &\leq
         \perm(\matrtheta).
  \end{align*}
  \exampleend
\end{Example}

\begin{Example}
  Let $n = 2$ and $M = 3$. We perform similar computations as in
  Example~\ref{example:degree:3:Bethe:permanent:all:one:matrix:1}, but for a
  general non-negative matrix $\matrtheta$. Towards computing
  $\permBetheM{3}(\matrtheta)$ as given
  in~\eqref{eq:degree:M:Bethe:permanent:simplified:average:1}, we make the
  following observations.
  \begin{itemize}

  \item The average
    in~\eqref{eq:degree:M:Bethe:permanent:simplified:average:1} is over $3! =
    6$ matrices. These matrices correspond to the triple covers of
    $\graphN(\matrtheta)$ shown in
    Fig.~\ref{fig:gc:graph:cover:interpretation:2:by:2:case:1}(b)--(g).

  \item For example, for the matrix $\matrtheta^{\uparrow (2)}$ corresponding
    to the triple cover in
    Fig.~\ref{fig:gc:graph:cover:interpretation:2:by:2:case:1}(c), we obtain
    \begin{align*}
      \perm\big( \matrtheta^{\uparrow (2)} \big)
        &= \theta_{1,1}^3 \theta_{2,2}^3
           +
           \theta_{1,1}^1 \theta_{1,2}^2 \theta_{2,1}^2 \theta_{2,2}^1 \\
        &\quad
           +
           \theta_{1,1}^2 \theta_{1,2}^1 \theta_{2,1}^1 \theta_{2,2}^2
           +
           \theta_{1,2}^3 \theta_{2,1}^3,
    \end{align*}
    where each (weighted) perfect matching in
    Fig.~\ref{fig:gc:graph:cover:interpretation:2:by:2:case:1:2} contributes
    one monomial to the above expression. One can verify that
    \begin{align*}
      \perm\big( \matrtheta^{\uparrow (2)} \big)
        &= \big(
             \theta_{1,1}^2 \theta_{2,2}^2 \! + \! \theta_{1,2}^2 \theta_{2,1}^2
           \big)
           \! \cdot \!
           (\theta_{1,1} \theta_{2,2} \! + \! \theta_{1,2} \theta_{2,1}) \\
        &\leq
           \big(
             \theta_{1,1} \theta_{2,2} \! + \! \theta_{1,2} \theta_{2,1}
           \big)^2
           \! \! \cdot
           (\theta_{1,1} \theta_{2,2} \! + \! \theta_{1,2} \theta_{2,1}) \\
        &= \big(
             \theta_{1,1} \theta_{2,2} \! + \! \theta_{1,2} \theta_{2,1}
           \big)^3  \\
        &= \big(
             \perm(\matrtheta)
           \big)^3.
    \end{align*}
    (The product expression in the first line is not surprising given the fact
    that graph in
    Fig.~\ref{fig:gc:graph:cover:interpretation:2:by:2:case:1}(c) contains
    two independent components, each contributing one factor to the above
    product.)
  \end{itemize}
  Similar observations can be made for the other five triple covers in
  Fig.~\ref{fig:gc:graph:cover:interpretation:2:by:2:case:1}(b)--(g), and so
  we obtain
  \begin{align*}
    \big( \permBetheM{3}(\matrtheta) \big)^3
      &\leq
         \big( \perm(\matrtheta) \big)^3,
  \end{align*}
  \ie,
  \begin{align*}
    \permBetheM{3}(\matrtheta)
      &\leq
         \perm(\matrtheta).
  \end{align*}
  \exampleend
\end{Example}

\begin{Example}
  Let $n = 2$ and $M \in \Zpp$. We perform similar computations as in
  Example~\ref{example:degree:M:Bethe:permanent:all:one:matrix:1}, but for a
  general non-negative matrix $\matrtheta$. The observations that we made
  there can be generalized (beyond the all-one matrix), and we obtain
  \begin{align*}
    \big( \permBetheM{M}(\matrtheta) \big)^M
      &= \sum_{\ell=0}^{M}
           (\theta_{1,1} \theta_{2,2})^{M-\ell}
           (\theta_{1,2} \theta_{2,1})^{\ell}.
  \end{align*}
  Because
  \begin{align*}
    \big( \perm(\matrtheta) \big)^M
      &= \sum_{\ell=0}^{M}
           {M \choose \ell}
           (\theta_{1,1} \theta_{2,2})^{M-\ell}
           (\theta_{1,2} \theta_{2,1})^{\ell},
  \end{align*}
  we see that
  \begin{align*}
    \big( \permBetheM{M}(\matrtheta) \big)^M
      &\leq
         \big( \perm(\matrtheta) \big)^M,
  \end{align*}
  \ie,
  \begin{align*}
    \permBetheM{M}(\matrtheta)
      &\leq
         \perm(\matrtheta).
  \end{align*}
  Moreover, in the limit $M \to \infty$, we have
  \begin{align*}
    \permBethe(\matrtheta)
      &= \limsup_{M \to \infty} \
           \permBetheM{M}(\matrtheta) \\
      &= \max(
              \theta_{1,1} \theta_{2,2}, \ 
              \theta_{2,1} \theta_{1,2}
             ).
  \end{align*}
  This confirms the result for $\permBethe(\matrtheta)$ in
  Lemma~\ref{lemma:permanent:Bethe:permanent:n:2:1}, which was obtained by
  analytical means.
  \exampleend
\end{Example}

For $n > 2$, we leave it as an open problem to obtain an ``explicit
expression'' for $\permBetheM{M}(\matrtheta)$, $M \in \Zpp$, either for the
all-one matrix case, or for the general non-negative matrix case.

In conclusion, the above examples shows that in general
$\permBethe(\matrtheta) \neq \perm(\matrtheta)$, however, they also show that
the Bethe permanent has the potential to give reasonably good estimates, in
particular in the cases where the ``coupling effect'' in the average graph
cover is not too strong. Heuristically, this ``coupling effect'' seems
actually to be the worst for $n = 2$ and to become weaker the larger $n$ is.

\subsection{Relevance of Finite Graph Covers}
\label{sec:relevance:finite:graph:covers:1}

If the NFG $\graphN(\matrtheta)$ had no cycles then the SPA could be used to
exactly compute the partition function. Namely, after a finite number of
iterations, the SPA would reach a fixed point and the partition function
$\ZGibbs\big( \graphN(\matrtheta) \big) = \perm(\matrtheta)$ could be computed
with the help of an expression like $\exp \big( - \FpdualBethe \big( \{
\udLambdaleftijt \}, \{ \udLambdarightijt \} \big) \big)$, where
$\FpdualBethe$ is defined in
Lemma~\ref{lemma:pseudo:dual:Bethe:free:energy:1}. However,
$\graphN(\matrtheta)$ has cycles: the use of this expression at a fixed point
of the SPA is still possible but usually it does not yield the correct
partition function. In this subsection, we would like to better understand the
source of this suboptimality.

To that end, observe that the SPA is an algorithm that processes information
locally on $\graphN(\matrtheta)$, \ie, messages are sent along edges, function
nodes take incoming messages from incident edges, do some computations, and
send out new messages along the incident edges. On the one hand, this locality
explains the main strengths of the SPA, namely its low complexity and its
parallelizability, two key factors for making the SPA a popular algorithm. On
the other hand, this locality explains also the main weakness of the
SPA. Namely, a locally operating like SPA ``cannot distinguish'' if it is
operating on $\graphN(\matrtheta)$ or any of its
covers~\cite{Koetter:Vontobel:03:1, Vontobel:Koetter:05:1:subm,
  Vontobel:10:7:subm}.

More precisely, let $\cgraph{N}$ be an $M$-cover $\cgraph{N}$ of
$\graphN(\matrtheta)$. Such an $M$-cover ``looks locally the same'' as
$\graphN(\matrtheta)$ in the sense that the local structure of $\cgraph{N}$ is
exactly the same as the one of $\graphN(\matrtheta)$. (Of course, globally
$\cgraph{N}$ and $\graphN(\matrtheta)$ are different because the former NFG
contains $M$ times as many function nodes and $M$ times as many edges.)
Consequently, if the SPA is run on $\cgraph{N}$ with the same initialization
as the SPA on $\graphN(\matrtheta)$ (every initial message is replicated $M$
times), we observe that, because both graphs look locally the same and because
the SPA is a locally operating algorithm, after every iteration the messages
on $\cgraph{N}$ are exactly the same as the messages on $\graphN(\matrtheta)$,
simply replicated $M$ times. In that sense, the SPA ``cannot distinguish'' if
it is operating on $\graphN(\matrtheta)$, or, implicitly, on $\cgraph{N}$, or
any other $M$-cover of $\graphN(\matrtheta)$. This observation allows us to
give the following interpretation of~\eqref{eq:degree:M:Bethe:permanent:1}
(which is reproduced here for the ease of reference)
\begin{align}
  \permBetheM{M}(\matrtheta)
    &\defeq
       \sqrt[M]{\Big\langle \!
                  \perm
                    \left(
                      \matrtheta^{\uparrow\cmatrP}
                    \right) 
                \! \Big\rangle_{\cmatrP \in \csetPsi_M}}.
                     \label{eq:degree:M:Bethe:permanent:1:copy:2}
\end{align}
Namely, because the SPA implicitly tries to compute in parallel the partition
function $\ZGibbs\big( \graphN(\matrtheta^{\uparrow\cmatrP}) \big) =
\perm(\matrtheta^{\uparrow\cmatrP})$ for all $M$-covers of
$\graphN(\matrtheta)$, yet it has to give back one real number only, the
``best it can do'' is to give back the average of these partition functions,
\ie, $\big\langle \perm \left( \matrtheta^{\uparrow\cmatrP} \right)
\big\rangle_{\cmatrP \in \csetPsi_M}$. (The $M$th root that appears
in~\eqref{eq:degree:M:Bethe:permanent:1:copy:2} is included so that the result
is properly normalized w.r.t.\ $\ZGibbs\big( \graphN(\matrtheta) \big) =
\perm(\matrtheta)$.)

Let us conclude this subsection by commenting on two recent papers.
\begin{itemize}

\item Translating the results of a paper by Greenhill, Janson, and
  Ruci{\'n}ski~\cite{Greenhill:Janson:Rucinski:10:1} to graphical models, it
  turns out that the authors compute a high-order approximation to the
  quantity $\big\langle \ZGibbs(\cgraph{N}') \big\rangle_{\cgraph{N}' \in
    \cset{N}'_{M}}$ for some NFG $\graphN'(\matrtheta)$ with
  $\ZGibbs(\graphN'(\matrtheta)) = \perm(\matrtheta)$. The NFG
  $\graphN'(\matrtheta)$ is in general different from $\graphN(\matrtheta)$,
  where the latter NFG was specified in
  Definition~\ref{def:permanent:normal:factor:graph:1}. We will elaborate on
  this interesting connection in
  Section~\ref{sec:connections:to:results:by:greenhill:janson:rucinsky:1}.

\item The paper~\cite{Barvinok:10:1} by Barvinok presents bounds on the number
  of zero/one matrices with prescribed row and column sums. (As already
  mentioned in Section~\ref{sec:related:work:1}, in statistical physics terms
  the approach taken therein can be considered as a mean-field approach.) In
  terms of NFGs, the quantity of interest is expressed as the partition
  function of an NFG that has the same topology as $\graphN(\matrtheta)$ but
  different function nodes.

  Section~3.1 of~\cite{Barvinok:10:1} then presents an interpretation of these
  bounds that has a similar flavor of the graph cover interpretation of the
  Bethe permanent, however, it also has stark differences. Namely, in terms of
  NFGs, Section~3.1 of \cite{Barvinok:10:1} presents an NFG where every
  function node of the base graph is replicated $M$ times and every edge is
  replicated $M^2$ times, \ie, all $Mn$ left-hand side function nodes are
  connected by exactly one edge to all the $Mn$ right-hand side function
  nodes. In order for this to make sense, the local functions are adapted so
  that they have $Mn$ arguments instead of $n$ arguments. It is then shown
  that the $M^2$th root of the partition function of this new NFG, $M \to
  \infty$, yields the relevant number in which the bounds are
  expressed. Despite all the similarities, the differences to finite graph
  covers are clear:
  \begin{itemize}
  
  \item There is only one such $M$-fold version of the base graph, whereas the
    number of $M$-covers of $\graphN(\matrtheta)$ is $(M!)^{(n^2)}$.
  
  \item The number of edges is $M^2n^2$, whereas the number of edges in an
    $M$-cover of $\graphN(\matrtheta)$ is $Mn^2$.
  
  \item The local functions need to be adapted in order to allow for $Mn$
    instead of $n$ arguments, whereas the local functions of an $M$-cover of
    $\graphN(\matrtheta)$ are the same as the local functions of
    $\graphN(\matrtheta)$.
  
  \end{itemize}

\end{itemize}

\newpage

\section{The Relationship between the Permanent \\
               and the Bethe Permanent}
\label{sec:permanent:Bethe:permanent:relationship:1}

In this section we explore the relationship between $\perm(\matrtheta)$ and
$\permBethe(\matrtheta)$, in particular, if and how $\perm(\matrtheta)$ can be
upper and lower bounded by expressions that are functions of
$\permBethe(\matrtheta)$. For an additional/complementary discussion on this
topic we refer to~\cite{Yedidia:Chertkov:11:1:subm}.

We start with a lemma that shows that there are non-negative square matrices
for which the Bethe permanent can give rather accurate estimates of the
permanent, thereby showing the overall potential of the Bethe permanent to be
the basis for good upper and lower bounds on the permanent of general
non-negative square matrices.

\begin{Lemma}
  \label{lemma:perm:Bethe:for:all:one:matrix:1}

  Let $\matrallone{n}$ be the all-one matrix of size $n \times n$. Then
  \begin{align*}
    \frac{\perm(\matrallone{n})}
         {\permBethe(\matrallone{n})}
      &= \sqrt{\frac{2 \pi n}{\e}}
           \cdot
           \big(
             1+o(1)
           \big),
  \end{align*}
  where $o(1)$ is w.r.t.~$n$.
\end{Lemma}

\begin{Proof}
  See Appendix~\ref{sec:proof:lemma:perm:Bethe:for:all:one:matrix:1}.
\end{Proof}

\mbox{}

Although the factor $\sqrt{2 \pi n / \e}$ is non-negligible, compared to
$\perm(\matrallone{n}) = n!$ it is rather small.

\subsection{Lower Bounds on the Permanent of the Matrix $\matrtheta$}

In this subsection we study lower bounds on $\perm(\matrtheta)$ based on
$\permBethe(\matrtheta)$.

\begin{Theorem}[Gurvits~\cite{Gurvits:11:1, Gurvits:11:2}]
  \label{theorem:Bethe:permanent:upper:bound:1}

  It holds that
  \begin{align*}
    \frac{\perm(\matrtheta)}
         {\permBethe(\matrtheta)}
      &\geq
         1.
  \end{align*}
\end{Theorem}

\begin{Proof}
  This result was recently shown by Gurvits~\cite{Gurvits:11:1,
    Gurvits:11:2}. Roughly speaking, its elegant proof is based on first
  expressing $\matrtheta$ in terms of a stationary point of
  $\FBethesub{\graphN(\matrtheta)}$ and then applying an inequality due to
  Schrijver~\cite{Schrijver:98:1}.
\end{Proof}

For more details, along with a discussion of this result's relationship to the
results in~\cite{Gurvits:08:1, Laurent:Schrijver:10:1}, we refer
to~\cite{Gurvits:11:1, Gurvits:11:2}.  For a somewhat different approach to
proving this theorem, we refer the interested reader
to~\cite{Yedidia:Chertkov:11:1:subm}.

\begin{Corollary}[Gurvits~\cite{Gurvits:11:1, Gurvits:11:2}]
  \label{cor:theorem:Bethe:permanent:upper:bound:1}

  For any $\matrgamma \in \setGamma{n}$ it holds that
  \begin{align*}
    \frac{\perm(\matrtheta)}
         {\exp\big( -\FBethesub{\graphN(\matrtheta)}(\matrgamma) \big)}
      &\geq
         1.
  \end{align*}
\end{Corollary}

\begin{Proof}
  This is a straightforward consequence of
  Theorem~\ref{theorem:Bethe:permanent:upper:bound:1} and
  Definitions~\ref{def:Bethe:partition:function:1}
  and~\ref{def:Bethe:permanent:1}.
\end{Proof}

\mbox{}

\noindent
Some comments on Theorem~\ref{theorem:Bethe:permanent:upper:bound:1} and
Corollary~\ref{cor:theorem:Bethe:permanent:upper:bound:1}:
\begin{itemize}

\item Corollary~\ref{cor:theorem:Bethe:permanent:upper:bound:1} has its
  significance when one is not willing to run the SPA algorithm, but one has a
  reasonably good estimate of the $\matrgamma \in \setGamma{n}$ that minimizes
  $\FBethesub{\graphN(\matrtheta)}$. This approach is for example interesting
  when one wants to obtain analytical lower bounds on the permanent of some
  parameterized class of non-negative square matrices.

\item Chertkov, Kroc, and Vergassola~\cite{Chertkov:Kroc:Vergassola:08:1}
  observed in 2008 that $\perm(\matrtheta) \geq \permBethe(\matrtheta)$ holds
  for all the matrices that they experimented with. They also outlined a
  potential approach to proving this inequality via the loop calculus
  technique by Chertkov and Chernyak~\cite{Chertkov:Chernyak:06:1}, which in
  the case of $\graphN(\matrtheta)$ states that $\perm(\matrtheta)$ equals
  $\permBethe(\matrtheta)$ plus certain correction terms
  (see~\cite{Forney:Vontobel:11:1} for a reformulation of the loop calculus in
  terms of NFGs). However, given the fact that for $\graphN(\matrtheta)$ these
  correction terms happen to be positive \emph{and} negative, it is at present
  unclear if Theorem~\ref{theorem:Bethe:permanent:upper:bound:1} can be proven
  with this technique.

\item In the Allerton 2010 version of this paper we stated the inequality that
  appears in Theorem~\ref{theorem:Bethe:permanent:upper:bound:1} as a
  theorem. However, while writing the present paper we realized that our
  ``proof'' had a flaw, which, so far, we have not been able to
  fix. Nevertheless, we still think that our proof strategy can work out and
  possibly give an alternative viewpoint of Schrijver's inequality that
  features prominently in~\cite{Gurvits:11:1, Gurvits:11:2}. In that respect,
  we list below some special cases of matrices $\matrtheta$ for which our
  proof strategy works, along with conjectures that, if true, would give an
  alternative proof of Theorem~\ref{theorem:Bethe:permanent:upper:bound:1} in
  its full generality.

\end{itemize}

\begin{Conjecture}
  \label{conj:perm:matrix:lifiting:bounds:1}

  For any $M \in \Zpp$ it holds that
  \begin{align*}
    \Big\langle \!
      \perm
      \left(
        \matrtheta^{\uparrow\cmatrP}
      \right) 
      \! \Big\rangle_{\cmatrP \in \csetPsi_M}
        &\leq
           \big(
             \perm(\matrtheta)
           \big)^M.
  \end{align*}
  Possibly also the following, stronger, statement is true: for any $M \in
  \Zpp$ and any $\cmatrP \in \csetPsi_M$ it holds that
  \begin{align*}
    \perm
      \left(
        \matrtheta^{\uparrow\cmatrP}
      \right)
        &\leq
           \big(
             \perm(\matrtheta)
           \big)^M.
  \end{align*}
  \conjectureend
\end{Conjecture}

\noindent
Theorem~\ref{theorem:Bethe:permanent:upper:bound:1} would then follow from
\begin{align*}
  \permBethe(\matrtheta)
    &\onestareq
       \limsup_{M \to \infty} \ 
         \permBetheM{M}(\matrtheta) \\
    &\twostarseq
       \limsup_{M \to \infty} \
         \sqrt[M]{\Big\langle \!
                    \perm
                      \left(
                        \matrtheta^{\uparrow\cmatrP}
                      \right) 
                  \! \Big\rangle_{\cmatrP \in \csetPsi_M}} \\
    &\threestarsleq
       \limsup_{M \to \infty} \
         \sqrt[M]{\perm(\matrtheta)^M} \\
    &= \limsup_{M \to \infty} \
          \perm(\matrtheta) \\
    &\fourstarseq
       \perm(\matrtheta),
\end{align*}
where at step~$\onestar$ we have used
Theorem~\ref{theorem:Behte:permanent:degree:M:partition:function:1}, where at
step~$\twostars$ we have used
Definition~\ref{def:Bethe:permanent:degree:M:partition:function:2}, where at
step~$\threestars$ we have used the weaker part of
Conjecture~\ref{conj:perm:matrix:lifiting:bounds:1}, and where
step~$\fourstars$ follows from evaluating the (now trivial) limit $M \to
\infty$.

We now list some special matrices $\matrtheta$ for which
Conjecture~\ref{conj:perm:matrix:lifiting:bounds:1} is true.

\begin{itemize}

\item Conjecture~\ref{conj:perm:matrix:lifiting:bounds:1} is true for
  $\matrtheta = \matrallone{n}$. (The proof is given in
  Appendix~\ref{sec:proof:cor:Bethe:permanent:all:one:matrix:bound:1}.)

\item Conjecture~\ref{conj:perm:matrix:lifiting:bounds:1} is true for all
  matrices $\matrtheta$ that were studied in
  Section~\ref{sec:finite:graph:cover:interpretation:Bethe:permanent:1}.

\end{itemize}

Actually, the results in
Section~\ref{sec:finite:graph:cover:interpretation:Bethe:permanent:1} suggest
the following, stronger version of
Conjecture~\ref{conj:perm:matrix:lifiting:bounds:1}.

\begin{Conjecture}
  \label{conj:perm:matrix:lifiting:bounds:1:stronger:1}

  Fix some $M \in \Zpp$ and consider the expressions
  \begin{align*}
    \Big\langle \!
      \perm
      \left(
        \matrtheta^{\uparrow\cmatrP}
      \right) 
      \! \Big\rangle_{\cmatrP \in \csetPsi_M}
        &\quad \text{ and } \quad
           \big(
             \perm(\matrtheta)
           \big)^M
  \end{align*}
  as polynomials in the indeterminates $\{ \theta_{i,j} \}_{i,j}$. We
  conjecture that the coefficient of every monomial of the first polynomial is
  upper bounded by the coefficient of the corresponding monomial of the second
  polynomial.

  Possibly also the following, stronger, statement is true. Fix some $M \in
  \Zpp$ and $\cmatrP \in \csetPsi_M$, and consider the expressions
  \begin{align*}
    \perm
      \left(
        \matrtheta^{\uparrow\cmatrP}
      \right)
        &\quad \text{ and } \quad
           \big(
             \perm(\matrtheta)
           \big)^M
  \end{align*}
  as polynomials in the indeterminates $\{ \theta_{i,j} \}_{i,j}$. We
  conjecture that the coefficient of every monomial of the first polynomial is
  upper bounded by the coefficient of the corresponding monomial of the second
  polynomial.
  \conjectureend
\end{Conjecture}

Let us conclude this subsection by noting that the inequalities
$\ZGibbs(\cgraph{N}) \leq \ZGibbs(\graphN)^M$, $M \in \Zpp$, $\cgraph{N} \in
\cset{N}_{M}$, \ie, inequalities of the type that appear in
Conjecture~\ref{conj:perm:matrix:lifiting:bounds:1}, have recently been used
to prove $\ZBethe(\graphN) \leq \ZGibbs(\graphN)$ for graphical models
$\graphN$ appearing in other contexts. We refer the interested reader to
\cite[Example~34 and Lemma~35]{Vontobel:10:7:subm} and~\cite{Ruozzi:12:1} for
details.

\subsection{Upper Bounds on the Permanent of the Matrix $\matrtheta$}

In this subsection we list conjectures and open problems w.r.t.\ upper bounds
on $\perm(\matrtheta)$ based on $\permBethe(\matrtheta)$.

\begin{Conjecture}[Gurvits~\cite{Gurvits:11:1, Gurvits:11:2}]
  \label{conj:Bethe:permanent:lower:and:upper:bound:1}

  Let $\matrtheta$ be an arbitrary non-negative matrix of size $n \times
  n$. For even $n$ it is conjectured that
  \begin{align}
    \frac{\perm(\matrtheta)}
         {\permBethe(\matrtheta)}
      &\leq
         \sqrt{2}^{\,n},
           \label{eq:upper:bound:conjecture:1}
  \end{align}
  with a similar conjecture for odd $n$. Note
  that~\eqref{eq:upper:bound:conjecture:1} holds with equality for the matrix
  $\matrtheta = \matr{I}_{(n/2) \times (n/2)} \otimes \matrallone{2}$, \ie,
  the Kronecker product of an identity matrix of size $(n/2) \times (n/2)$ and
  the all-one matrix of size $2 \times 2$.
  \conjectureend
\end{Conjecture}

We refer the interested reader to~\cite{Gurvits:11:1, Gurvits:11:2} for a
discussion of families of non-negative matrices for which the above conjecture
has been verified.

Note that Conjecture~\ref{conj:Bethe:permanent:lower:and:upper:bound:1}
replaces the conjecture that we made in the Allerton 2010 version of this
paper where, for fixed $n$, the largest ratio $\perm(\matrtheta) /
\permBethe(\matrtheta)$ was thought to be obtained for the all-one matrix of
size $n \times n$.

Besides proving the bound in
Conjecture~\ref{conj:Bethe:permanent:lower:and:upper:bound:1}, it would be
desirable to prove statements of the form
\begin{align*}
  \operatorname{Pr}
    \left\{
      \matrtheta \in \boldsymbol{\Theta}:
      \frac{\perm(\matrtheta)}
         {\permBethe(\matrtheta)}
      \leq \tau
    \right\}
    \geq
      1 - \varepsilon,
\end{align*}
where $\boldsymbol{\Theta}$ is some ensemble of random matrices of size $n
\times n$, where $\tau$ is some positive real number, and where $\varepsilon$
is some small positive number. For example, for the ensemble of $n \times n$
matrices where the matrix entries are chosen uniformly and independently
between $0$ and $1$, we conjecture that $\perm(\matrtheta) /
\permBethe(\matrtheta)$ is, with high probability, upper bounded by the ratio
that appears in Lemma~\ref{lemma:perm:Bethe:for:all:one:matrix:1}. (Note that
this ratio is much smaller than the ratio that appears in
Conjecture~\ref{conj:Bethe:permanent:lower:and:upper:bound:1}.)

\subsection{Closeness of the Permanent to the Bethe Permanent}

In this subsection we list some cases where $\perm(\matrtheta)$ is relatively
close to $\permBethe(\matrtheta)$. We start with an auxiliary result that
relates the Bethe permanent of a lifted matrix to the Bethe permanent of the
base matrix.

\begin{Lemma}
  \label{lemma:Bethe:permanent:of:cover:matrix:1}

  For any $M \in \Zpp$ and any $\cmatrP \in \csetPsi_M$ it holds that
  \begin{align*}
    \permBethe
      \left(
        \matrtheta^{\uparrow\cmatrP}
      \right)
        &= \big(\!
             \permBethe(\matrtheta)
           \big)^M.
  \end{align*}
\end{Lemma}

\begin{Proof}
  See Appendix~\ref{sec:proof:lemma:Bethe:permanent:of:cover:matrix:1}.
\end{Proof}

\begin{Theorem}
  \label{theorem:permanenttypical:lifted:matrix:1}

  For any $\alpha > 1$ and any $M \geq M_{\alpha}$, the majority of the
  matrices in $\bigl\{ \matrtheta^{\uparrow\cmatrP} \bigr\}_{\cmatrP \in
    \csetPsi_M}$ satisfies
  \begin{align*}
    1
      &\leq
         \frac{\perm
                 \left(
                   \matrtheta^{\uparrow\cmatrP}
                 \right)
              }
              {\permBethe
                 \left(
                   \matrtheta^{\uparrow\cmatrP}
                 \right)
              }
       <
         \alpha^M.
  \end{align*}
  Here $M_{\alpha}$ is a parameter that depends on $\alpha$.
\end{Theorem}

\begin{Proof}
  The first inequality follows from
  Theorem~\ref{theorem:Bethe:permanent:upper:bound:1}. We prove the second
  inequality by contradiction. So, assume that there is an $\alpha > 1$ and a
  constant $M_{\alpha}$ such that for all $M \geq M_{\alpha}$ the set
  $\csetPsi'_M \subseteq \csetPsi_M$ of all lifted matrices
  $\matrtheta^{\uparrow\cmatrP}$ that satisfy $\perm \bigl(
  \matrtheta^{\uparrow\cmatrP} \bigr) \geq \alpha^M \cdot \permBethe \bigl(
  \matrtheta^{\uparrow\cmatrP} \bigr)$ has size at least $|\csetPsi_M|/2$. Then
  \begin{align*}
    \permBetheM{M}(\matrtheta)
      &\onestareq
         \sqrt[M]{\Big\langle \!
                    \perm
                      \left(
                        \matrtheta^{\uparrow\cmatrP}
                      \right) 
                  \! \Big\rangle_{\cmatrP \in \csetPsi_M}} \\
      &\twostarseq
         \sqrt[M]{\frac{1}{|\csetPsi_M|}
                  \sum_{\cmatrP \in \csetPsi_M}
                    \perm
                      \left(
                        \matrtheta^{\uparrow\cmatrP}
                      \right)
                 } \\
      &\geq
         \sqrt[M]{\frac{1}{|\csetPsi_M|}
                  \sum_{\cmatrP \in \csetPsi'_M}
                    \perm
                      \left(
                        \matrtheta^{\uparrow\cmatrP}
                      \right)
                 } \\
      &\threestarsgeq
         \sqrt[M]{\frac{1}{|\csetPsi_M|}
                  \sum_{\cmatrP \in \csetPsi'_M}
                    \alpha^M
                    \cdot 
                    \permBethe
                      \left(
                        \matrtheta^{\uparrow\cmatrP}
                      \right)
                 } \\
      &\fourstarseq
         \sqrt[M]{\frac{1}{|\csetPsi_M|}
                  \sum_{\cmatrP \in \csetPsi'_M}
                    \alpha^M
                    \cdot 
                    \big( \permBethe(\matrtheta) \big)^M
                 } \\
      &= \sqrt[M]{\frac{|\csetPsi'_M|}{|\csetPsi_M|}}
         \cdot
         \alpha
         \cdot
         \permBethe(\matrtheta) \\
      &\fivestarsgeq
         2^{-1/M}
         \cdot
         \alpha
         \cdot
         \permBethe(\matrtheta),
  \end{align*}
  where at step~$\onestar$ we have used
  Definition~\ref{def:Bethe:permanent:degree:M:partition:function:2}, where at
  step~$\twostars$ we have replaced the angular brackets by the corresponding
  normalized sum, where at step~$\threestars$ we have used the assumption, where
  at step~$\fourstars$ we have used
  Lemma~\ref{lemma:Bethe:permanent:of:cover:matrix:1}, and where at
  step~$\fivestars$ we have again used the assumption. However, taking
  $\limsup_{M \to \infty}$ on both sides of the above expression, we see that we
  obtain a contradiction w.r.t.\
  Theorem~\ref{theorem:Behte:permanent:degree:M:partition:function:1}.
\end{Proof}

The following example partially corroborates
Theorem~\ref{theorem:permanenttypical:lifted:matrix:1}.

\begin{Example}
  For some positive integer $M$, consider the matrix
  \begin{align*}
    \matrtheta^{\uparrow\cmatrP}
      &= \begin{pmatrix}
           \theta_{1,1} \cmatrI & \theta_{1,2} \cmatrI \\
           \theta_{2,1} \cmatrI & \theta_{2,2} \cmatrP'_{2,2}
         \end{pmatrix},
  \end{align*}
  where $\cmatrI$ is the identity matrix of size $M \times M$ and where
  $\cmatrP'_{2,2}$ is a once cyclically left-shifted identity matrix of size
  $M \times M$. Then
  \begin{align*}
    \perm
      \big(
        \matrtheta^{\uparrow\cmatrP}
      \big)
      &= \theta_{1,1}^M \theta_{2,2}^M
         +
         \theta_{1,2}^M \theta_{2,1}^M, \\
    \permBethe
      \big(
        \matrtheta^{\uparrow\cmatrP}
      \big)
      &= \big(\!
           \permBethe(\matrtheta)
         \big)^M \\
      &= \big(\!
           \max(\theta_{1,1} \theta_{2,2}, \, 
                \theta_{1,2} \theta_{2,1})
         \big)^M,
  \end{align*}
  where the first result is a consequence of the observation that the
  underlying graph has exactly one cycle, \ie, only two perfect matchings, and
  where the second result follows from
  Lemmas~\ref{lemma:permanent:Bethe:permanent:n:2:1}
  and~\ref{lemma:Bethe:permanent:of:cover:matrix:1}. Therefore,
  \begin{align*}
    1
      &\leq
         \frac{\perm
                 \big(
                   \matrtheta^{\uparrow\cmatrP}
                 \big)
              }
              {\permBethe
                 \big(
                   \matrtheta^{\uparrow\cmatrP}
                 \big)
              }
        \leq
          2.
  \end{align*}
  Note that the right-hand side of the above expression does not only grow
  sub-exponentially in $M$, it does not grow at all.
  \exampleend
\end{Example}

Let us conclude this subsection with the following remark. As already
mentioned, the proof of Theorem~\ref{theorem:Bethe:permanent:upper:bound:1}
takes advantage of an inequality by Schrijver~\cite{Schrijver:98:1}, and
therefore the closeness of $\perm(\matrtheta)$ to $\permBethe(\matrtheta)$ is
linked with the tightness of Schrijver's inequality. Now, interestingly
enough, when Schrijver demonstrates a certain asymptotic tightness of his
inequality, see \cite[Section~3]{Schrijver:98:1}, he \emph{implicitly}
evaluates and compares both sides of his inequality for some finite cover of a
certain graph.

\subsection{Open Problems on the Relationship between the Permanent
                    and the Bethe Permanent}

There are also classes of structured matrices for which it would be
interesting to better understand the relationship between the permanent and
the Bethe permanent. For example, the permanent of the matrix
\begin{align*}
  \matrtheta
    &= \begin{pmatrix}
          \alpha_1^{\mu_1} & \alpha_1^{\mu_2} & \cdots & \alpha_1^{\mu_m}
                     & 1          & \cdots & 1 \\ 
          \alpha_2^{\mu_1} & \alpha_2^{\mu_2} & \cdots & \alpha_2^{\mu_m}
                     & 1          & \cdots & 1 \\ 
          \vdots     & \vdots     &        & \vdots    
                     & \vdots     &        & \vdots \\ 
          \alpha_n^{\mu_1} & \alpha_n^{\mu_2} & \cdots & \alpha_n^{\mu_m}
                     & 1          & \cdots & 1
       \end{pmatrix},
\end{align*}
with $0 \leq m \leq n$, real numbers $\alpha_{\ell} \geq 0$, $\ell \in [n]$,
and real numbers $\mu_{\ell}$, $\ell \in [m]$, turns up in a variety of
contexts.
\begin{itemize}

\item When $\sum_{\ell \in [n]} \alpha_{\ell} = 1$ and $\mu_{\ell}$ are
  non-negative integers then $\perm(\matrtheta)$ corresponds to the
  probability of the pattern of a sequence (see, \eg,
  \cite{Viswanathan:09:Talk:1, Vontobel:12:1}).

\item When $m = n$ and $\mu_{\ell} = n - 1 - \ell$, $\ell \in [n]$, then
  $\perm(\matrtheta)$ appears in the analysis of list ordering algorithms
  (see, \eg, \cite{Rivest:76:1}) or in the analysis of source coding
  algorithms (see, \eg, \cite{Sayir:96:1}). Note that in this case,
  $\matrtheta$ is a Vandermonde matrix.

\end{itemize}
Moreover, given the fact that the above $\matrtheta$ depends only on (at most)
$2n$ parameters (and not on $n^2$ parameters as $\matrtheta$
in~\eqref{eq:def:permanent:1}), one wonders if speed-ups in the SPA-based
computation of $\permBethe(\matrtheta)$ are possible.

In some applications one is not interested in the absolute value of the
permanent, only the relative value in the sense that for two matrices
$\matrtheta$ and $\matrtheta'$ one wants to know which one has the larger
permanent. Therefore, for some suitable stochastic setting it would be
desirable to state with what probability $\perm(\matrtheta) \leq
\perm(\matrtheta')$ is equivalent to $\permBethe(\matrtheta) \leq
\permBethe(\matrtheta')$. Some very encouraging initial investigations of this
topic have been presented in~\cite[Section~4.2]{Huang:Jebara:09:1}.

\subsection{Connections to Results by
                       Greenhill, Janson, and Ruci{\'n}ski}

\label{sec:connections:to:results:by:greenhill:janson:rucinsky:1}

After the initial submission of the present paper, we became aware of the
paper by Greenhill, Janson, and
Ruci{\'n}ski~\cite{Greenhill:Janson:Rucinski:10:1} on counting perfect
matchings in random graph covers. Using the findings
of~\cite{Vontobel:10:7:subm} and the present paper, their results can, once
they have been translated to factor graphs, be seen as defining an NFG
$\graphN' \defeq \graphN'(\matrtheta)$ with $\ZGibbs(\graphN') =
\perm(\matrtheta)$ and computing $\ZBethe(\graphN')$, along with approximately
computing $\ZBetheM{M}(\graphN')$. The NFG $\graphN'$ is in general different
from $\graphN \defeq \graphN(\matrtheta)$, where the latter NFG was specified
in Definition~\ref{def:permanent:normal:factor:graph:1} and shown in
Figure~\ref{fig:ffg:permanent:1}.

The advantage of $\graphN'$ is that minimizing its Bethe free energy function
towards determining $\ZBethe(\graphN')$ is quite straightforward. Moreover,
high-order approximations to $\ZBetheM{M}(\graphN')$ can be given. The
disadvantage of $\graphN'$ is that $\ZBethe(\graphN')$ is a weaker lower bound
to $\perm(\matrtheta)$ than $\permBethe(\matrtheta) = \ZBethe(\graphN)$.

\begin{figure}
  \begin{center}
    \subfigure[NFG $\graphN \defeq \graphN(\matrtheta)$.]
      {\epsfig{file=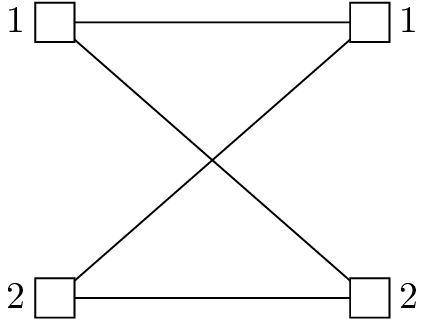, width=0.3\linewidth}}
    \quad
    \quad
    \quad
    \quad
    \subfigure[NFG $\graphN' \defeq \graphN'(\matrtheta)$.]
      {\epsfig{file=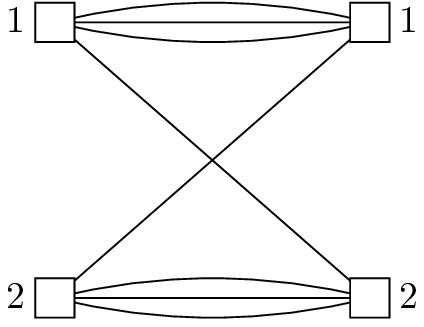, width=0.3\linewidth}}
  \end{center}
  \caption{NFGs used in
    Section~\ref{sec:connections:to:results:by:greenhill:janson:rucinsky:1}.}
  \label{fig:alternative:NFG:1}
\end{figure}

Let us elaborate on these comments. Namely, consider a matrix like
\begin{align}
  \matrtheta
    &\defeq
       \begin{pmatrix}
         3 & 1 \\
         1 & 3
       \end{pmatrix},
         \label{eq:matrtheta:NFG:comparison:1}
\end{align}
where all entries are non-negative integers and where all row and all column
sums are equal to some constant $d$. Here, $n = 2$, $d = 4$, and
$\perm(\matrtheta) = 10$. Its NFG $\graphN \defeq \graphN(\matrtheta)$ as
specified in Definition~\ref{def:permanent:normal:factor:graph:1} is shown in
Figure~\ref{fig:alternative:NFG:1}~(a). In terms of factor graphs, the
paper~\cite{Greenhill:Janson:Rucinski:10:1} considers the NFG $\graphN' \defeq
\graphN'(\matrtheta)$ shown in Figure~\ref{fig:alternative:NFG:1}~(b): like
$\graphN$ it has $n$ function nodes on the left-hand side and $n$ function
nodes on the right-hand side. However, for every $(i,j) \in \setI \times
\setJ$, there are $d \cdot \theta_{i,j}$ edges connecting function node $i$ on
the left-hand side to function node $j$ on the right-hand side. The variable
associated with an edge of $\graphN'$ takes on values in the set $\{ 0, 1
\}$. Moreover, a local function takes on the value $1$ if exactly one of the
variables associated with the incident edges is $1$, and takes on the value
$0$ otherwise. One can show that these definitions yield $Z(\graphN') =
\perm(\matrtheta)$. Indeed, this result follows from observing that valid
configurations of $\graphN'$ correspond to perfect matchings of the graph
underlying $\graphN'$, that the global function value of every valid
configurations of $\graphN'$ is $1$, and that the graph underlying $\graphN'$
has $\perm(\matrtheta)$ perfect matchings.

Note that in the case of $\graphN$, the graph structure is \emph{independent}
of $\matrtheta$ but the local function values \emph{depend} on $\matrtheta$,
whereas in the case of $\graphN'$, the graph structure \emph{depends} on
$\matrtheta$ but the local function node values are \emph{independent} of
$\matrtheta$.

The Bethe free energy function of $\graphN'$ is minimized by $(\bel'_{e,0},
\bel'_{e,1}) = (1 \! - \! 1/d, 1/d)$, $e \in \setE(\graphN')$, with
corresponding beliefs for the function nodes. (This can, \eg, be verified with
the help of symmetry arguments, along with suitably generalizing the convexity
results of Corollary~\ref{cor:convexity:Bethe:free:energy:function:1} from
$\graphN$ to $\graphN'$.) With this, after a few manipulations,
\begin{align}
  \ZBethe(\graphN')
    &= \left(
         \frac{(d-1)^{d-1}}{d^{d-2}}
       \right)^{\!\!n}.
         \label{eq:Z:Bethe:modified:1}
\end{align} 
Interestingly, the expression on the right-hand side
of~\eqref{eq:Z:Bethe:modified:1} appears also in Corollary~1a
in~\cite{Schrijver:98:1}. (One of the main results of Schrijver's
paper~\cite{Schrijver:98:1} is to show that this expression is a lower bound
on $\perm(\matrtheta)$.)

Clearly, the advantage of $\graphN'$ is that we can explicitly compute
$\ZBethe(\graphN')$. However, $\ZBethe(\graphN')$ is a weaker lower bound on
$\perm(\matrtheta)$ than $\permBethe(\matrtheta) = \ZBethe(\graphN)$. (For
example, for the matrix $\matrtheta$ in~\eqref{eq:matrtheta:NFG:comparison:1}
we obtain $\perm(\matrtheta) = 10 \geq \permBethe(\matrtheta) =
\ZBethe(\graphN) = 9 \geq \ZBethe(\graphN') = 729/256 = 2.848\ldots$\, .) This
is not totally surprising given the fact that the right-hand side
of~\eqref{eq:Z:Bethe:modified:1} depends only on $\matrtheta$ inasmuch as
$\matrtheta$ determines $n$ and $d$. Indeed, observing that $\frac{1}{d} \cdot
\matrtheta$ is a doubly stochastic matrix, we get
\begin{align*}
  \log
    &
     \big(
       \ZBethe(\graphN)
     \big) \\
    &\onestargeq
       \big.
         -
         \FBethesub{\graphN}(\matrgamma)
       \big|_{\matrgamma = \frac{1}{d} \cdot \matrtheta} \\
    &\twostarseq
      \big.
         -
         \UBethesub{\graphN}(\matrgamma)
         +
         \HBethesub{\graphN}(\matrgamma)
       \big|_{\matrgamma = \frac{1}{d} \cdot \matrtheta} \\
    &\threestarseq
       \sum_{i,j}
         \frac{\theta_{i,j}}{d}
         \log(\theta_{i,j}) \\
    &\quad
       -
       \sum_{i,j}
         \frac{\theta_{i,j}}{d}
         \log\left( \frac{\theta_{i,j}}{d} \right)
       +
       \sum_{i,j}
         \left( 1 \! - \! \frac{\theta_{i,j}}{d} \right)
         \log\left( 1 \! - \! \frac{\theta_{i,j}}{d} \right) \\
    &= n \log(d)
       +
       \sum_{i,j}
         \left( 1 \! - \! \frac{\theta_{i,j}}{d} \right)
         \log\left( 1 \! - \! \frac{\theta_{i,j}}{d} \right) \\
    &\fourstarseq
       n \log(d)
       +
       \sum_i
         \left(
           \sum_{j=1}^{n}
             u\left( \frac{\theta_{i,j}}{d} \right)
           +
           \sum_{j=n+1}^{\max(n,d)}
             u(0)
         \right) \\
    &\fivestarsgeq
       n \log(d)
       +
       \sum_i
         \left(
           \sum_{j=1}^{d}
             u\left( \frac{1}{d} \right)
           +
           \sum_{j=d+1}^{\max(n,d)}
             u(0)
         \right) \\
     &= n (d \! - \! 1) \log(d \! - \! 1)
        -
        n (d \! - \! 2) \log(d) \\
     &\sixstarseq
        \log
          \big(
            \ZBethe(\graphN')
          \big),
\end{align*}
where at step~$\onestar$ we have used
Definition~\ref{def:Bethe:partition:function:1}, where at steps~$\twostars$
and~$\threestars$ we have used
Lemma~\ref{lemma:Bethe:free:energy:as:function:of:gamma:1}, where at
step~$\fourstars$ we have used the function $u: [0,1] \to \R, \ \xi \mapsto
(1-\xi) \log (1-\xi)$, where at step~$\fivestars$ we have used Karamata's
inequality~\cite{Marshall:Olkin:79:1} (note that $u$ is convex and that, after
sorting, $(\theta_{i,1}/d, \ldots, \theta_{i,n}/d, 0, \ldots, 0)$ majorizes
$(1/d, \ldots, 1/d, 0, \ldots, 0)$), and where at step~$\sixstars$ we have
used~\eqref{eq:Z:Bethe:modified:1}. (See also~\cite[Section~3]{Gurvits:11:2}
for similar inequalities as in the above display equation.)

Interestingly enough, as shown by the authors
of~\cite{Greenhill:Janson:Rucinski:10:1}, for any $M \in \Zpp$ one can give a
high-order approximation of $\big\langle \ZGibbs(\cgraph{N}')
\big\rangle_{\cgraph{N}' \in \cset{N}'_{M}}$, and therefore of the degree-$M$
Bethe partition function~\cite{Vontobel:10:7:subm} $\ZBetheM{M}(\graphN') =
\big( \big\langle \ZGibbs(\cgraph{N}') \big\rangle_{\cgraph{N}' \in
  \cset{N}'_{M}} \big)^{1/M}$. For the corresponding expressions we refer the
interested reader to~\cite{Greenhill:Janson:Rucinski:10:1}.

Near the beginning of this subsection we assumed that $\matrtheta$ is a
non-negative integral matrix where all row and all column sums are equal to
some constant $d$. This is less restrictive than it appears. Namely,
Sinkhorn's theorem states that any positive $n \times n$ matrix $\matrtheta$
can be written as $\matrtheta = \matr{D}_1 \cdot \matrtheta' \cdot \matr{D}_2$
where $\matrtheta'$ is doubly stochastic and where $\matr{D}_1$ and
$\matr{D}_2$ are diagonal matrices with strictly positive diagonal elements
(see, \eg, \cite{Marshall:Olkin:68:1}, which presents also some
generalizations of this statement). If there is a positive integer $d$ such
that $d \cdot \matrtheta'$ has only integral entries, then we can write
$\matrtheta = \frac{1}{d} \cdot \matr{D}_1 \cdot (d \cdot \matrtheta') \cdot
\matr{D}_2$. (If there is no such $d$, then $d$ can be chosen large enough so
that $d \cdot \matrtheta'$ is as close to an integral matrix as desired.) With
this, $\perm(\matrtheta) = \frac{1}{d^n} \cdot \big( \prod_{i \in [n]}
(\matr{D}_1)_{i,i} \big) \cdot \perm(d \cdot \matrtheta') \cdot \big( \prod_{i
  \in [n]} (\matr{D}_2)_{i,i} \big)$, and we have reduced the problem of
(approximately) computing the permanent of $\matrtheta$ to (approximately)
computing the permanent of $d \cdot \matrtheta'$, a non-negative integral
matrix where all row and all column sums are equal to some constant $d$. The
complexity of (approximately) computing the decomposition $\matrtheta =
\matr{D}_1 \cdot \matrtheta' \cdot \matr{D}_2$ is discussed
in~\cite{Linial:Samorodnitsky:Wigderson:00:1}.

\section{The Fractional Bethe Permanent}
\label{sec:fractional:Bethe:entropy:1}

The terms that appear in $\HBethe(\vbel)$ in
Definition~\ref{def:Bethe:free:energy:1} all have either coefficient $+1$ or
$-1$, with obvious implications for the coefficients of the terms of
$\HBethe(\matrgamma)$ in
Lemma~\ref{lemma:Bethe:free:energy:as:function:of:gamma:1}. The main idea
behind the fractional Bethe entropy function is to allow these coefficients to
take on also other values. This is done towards the goal of obtaining a
modified Bethe free energy function whose minimum resembles the minimum of the
Gibbs free energy function even more.\footnote{One might also modify
  $\UBethe(\matrgamma)$, however, we do not pursue this option here.}  Such
generalizations of the Bethe entropy function were for example considered
in~\cite{Wiegerinck:Heskes:03:1, Wainwright:Jaakkola:Willsky:05:2,
  Heskes:06:1, Weiss:Meltzer:Yanover:07:1, Hazan:Shashua:10:1,
  Ruozzi:Tatikonda:10:1:subm} and a combinatorial characterization of the
fractional Bethe entropy function was discussed in~\cite{Vontobel:11:1}. In
particular, for the permanent estimation problem such generalizations are
extensively studied in the very recent paper by A.~B.~Yedidia and
Chertkov~\cite{Yedidia:Chertkov:11:1:subm}, to which we refer for additional
discussion on this topic.

As we will see in this section, if the modifications to the Bethe entropy
function are applied within some suitable limits, the concavity of the
modified Bethe entropy function (and therefore the convexity of the modified
Bethe free energy function) will be maintained.

\begin{Definition}
  \label{def:fractional:Bethe:entropy:function:1}

  Let
  \begin{align*}
    \vkappa
      &\defeq
         \bigl\{
           \{ \kappa_i \}_{i \in \setI},
           \{ \kappa_j \}_{j \in \setJ}, 
           \{ \kappa_{i,j} \}_{(i,j) \in \setI \times \setJ}
         \bigr\}
  \end{align*}
  be a collection of real values. We define the $\vkappa$-fractional Bethe
  entropy function to be
  \begin{alignat*}{2}
      \fracHBethe{\vkappa}: \ 
        &\setGamma{n}
           &\, \to&\  \R, \\
        &\matrgamma
        &\, \mapsto&\ 
           \sum_i
             \kappa_i
             \cdot
             \HBethesub{i}(\matrgamma_i)
           +
           \sum_j
             \kappa_j
             \cdot
             \HBethesub{j}(\matrgamma_i) \\
        &&\, &\ 
           -
           \sum_{i,j}
             \kappa_{i,j}
             \cdot
             \HBethesub{(i,j)}(\gamma_{i,j}).
  \end{alignat*}
  (Clearly, if all values in $\vkappa$ equal $1$ then
  $\fracHBethe{\vkappa}(\matrgamma) = \HBethe(\matrgamma)$, with
  $\HBethe(\matrgamma)$ as shown in
  Lemma~\ref{lemma:Bethe:free:energy:as:function:of:gamma:1}.)
  \defend
\end{Definition}

\begin{Lemma}
  \label{lemma:fractional:Bethe:entropy:function:reexpressed:1}

  The fractional Bethe entropy function from
  Definition~\ref{def:fractional:Bethe:entropy:function:1} can also be
  expressed as follows
  \begin{align*}
    \fracHBethe{\vkappa}(\matrgamma)
      &= -
         \sum_{i,j}
           (\kappa_i \! + \! \kappa_j \! - \! \kappa_{i,j})
           \cdot
           \gamma_{i,j} \log(\gamma_{i,j}) \\
      &\quad\ 
         +
         \sum_{i,j}
           \kappa_{i,j}
           \cdot
           (1 \! - \! \gamma_{i,j}) \log(1 \! - \! \gamma_{i,j}).
  \end{align*}
  (If all values in $\vkappa$ equal $1$ then $\fracHBethe{\vkappa}(\matrgamma)
  = \HBethe(\matrgamma)$, with $\HBethe(\matrgamma)$ as shown in
  Corollary~\ref{cor:FBethe:as:function:of:gamma:1}.)
\end{Lemma}

\begin{Proof}
  Follows from combining
  Definition~\ref{def:fractional:Bethe:entropy:function:1} and
  Lemma~\ref{lemma:Bethe:free:energy:as:function:of:gamma:1}.
\end{Proof}

The following definition generalizes
Definitions~\ref{def:Bethe:partition:function:1}
and~\ref{def:Bethe:permanent:1} and
Corollary~\ref{cor:FBethe:as:function:of:gamma:1}.

\begin{Definition}
  \label{def:fractional:Bethe:free:energy:function:1}
  \label{def:fractional:Bethe:permanent:1}

  We define the $\vkappa$-fractional Bethe free energy function to be
  \begin{alignat*}{2}
      \fracFBethe{\vkappa}: \ 
        &\setGamma{n}
           &\, \to&\  \R, \\
        &\matrgamma
        &\, \mapsto&\
           \UBethe(\matrgamma)
           -
           \fracHBethe{\vkappa}(\matrgamma),
  \end{alignat*}
  and the $\vkappa$-fractional Bethe permanent to be
  \begin{align*} 
    \fracpermBethe{\vkappa}(\matrtheta)
      &\defeq
         \exp
           \left(
             -
             \min_{\vbel \in \lmpB}
               \fracFBethe{\vkappa}(\vbel)
           \right).
  \end{align*}
  \defend
\end{Definition}

The following theorem gives a sufficient condition on $\vkappa$ so that the
$\vkappa$-fractional Bethe entropy function is concave in $\matrgamma$,
thereby generalizing Theorem~\ref{theorem:concavity:Bethe:entropy:1:details}.

\begin{Theorem}
  \label{theorem:concavity:fractional:Bethe:entropy:1}

  If $\vkappa$ is such that
  \begin{alignat*}{2}
    \kappa_i
      &\geq 0 \quad && (i \in \setI), \\
    \kappa_j
      &\geq 0 \quad && (j \in \setJ), \\
    \kappa_i + \kappa_j
      &\geq 2 \kappa_{i,j} \quad && ((i,j) \in \setI \times \setJ).
  \end{alignat*}
  then $\fracHBethe{\vkappa}(\matrgamma)$ is a concave function of
  $\matrgamma$ and $\fracFBethe{\vkappa}(\matrgamma)$ is a convex function of
  $\matrgamma$.
\end{Theorem}

\begin{Proof}
  We have
  \begin{align*}
    &
    \fracHBethe{\vkappa}(\matrgamma) \\
      &\onestareq
         -
         \sum_{i,j}
           \left(
             \frac{\kappa_i \! + \! \kappa_j}{2}
             +
             \frac{\kappa_i \! + \! \kappa_j}{2}
             -
             \kappa_{i,j}
           \right)
           \! \cdot \!
           \gamma_{i,j} \log(\gamma_{i,j}) \\
      &\quad\ 
         +
         \sum_{i,j}
           \left(
             \frac{\kappa_i \! + \! \kappa_j}{2}
             -
             \frac{\kappa_i \! + \! \kappa_j}{2}
             + 
             \kappa_{i,j}
           \right)
           \! \cdot \!
           (1 \! - \! \gamma_{i,j}) \log(1 \! - \! \gamma_{i,j}) \\
      &\twostarseq
         \sum_i
           \frac{\kappa_i}{2}
           \cdot
           S(\matrgamma_i)
         +
         \sum_j
           \frac{\kappa_j}{2}
           \cdot
           S(\matrgamma_j) \\
      &\quad\ 
         +
         \sum_{i,j}
           \left(
             \frac{\kappa_i \! + \! \kappa_j}{2}
             -
             \kappa_{i,j}
           \right)
           \cdot
           h(\gamma_{i,j}),
  \end{align*}
  where at step~$\onestar$ we have used
  Lemma~\ref{lemma:fractional:Bethe:entropy:function:reexpressed:1}, and where
  at step~$\twostars$ we have used the $S$-function as specified in
  Definition~\ref{def:md:ess:function:1} and have introduced the binary
  entropy function $h:\ [0,1] \to \R, \ \xi \mapsto - \xi \log(\xi) - (1 \! -
  \! \xi) \log(1 \!  - \! \xi)$. If $\kappa_i \geq 0$, $\kappa_j \geq 0$, and
  $\frac{\kappa_i \! + \!  \kappa_j}{2} - \kappa_{i,j} \geq 0$ (the latter
  being equivalent to $\kappa_i + \kappa_j \geq 2 \kappa_{i,j}$), then the
  concavity of $\fracHBethe{\vkappa}(\matrgamma)$ in $\matrgamma$ follows from
  Theorem~\ref{theorem:md:ess:function:properties:1}, the well-known concavity
  of the binary entropy function, and the fact that the sum of concave
  functions is a concave function.

  The convexity of $\fracFBethe{\vkappa}(\matrgamma)$ in $\matrgamma$ follows
  from the concavity of $\fracHBethe{\vkappa}(\matrgamma)$ in $\matrgamma$ and
  the linearity of $\UBethe(\matrgamma)$ in $\matrgamma$.
\end{Proof}

\begin{Lemma}
  \label{lemma:frac:perm:Bethe:for:all:one:matrix:1}

  An interesting choice for $\vkappa$ is
  \begin{alignat*}{2}
    \kappa_i
      &= 1 \quad && (i \in \setI), \\
    \kappa_j
      &= 1 \quad && (j \in \setJ), \\
    \kappa_{i,j}
      &= 1 - \frac{1}{2n} \quad && ((i,j) \in \setI \times \setJ).
  \end{alignat*}
  The resulting $\fracHBethe{\vkappa}(\matrgamma)$ is a concave function of
  $\matrgamma$ and the resulting $\fracFBethe{\vkappa}(\matrgamma)$ is a
  convex function of $\matrgamma$. Moreover, letting $\matrallone{n}$ be the
  all-one matrix of size $n \times n$, we obtain
  \begin{align*}
    \frac{\perm(\matrallone{n})}
         {\fracpermBethe{\vkappa}(\matrallone{n})}
      &= \frac{\sqrt{2 \pi}}{\e}
           \! \cdot \!
           \big(
             1+o(1)
           \big)
       = 0.922\ldots
           \! \cdot \!
           \big(
             1 \! + \! o(1)
           \big),
  \end{align*}
  where $o(1)$ is w.r.t.\ $n$. (Note that, in contrast to
  Lemma~\ref{lemma:perm:Bethe:for:all:one:matrix:1}, there is no
  $\sqrt{n}$-factor on the right-hand side of the above expression.)
\end{Lemma}

\begin{Proof}
  See Appendix~\ref{sec:proof:lemma:frac:perm:Bethe:for:all:one:matrix:1}.
\end{Proof}

\begin{figure}
  \begin{center}
    \psfrag{xlabel}{$n$}
    \psfrag{ylabel}{\hskip-2cm $\perm(\matrallone{n}) \ / \ 
                          \fracpermBethe{\vkappa}(\matrallone{n})$}
    \epsfig{file=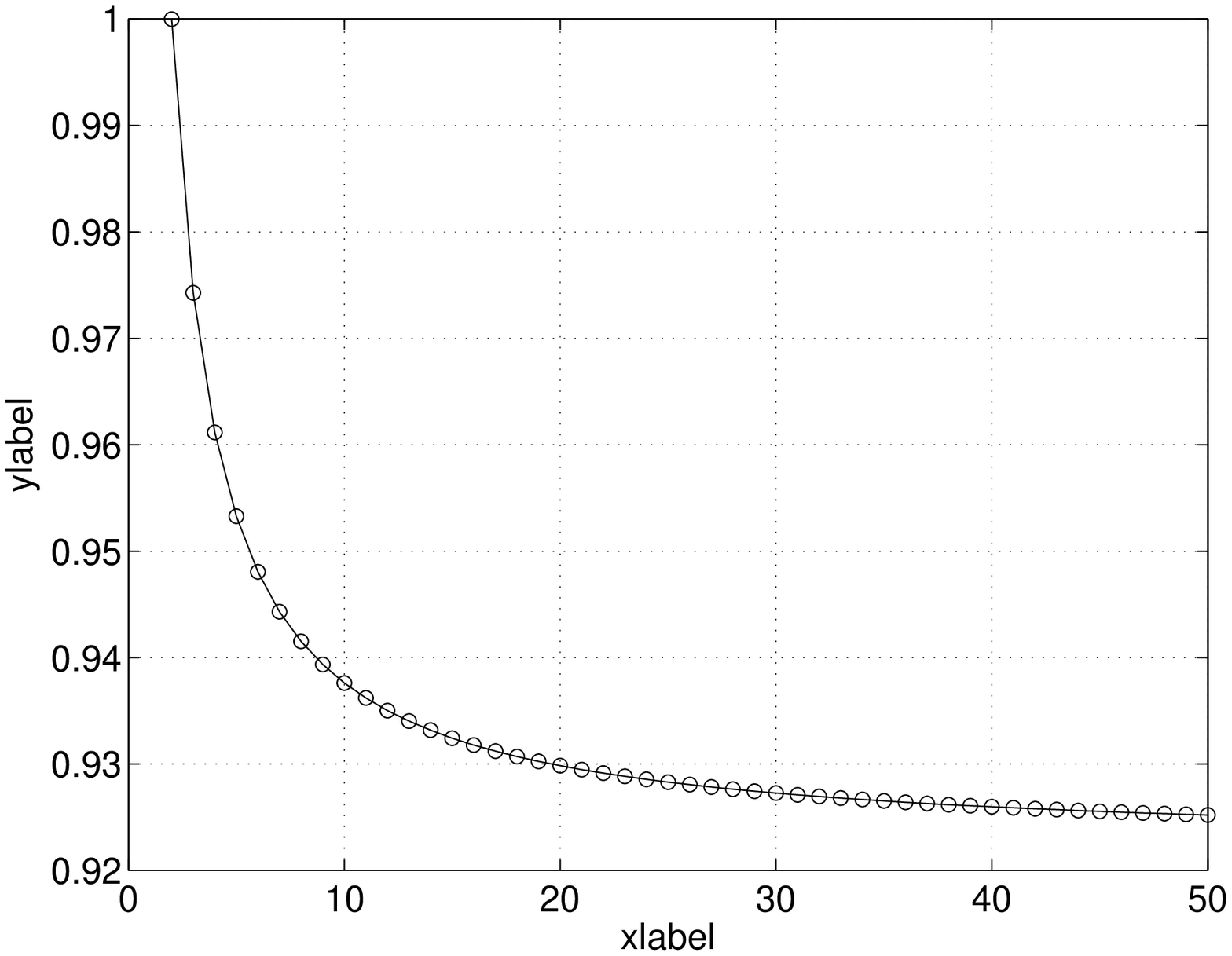, width=\linewidth}
  \end{center}
  \caption{Illustration of the ratio $\perm(\matrallone{n}) /
    \fracpermBethe{\vkappa}(\matrallone{n})$ for the special choice of
    $\vkappa$ in Lemma~\ref{lemma:frac:perm:Bethe:for:all:one:matrix:1}, when
    $n$ varies from $2$ to $50$.}
  \label{fig:correction:factor:frac:Bethe:perm:1}
\end{figure}

Let us make a few comments about the choice of $\vkappa$ in
Lemma~\ref{lemma:frac:perm:Bethe:for:all:one:matrix:1}.
\begin{itemize}

\item Fig.~\ref{fig:correction:factor:frac:Bethe:perm:1} shows the exact
  ratios for $n$ from $2$ to $50$. In particular, note that for $n = 2$ we
  have
  \begin{align*}
    \frac{\perm(\matrallone{2})}
         {\fracpermBethe{\vkappa}(\matrallone{2})}
      &= 1.
  \end{align*}

\item For even integers $n$ and for the choice of $\vkappa$ from
  Lemma~\ref{lemma:frac:perm:Bethe:for:all:one:matrix:1}, the matrix
  $\matrtheta = \matr{I}_{(n/2) \times (n/2)} \otimes \matrallone{2}$ yields
  the ratio $\frac{\perm(\matrtheta)}{\fracpermBethe{\vkappa}(\matrtheta)} =
  1$. This is in stark contrast to
  Conjecture~\ref{conj:Bethe:permanent:lower:and:upper:bound:1} where
  $\matrtheta$ represents the conjectured ``worst-case'' matrix for the ratio
  $\frac{\perm(\matrtheta)}{\permBethe(\matrtheta)}$.
  
\item For integers $n$ and $k$ such that $k$ divides $n$ we have
  \begin{align*}
    (0.922\ldots)^{n/k}
     &\leq
        \frac{\perm(\matrtheta)}
             {\fracpermBethe{\vkappa}(\matrtheta)}
      \leq
        1
  \end{align*}
  for the matrix $\matrtheta \defeq \matr{I}_{(n/k) \times (n/k)} \otimes
  \matrallone{k}$.

\end{itemize}

Let us conclude this section on the fractional Bethe entropy function with a
few comments.
\begin{itemize}

\item The SPA message update equations in Section~\ref{sec:factor:graph:spa:1}
  need to be modified so that its fixed points correspond to stationary points
  of the fractional Bethe free energy, \ie, so that a modified version of the
  theorem by Yedidia, Freeman, and Weiss~\cite{Yedidia:Freeman:Weiss:05:1}
  holds. In contrast to the SPA message update equations in
  Section~\ref{sec:factor:graph:spa:1}, the modified SPA message update
  equations will be such that the right-going messages depend not only on the
  previous left-going messages but also on the previous right-going messages,
  and such that the left-going messages depend not only on the previous
  right-going messages but also on the previous left-going messages.  (We omit
  the details.) Moreover, the convergence analysis in
  Section~\ref{sec:factor:graph:spa:1} has to be revisited.

\item We leave it as an open problem to explore the $\vkappa$ parameter space
  and to find fractional Bethe permanents for which interesting statements can
  be made, in particular for which a statement like the one in
  Theorem~\ref{theorem:Bethe:permanent:upper:bound:1} can be made.

\end{itemize}

\section{Comments and Conjectures}
\label{sec:comments:and:conjectures:1}

It is an interesting challenge to look at theorems involving permanents and to
prove that the theorems still hold if the permanents in these theorems are
replaced by Bethe permanents. Let us mention two conjectures along these
lines that were listed in~\cite{Vontobel:10:2}.

\subsection{Perm-Pseudo-Codewords}
\label{sec:perm:pseudo:codewords:1}

The following conjecture is based on a theorem
in~\cite{Smarandache:Vontobel:09:2} involving permanents of submatrices of a
parity-check matrix.

\begin{Definition}
  \label{def:Behte:perm:vector:1} 
 
  Let $\code{C}$ be a binary linear code described by a parity-check matrix
  $\matr{H}\in \Ftwo^{m \times n}$, $m < n$. For a size-$(m{+}1)$ subset
  $\setS$ of the column index set $\setI(\matr{H})$ we define the Bethe
  perm-vector based on $\setS$ to be the vector $\vomega \in \Z^n$ with
  components
  \begin{align*}
    \omega_i
      &\defeq
         \begin{cases} 
           \permBethe\big( \matr{H}_{\setS \setminus i} \big) 
             & \text{if $i \in \setS$} \\
           0                                     
             & \text{otherwise}
         \end{cases},
  \end{align*}
  where $\matr{H}_{\setS \setminus i}$ is the submatrix of $\matr{H}$
  consisting of all the columns of $\matr{H}$ whose index is in the set $\setS
  \setminus \{ i \}$.
  \defend
\end{Definition}

\begin{Conjecture}
  \label{conj:Bethe:perm:vector:1}
 
  Let $\code{C}$ be a binary linear code described by the parity-check matrix
  $\matr{H}\in \Ftwo^{m \times n}$, $m < n$, let $\fch{K}{H}$ be the
  fundamental cone associated with $\matr{H}$~\cite{Koetter:Vontobel:03:1,
    Vontobel:Koetter:05:1:subm}, and let $\setS$ be a size-$(m{+}1)$ subset of
  $\setI(\matr{H})$. The Bethe perm-vector $\vomega$ based on $\setS$ is a
  pseudo-codeword of $\matr{H}$, \ie,
  \begin{align}
    \vomega 
      &\in \fch{K}{H}
      \label{eq:lemma:perm:vector:1:1},
  \end{align}
  \conjectureend
\end{Conjecture}

A proof of this conjecture has recently been presented by
Smarandache~\cite{Smarandache:11:1:subm}.

\subsection{Permanent-Based Kernels}
\label{sec:perm:kernels:1}

Based on a result by Cuturi~\cite{Cuturi:07:1}, Huang and
Jebara~\cite{Huang:Jebara:09:1} made the following conjecture.

\begin{Conjecture}[Huang and Jebara~\cite{Huang:Jebara:09:1}]
  \label{con:Bethe:perm:based:kernel:1}
 
  Let $n$ be a positive integer and let $\set{X}$ be a set endowed with a
  kernel $\kappa$. Let $X = \{ x_1, \ldots, x_n \} \in \set{X}^n$ and $Y = \{
  y_1, \ldots, y_n \} \in \set{X}^n$. Then
  \begin{align*}
    \kappa_{\permBethe}: \
     (X,Y)
       \mapsto \permBethe
                 \Big(
                   \big[
                     \kappa(x_i,y_j)
                   \big]_{1 \leq i \leq n, \, 1 \leq j \leq n}
                 \Big)
  \end{align*}
  is a positive definite kernel on $\set{X}^n \times \set{X}^n$.
  \conjectureend
\end{Conjecture}

\section{Conclusions}
\label{sec:conclusions:1}

In this paper, we have pursued a graphical-model-based approach to
approximating the permanent of a non-negative square matrix, the resulting
approximation being called the Bethe permanent. We have seen that the
associated functions, like the Bethe entropy function and the Bethe free
energy function, are remarkably well behaved for a graphical model with a
non-trivial cycle structure. In that respect, an important part is played by a
theorem by Birkhoff and von Neumann (see
Theorem~\ref{theorem:Birkhoff:von:Neumann:1}).  Moreover, the SPA can be used
to efficiently find the minimum of the Bethe free energy function and thereby
the Bethe permanent. We have also presented a graph-cover-based analysis that
gives additional insights into the inner workings of the Bethe permanent, its
strengths, and its weaknesses, and we have commented on Bethe-permanent-based
upper and lower bounds on the permanent. Along the way we have stated several
conjectures and open problems, that, if answered one way or the other, could
further elucidate the relationship between the permanent and the Bethe
permanent.

\section*{Acknowledgments}

We gratefully acknowledge Farzad Parvaresh for pointing out to us the
papers~\cite{Barvinok:10:1, Barvinok:Samorodnitsky:11:1}, Krishna Viswanathan
for discussions on the permanent of structured matrices, Adam Yedidia and
Misha Chertkov for sharing an early version of their
paper~\cite{Yedidia:Chertkov:11:1:subm}, and Leonid Gurvits for general
discussions about permanents. Moreover, we very much appreciate the helpful
comments that were made by the reviewers.

\appendices

\section{Proof of
  Theorem~\ref{theorem:md:ess:function:properties:1}}
\label{sec:proof:theorem:md:ess:function:properties:1}

Observe that once the concavity of $S$ is established, it is straightforward
to verify the claim in the theorem statement that $S(\vxi) \geq 0$ for all
$\vxi \in \setPi_{[n]}$. Indeed, because $\setPi_{[n]}$ is a polytope with $n$
vertices, because $S$ takes on the value $0$ at each of these vertices, and
because $S$ is concave, this statement is true.

Therefore, let us focus on the concavity statement. Clearly, for $n = 2$ the
statement can easily be verified and so the rest of this appendix will only
discuss the case $n \geq 3$.

By definition, a multi-dimensional function is concave if it is a concave
function along any straight line in its domain. Towards showing that this is
indeed the case for $S$, let us fix an arbitrary point $\vxi \in \setPi_{[n]}$
and an arbitrary direction $\hvxi \in \R^n \setminus \{ \vect{0} \}$ such that
the function $\vxi(\temp) \defeq \vxi + \temp \cdot \hvxi$ satisfies
$\vxi(\temp) \in \setPi_{[n]}$ for a suitable $\temp$-interval around $0$ (to
be defined later). We need to distinguish three different cases that will be
discussed separately in the following subsections:
\begin{enumerate}

\item The point $\vxi$ is in the interior of $\setPi_{[n]}$.

\item The point $\vxi$ is at a vertex of $\setPi_{[n]}$.

\item The point $\vxi$ is neither in the interior nor at a vertex
  of~$\setPi_{[n]}$.

\end{enumerate}

\subsection{The Point $\vxi$ is in the Interior of $\setPi_{[n]}$}
\label{sec:vxi:interior:1}

It is straightforward to see that the direction vector $\hvxi$ must satisfy
\begin{align}
  \sum_{\ell}
    \hxi_{\ell}
    &= 0,
         \label{eq:tvxi:property:1}
\end{align}
otherwise $\vxi(\temp) \in \setPi_{[n]}$ holds only for $\temp =0$. Therefore,
we assume that~\eqref{eq:tvxi:property:1} is satisfied. Moreover, because
$\vxi \in \interior(\setPi_{[n]})$, we have $0 < \xi_{\ell} <1$, $\ell \in
[n]$, and we can find an $\varepsilon > 0$ such that $\vxi(\temp) \in
\setPi_{[n]}$ for $- \varepsilon \leq \temp \leq \varepsilon$. We will now
show that the function $\tau \mapsto S\bigl( \vxi(\temp) \bigr)$ is concave at
$\temp = 0$.

We start by computing the first-order derivative
\begin{align*}
  \rdiff{\temp}
  S\big( \vxi(\temp) \big)
    &= -
       \sum_{\ell}
         \rdiff{\xi_{\ell}(\temp)}
           s\bigl( \xi_{\ell}(\temp) \bigr)
         \cdot
         \hxi_{\ell},
\end{align*}
and the second-order derivative
\begin{align*}
  \rddiff{\temp}
  S\big( \vxi(\temp) \big)
    &= \sum_{\ell}
         \rddiff{\xi_{\ell}(\temp)}
           s\bigl( \xi_{\ell}(\temp) \bigr)
         \cdot
         \hxi_{\ell}^2 \\
    &\onestareq
       -
       \sum_{\ell}
         \frac{\hxi_{\ell}^2}{\xi_{\ell}(\temp)}
       +
       \sum_{\ell}
         \frac{\hxi_{\ell}^2}{1-\xi_{\ell}(\temp)},
\end{align*}
where at step~$\onestar$ we have used
Lemma~\ref{lemma:ess:function:properties:1}. In particular, at $\temp =0$ we have
\begin{align*}
  \left.
  \rddiff{\temp}
  S\big( \vxi(\temp) \big)
  \right|_{\temp=0}
    &= \sum_{\ell}
         \delta_{\ell},
\end{align*}
where $\delta_{\ell}$, $\ell \in [n]$, is defined as
\begin{align}
   \delta_{\ell}
     &\defeq
        -
        \sum_{\ell}
          \frac{\hxi_{\ell}^2}
               {\xi_{\ell}}
        +
        \sum_{\ell}
          \frac{\hxi_{\ell}^2}
               {1-\xi_{\ell}}
      = -
        \sum_{\ell}
          \hxi_{\ell}^2
          \cdot
          \frac{1 - 2\xi_{\ell}}
               {\xi_{\ell} (1-\xi_{\ell})}.
                 \label{eq:HBethe:concavity:delta:def:1}
\end{align}
The proof will be finished once we have shown that $\rddiff{\temp} S\big( \vxi(\temp)
\big) \leq 0$ at $\temp =0$, which is equivalent to the condition that
\begin{align}
  \sum_{\ell}
    \delta_{\ell}
    &\leq 0.
      \label{eq:HBethe:concavity:condition:1}
\end{align}

We show this by separately considering two cases, the first case being $\vxi
\in \interior(\setPi_{[n]}) \, \cap \, [0, 1/2]^n$, the second case being
$\vxi \in \interior(\setPi_{[n]}) \setminus [0, 1/2]^n$.

The first case, $\vxi \in \interior(\setPi_{[n]}) \, \cap \, [0, 1/2]^n$, is
relatively straightforward. Namely, for all $\ell \in [n]$ we have $0 <
\xi_{\ell} \leq 1/2$, which implies $1 - 2\xi_{\ell} \geq 0$, which in turn
implies $\delta_{\ell} \leq 0$, and so~\eqref{eq:HBethe:concavity:condition:1}
is satisfied.

The second case, $\vxi \in \interior(\setPi_{[n]}) \setminus [0, 1/2]^n$,
needs somewhat more work. We start by observing that there is a unique $\ells
\in [n]$ such that $\xi_{\ells} > 1/2$. (Note that there can only be one such
$\ells \in [n]$ because $\sum_{\ell} \xi_{\ell} = 1$.) Consequently, $1 - 2
\xi_{\ells} < 0$ and $1 - 2\xi_{\ell} > 0$, $\ell \neq \ells$.

In the following, it is sufficient to consider only directions $\hvxi$ that
satisfy $\hxi_{\ells} > 0$ and $\hxi_{\ell} \leq 0$, $\ell \neq \ells$, or
that satisfy $\hxi_{\ells} < 0$ and $\hxi_{\ell} \geq 0$, $\ell \neq
\ells$. This follows from contemplating~\eqref{eq:tvxi:property:1}
and~\eqref{eq:HBethe:concavity:delta:def:1} and from observing that for a
given $\vxi$ and given directional magnitudes $\bigl\{ |\hxi_{\ell}|
\bigr\}_{\ell \neq \ells}$, the left-hand side
of~\eqref{eq:HBethe:concavity:condition:1} is maximized by a $\hvxi$ that
satisfies the conditions that we have just mentioned.\footnote{In other words,
  such a $\hvxi$ produces the ``worst-case'' left-hand side
  in~\eqref{eq:HBethe:concavity:condition:1}: if we can show non-positivity
  for such direction vectors, we have implicitly shown non-positivity for any
  other direction vector.}  From~\eqref{eq:tvxi:property:1} it follows that
such direction vectors $\hvxi$ satisfy
\begin{align}
  |\hxi_{\ells}|
    &= \sum_{\ell \neq \ells}
         |\hxi_{\ell}|.
         \label{eq:tvxi:normal:factorization:2}
\end{align}

Before continuing, let us introduce
\begin{align*}
   \delta'
     &\defeq
        -
        \frac{\hxi_{\ells}^2}
               {\xi_{\ells}}
        +
        \sum_{\ell \neq \ells}
          \frac{\hxi_{\ell}^2}
               {1-\xi_{\ell}}, \\
   \delta''
     &\defeq
        +
        \frac{\hxi_{\ells}^2}
             {1-\xi_{\ells}}
        -
        \sum_{\ell \neq \ells}
        \frac{\hxi_{\ell}^2}
               {\xi_{\ell}}.
\end{align*}
Note that $\sum_{\ell} \delta_{\ell} = \delta' + \delta''$, and so, if we can
show that $\delta' \leq 0$ and $\delta'' \leq 0$ then we have verified the
desired result~\eqref{eq:HBethe:concavity:condition:1}.

The fact $\delta' \leq 0$ is a consequence of the equation
\begin{align*}
  \sum_{\ell \neq \ells}
    \frac{\hxi_{\ell}^2}
         {1-\xi_{\ell}}
    &\overset{\onestar}{\leq}
      \frac{1}{\xi_{\ells}}
       \sum_{\ell \neq \ells}
         \hxi_{\ell}^2 
     \overset{\twostars}{\leq}
       \frac{1}{\xi_{\ells}}
       \cdot
       \left(
         \sum_{\ell \neq \ells}
           |\hxi_{\ell}|
       \right)^2
     \threestarseq
       \frac{\hxi_{\ells}^2}
            {\xi_{\ells}},
\end{align*}
where step~$\onestar$ follows from $\vxi$ being in $\setPi_{[n]}$, which
implies that $\xi_{\ells} = 1 - \sum_{\ell' \neq \ells} \xi_{\ell'}$, which in
turn implies that $\xi_{\ells} \leq 1 - \xi_{\ell}$ for all $\ell \neq
\ells$. Moreover, step~$\twostars$ follows from a simple inequality and
step~$\threestars$ follows from~\eqref{eq:tvxi:normal:factorization:2}.

The fact $\delta'' \leq 0$ is shown as follows. We start by observing that
\begin{align*}
  \left(
    1
    -
    \xi_{\ells}
  \right)
  \cdot
  \left(
    \sum_{\ell \neq \ells}
      \frac{\hxi_{\ell}^2}
           {\xi_{\ell}}
  \right)
    &\onestareq
       \left(
         \sum_{\ell \neq \ells}
           \xi_{\ell}
       \right)
       \cdot
       \left(
         \sum_{\ell \neq \ells}
           \frac{\hxi_{\ell}^2}
                {\xi_{\ell}}
       \right) \\
    &= \left(
         \sum_{\ell \neq \ells}
           \sqrt{\xi_{\ell}}^2
       \right)
       \cdot
       \left(
         \sum_{\ell \neq \ells}
           \left(
             \frac{|\hxi_{\ell}|}
                  {\sqrt{\xi_{\ell}}}
           \right)^{\!2}
       \right) \\
    &\overset{\twostars}{\geq}
       \left(
         \sum_{\ell \neq \ells}
           |\hxi_{\ell}|
       \right)^2
     \threestarseq
       \hxi_{\ells}^2,
\end{align*}
where step~$\onestar$ follows from $\vxi$ being in $\setPi_{[n]}$ (which
implies that $\xi_{\ells} = 1 - \sum_{\ell \neq \ells} \xi_{\ell}$), where at
step~$\twostars$ we use the Cauchy-Schwarz inequality, and where at
step~$\threestars$ we use~\eqref{eq:tvxi:normal:factorization:2}. Rearranging
this inequality, we see that it is equivalent to the inequality
$\delta''_{\ell} \leq 0$.

\subsection{The Point $\vxi$ is at a Vertex of $\setPi_{[n]}$}
\label{sec:vxi:at:vertex:1}

Clearly, the direction vector $\hvxi$ must
satisfy~\eqref{eq:tvxi:property:1}. Moreover, because $\vxi$ is at a vertex of
$\setPi_{[n]}$, there is an $\ells \in [n]$ such that $\xi_{\ells} = 1$ and
$\xi_{\ell} = 0$, $\ell \neq \ells$, and such that $\hxi_{\ells} < 0$ and
$\hxi_{\ell} \geq 0$, $\ell \neq \ells$. Then we can find an $\varepsilon > 0$
such that $\vxi(\temp) \in \setPi_{[n]}$ for $0 \leq \temp \leq \varepsilon$. We
will now show that the function $\tau \mapsto S\bigl( \vxi(\temp) \bigr)$ is
concave at $\temp = 0$.

We start by plugging in the definition of $\vxi(\temp)$ into $S\bigl(
\vxi(\temp) \bigr)$, \ie,
\begin{align}
  S\big( \vxi(\temp) \big)
    &= -
       \sum_{\ell}
         \xi_{\ell}(\temp) \log\bigl( \xi_{\ell}(\temp) \bigr) \nonumber \\
    &\quad
       +
       \sum_{\ell}
         \bigl( 1-\xi_{\ell}(\temp) \bigr) \log\bigl( 1-\xi_{\ell}(\temp) \bigr)
           \nonumber  \\
    &= - \,
       (1+\temp\hxi_{\ells}) \log(1+\temp\hxi_{\ells})
       -
       \sum_{\ell \neq \ells}
         (\temp\hxi_{\ell}) \log(\temp\hxi_{\ell}) \nonumber \\
    &\quad
       +
       (-\temp\hxi_{\ells}) \log(-\temp\hxi_{\ells})
       +
       \sum_{\ell \neq \ells}
         (1-\temp\hxi_{\ell}) \log(1-\temp\hxi_{\ell}) .\nonumber
\end{align}
From this we compute the first-order derivative
\begin{align}
  \rdiff{\temp}
    S\big( \vxi(\temp) \big)
    &= - \,
       \hxi_{\ells}
       \log(1+\temp\hxi_{\ells})
       - \hxi_{\ells} \nonumber \\
    &\quad
       -
       \sum_{\ell \neq \ells}
         \hxi_{\ell} \log(\temp)
       -
       \sum_{\ell \neq \ells}
         \hxi_{\ell} \log(\hxi_{\ell})
       -
       \sum_{\ell \neq \ells}
         \hxi_{\ell} \nonumber \\
    &\quad
       -
       \hxi_{\ells} \log(\temp)
       -
       \hxi_{\ells} \log(-\hxi_{\ells})
       -
       \hxi_{\ells} \nonumber \\
    &\quad
       -
       \sum_{\ell \neq \ells}
         \hxi_{\ell}
         \log(1-\temp\hxi_{\ell})
       -
       \sum_{\ell \neq \ells}
         \hxi_{\ell} \nonumber \\
    &\onestareq
       - \,
       \hxi_{\ells}
       \log(1+\temp\hxi_{\ells})
       -
       \sum_{\ell \neq \ells}
         \hxi_{\ell} \log(\hxi_{\ell}) \nonumber \\
    &\quad
       -
       \hxi_{\ells} \log(-\hxi_{\ells})
       -
       \sum_{\ell \neq \ells}
         \hxi_{\ell}
         \log(1-\temp\hxi_{\ell}),
           \label{eq:S:first:orderderivative:1}
\end{align}
where at step~$\onestar$ we have used $\sum_{\ell} \hxi_{\ell} = 0$ multiple
times. The second-order derivative is then
\begin{align*}
  \rddiff{\temp}
    S\big( \vxi(\temp) \big)
    &= -
       \frac{\hxi_{\ells}^2}
            {1+\temp\hxi_{\ells}}
       +
       \sum_{\ell \neq \ells}
         \frac{\hxi_{\ell}^2}
              {1-\temp\hxi_{\ell}}.
\end{align*}
For $\temp \downarrow 0$ we obtain
\begin{align*}
  \left.
    \rddiff{\temp}
      S\big( \vxi(\temp) \big)
  \right|_{\temp \downarrow 0}
    &= -
       \hxi_{\ells}^2
       +
       \sum_{\ell \neq \ells}
         \hxi_{\ell}^2 \\
    &\onestareq
       -
       \left(
         -
         \sum_{\ell \neq \ells}
         \hxi_{\ell}
       \right)^2
       +
       \sum_{\ell \neq \ells}
         \hxi_{\ell}^2 \\
    &\twostarsleq
       0,
\end{align*}
where at step~$\onestar$ we have used~\eqref{eq:tvxi:property:1} and where
step~$\twostars$ follows from a simple inequality and the fact that
$\hxi_{\ell} \geq 0$ for $\ell \neq \ells$. Therefore, the function $\tau \mapsto
S\bigl( \vxi(\temp) \bigr)$ is concave at $\temp =0$.

\subsection{The Point $\vxi$ is Neither in the Interior nor at a 
                    Vertex of~$\setPi_{[n]}$}
\label{sec:vxi:neigther:interior:nor:vertex:1}

The fact that $\vxi$ is neither in the interior nor at a vertex of
$\setPi_{[n]}$ means that there is an $\ells \in [n]$ such that $0 <
\xi_{\ells} < 1$. Clearly, the direction vector $\hvxi$ must
satisfy~\eqref{eq:tvxi:property:1}, plus some additional constraints that are
irrelevant for the discussion here. Then we can find an $\varepsilon > 0$ such
that $\vxi(\temp) \in \setPi_{[n]}$ for $0 \leq \temp \leq \varepsilon$. The
concavity of the function $\tau \mapsto S\bigl( \vxi(\temp) \bigr)$ at $\temp
=0$ follows then from the observation that, for small non-negative $\temp$,
the second-order derivative of $S\big( \vxi(\temp) \big)$ w.r.t.~$\temp$ is
dominated by the second-order derivative of the expression $- \sum_{\ell: \,
  \xi_{\ell} = 0, \, \hxi_{\ell} > 0} \xi_{\ell}(\temp) \log\bigl(
\xi_{\ell}(\temp) \bigr)$, a function that is concave in~$\temp$.

\section{Proof of Lemma~\ref{lemma:linear:behavior:md:ess:1}}
\label{sec:proof:lemma:linear:behavior:md:ess:1}

We obtain the expression in the lemma statement by evaluating $S\big(
\vxi(\temp) \big)$ and the first-order derivative of $S\big( \vxi(\temp)
\big)$ w.r.t.~$\temp$ at $\temp =0$. Clearly, $S\big( \vxi(\temp) \big) = 0$
and so we can focus on computing the first-order derivative.

Fortunately, in Appendix~\ref{sec:vxi:at:vertex:1} we have already computed
the first-order derivative for exactly the same setup. Namely,
from~\eqref{eq:S:first:orderderivative:1} we obtain
\begin{align*}
  \rdiff{\temp}
    S\big( \vxi(\temp) \big)
    &= -
       \hxi_{\ells}
       \log(1+\temp\hxi_{\ells})
       -
       \sum_{\ell \neq \ells}
         \hxi_{\ell} \log(\hxi_{\ell}) \\
    &\quad
       -
       \hxi_{\ells} \log(-\hxi_{\ells})
       -
       \sum_{\ell \neq \ells}
         \hxi_{\ell}
         \log(1-\temp\hxi_{\ell}).
\end{align*}
In the limit $\temp \downarrow 0$ this simplifies to
\begin{align}
  \hskip-0.25cm
  \left.
    \rdiff{\temp}
      S\big( \vxi(\temp) \big)
  \right|_{\temp \downarrow 0}
    &= -
       \sum_{\ell \neq \ells}
         \hxi_{\ell}
         \log(\hxi_{\ell})
       +
       (-\hxi_{\ells})
         \log(-\hxi_{\ells}).
         \label{eq:linear:behavior:md:ess:2}
\end{align}
This can be rewritten as follows
\begin{align*}
  \left.
    \rdiff{\temp}
      S\big( \vxi(\temp) \big)
  \right|_{\temp \downarrow 0}
    &= |\hxi_{\ells}|
       \cdot 
       \left(
         -
         \sum_{\ell \neq \ells}
           \frac{|\hxi_{\ell}|}
                {|\hxi_{\ells}|}
           \log
             \left(
               \frac{|\hxi_{\ell}|}
                    {|\hxi_{\ells}|}
             \right)
       \right),
\end{align*}
where we have used $-\hxi_{\ells} = |\hxi_{\ells}|$, $\hxi_{\ell} =
|\hxi_{\ell}|$, $\ell \neq \ells$, and $|\hxi_{\ells}| = \sum_{\ell \neq
  \ells} |\hxi_{\ell}|$, \ie, $\sum_{\ell \neq \ells} |\hxi_{\ell}| /
|\hxi_{\ells}| = 1$. This verifies the expressions for $S\big( \vxi(\temp) \big)$
in the lemma statement.

Finally, the non-negativity of the coefficient of $\temp$
in~\eqref{eq:S:value:near:vertex:1} follows from $|\hxi_{\ells}| \geq
|\hxi_{\ell}|$, $\ell \neq \ells$, which is a consequence of the
above-mentioned relation $|\hxi_{\ells}| = \sum_{\ell \neq \ells}
|\hxi_{\ell}|$.

\section{Proof of Lemma~\ref{lemma:linear:behavior:HBethe:1}}
\label{sec:proof:lemma:linear:behavior:HBethe:1}

Clearly we have $\gamma_{i,j} = 1$ if $j = \sigma(i)$ and $\gamma_{i,j} = 0$
otherwise. From the condition that $\hmatrgamma$ is such that $\vgamma(\temp)
\in \setGamma{n}$ for small non-negative $\temp$, it follows that $\sum_{j}
\hgamma_{i,j} = 0$ for all $i \in \setI$ and $\sum_i \hgamma_{i,j} = 0$ for
all $j \in \setJ$. Moreover, for every $i \in \setI$ we have $\hgamma_{i,j}
\leq 0$ if $j = \sigma(i)$ and $\hgamma_{i,j} \geq 0$ otherwise. Then
\begin{align*}
  &
  \HBethe\big( \matrgamma(\temp) \big) \\
    &\onestareq
       \frac{1}{2}
       \sum_i
         S\big( \matrgamma_i(\temp) \big)
       +
       \frac{1}{2}
       \sum_j
         S\big( \matrgamma_j(\temp) \big) \\
    &\twostarseq
       -
       \frac{\temp}{2}
         \sum_i \!\!
           \sum_{j \neq \sigma(i)}
             \hgamma_{i,j} \log(\hgamma_{i,j})
       +
       \frac{\temp}{2}
         \sum_i
           (-\hgamma_{i,\sigma(i)}) \log(-\hgamma_{i,\sigma(i)}) \\
     &\quad
       -
       \frac{\temp}{2}
         \sum_j \!\!
           \sum_{i \neq \bsigma(j)}
             \hgamma_{i,j} \log(\hgamma_{i,j}) 
       +
       \frac{\temp}{2}
         \sum_j
           (-\hgamma_{\bsigma(j),j}) \log(-\hgamma_{\bsigma(j),j}) \\
     &\quad\ 
       +
       O(\temp^2),
\end{align*}
where step~$\onestar$ follows from
Lemma~\ref{lemma:Bethe:entropy:function:in:terms:of:S:1} and where at
step~$\twostars$ we have used $S(\vgamma_i) = 0$, $S(\vgamma_j) = 0$,
and~\eqref{eq:linear:behavior:md:ess:2}.

We observe that in the above expression there are exactly two terms for every
edge $e = (i,j) \in \setI \times \setJ$. Rewriting these summations such that
all the main summations are over $i \in \setI$, we obtain
\begin{align*}
  &
  \HBethe\big( \matrgamma(\temp) \big) \\
    &= -
       \temp
         \sum_i \!\!
           \sum_{j \neq \sigma(i)}
             \hgamma_{i,j} \log(\hgamma_{i,j})
       +
       \temp
         \sum_i
           (-\hgamma_{i,\sigma(i)}) \log(-\hgamma_{i,\sigma(i)}) \\
    &\quad\ 
       +
       O(\temp^2) \\
    &\onestareq
       \temp
       \sum_i
         |\hgamma_{i,\sigma(i)}|
         \cdot 
         \left(
           - \!\!\!
           \sum_{j \neq \sigma(i)}
           \frac{|\hgamma_{i,j}|}
                {|\hgamma_{i,\sigma(i)}|}
           \log
             \left(
               \frac{|\hgamma_{i,j}|}
                    {|\hgamma_{i,\sigma(i)}|}
             \right)
         \right)
       +
       O(\temp^2),
\end{align*}
which is the first display equation in the lemma statement. Here, at
step~$\onestar$ we have used $-\hgamma_{i,\sigma(i)} =
|\hgamma_{i,\sigma(i)}|$, $\hgamma_{i,j} = |\hgamma_{i,j}|$, $j \neq
\sigma(i)$, and $|\hgamma_{i,\sigma(i)}| = \sum_{j \neq \sigma(i)}
|\hgamma_{i,j}|$, \ie, $\sum_{j \neq \sigma(i)} |\hgamma_{i,j}| /
|\hgamma_{i,\sigma(i)}| = 1$.

The non-negativity of the coefficient of $\temp$ in the above expression
follows from $|\hgamma_{i,\sigma(i)}| \geq |\hgamma_{i,j}|$, $j \neq
\sigma(i)$, which is a consequence of the above-mentioned relation
$|\hgamma_{i,\sigma(i)}| = \sum_{j \neq \sigma(i)} |\hgamma_{i,j}|$.

On the other hand, rewriting these summations such that all the main
summations are over $j \in \setJ$, we obtain the second display equation in
the lemma statement.

\section{Proof of Theorem~\ref{theorem:FBethe:extremal:behavior:1}}
\label{sec:proof:theorem:FBethe:extremal:behavior:1}

From the assumptions in the theorem statement it follows that
$|\hgamma_{i,\sigma(i)}| = - \hgamma_{i,\sigma(i)}$ for all $i \in \setI$ and
that $|\hgamma_{i,j}| = \hgamma_{i,j}$ for all $i \in \setI$, $j \in \setJ
\setminus \{ \sigma(i) \}$ (see also the proof of
Lemma~\ref{lemma:linear:behavior:HBethe:1} in
Appendix~\ref{sec:proof:lemma:linear:behavior:HBethe:1}). Then,
\begin{align*}
  &
  \UBethe\big( \matrgamma(\temp) \big) \\
    &\onestareq
       -
       \sum_{i}
         (1 \! + \! \temp \hgamma_{i,\sigma(i)})
         \log( \theta_{i,\sigma(i)} )
       -
       \sum_{i}
         \sum_{j \neq \sigma(i)}
           (\temp \hgamma_{i,j})
           \log( \theta_{i,j} ) \\
    &\twostarseq
       -
       \sum_{i}
         \log( \theta_{i,\sigma(i)} )
       -
       \temp
       \sum_{i}
         \sum_{j \neq \sigma(i)}
           |\hgamma_{i,j}|
           \log
             \left(
               \frac{\theta_{i,j}}
                    {\theta_{i,\sigma(i)}}
             \right) \\
    &\threestarseq
       C
       -
       \temp
       \sum_{i}
         \sum_{j \neq \sigma(i)}
           |\hgamma_{i,j}|
           \log
             \left(
               \frac{\theta_{i,j}}
                    {\theta_{i,\sigma(i)}}
             \right),
\end{align*}
where at step~$\onestar$ we have used
Corollary~\ref{cor:FBethe:as:function:of:gamma:1}, where at step~$\twostars$
we have used that $\sum_j \hgamma_{i,j} = 0$ holds for every $i \in \setI$,
\ie, that $- \hgamma_{i,\sigma(i)} = \sum_{j \neq \sigma(i)} \hgamma_{i,j} =
\sum_{j \neq \sigma(i)} |\hgamma_{i,j}|$ holds for every $i \in \setI$, and
where at step~$\threestars$ we have defined $C \defeq - \sum_{i} \log(
\theta_{i,\sigma(i)} )$. (Note that there is no $O(\temp^2)$ term in the above
expressions.) Then
\begin{align}
  \FBethe&\big( \matrgamma(\temp) \big) \nonumber \\
    &\onestareq
       \UBethe(\matrgamma) - \HBethe(\matrgamma) \nonumber \\
    &\twostarseq
       C
       -
       \temp
       \sum_{i}
         \sum_{j \neq \sigma(i)}
           |\hgamma_{i,j}|
           \log
             \left(
               \frac{\theta_{i,j}}
                    {\theta_{i,\sigma(i)}}
             \right) \nonumber \\
    &\quad\ 
       -
       \temp
       \sum_i
       |\hgamma_{i,\sigma(i)}|
         \cdot 
         \left(
           -
           \sum_{j \neq \sigma(i)}
           \frac{|\hgamma_{i,j}|}
                {|\hgamma_{i,\sigma(i)}|}
           \log
             \left(
               \frac{|\hgamma_{i,j}|}
                    {|\hgamma_{i,\sigma(i)}|}
             \right)
         \right) \nonumber \\
    &\quad\ 
       +
       O(\temp^2) \nonumber \\
    &\threestarseq
       C
       -
       \temp
       \sum_{i}
         \sum_{i' \neq i}
           |\hgamma_{i,\sigma(i')}|
           \log
             \left(
               \frac{\theta_{i,\sigma(i')}}
                    {\theta_{i,\sigma(i)}}
             \right) \nonumber \\
     &\quad\ 
       -
       \temp
       \sum_i
       |\hgamma_{i,\sigma(i)}|
         \cdot 
         \left(
           -
           \sum_{i' \neq i}
           \frac{|\hgamma_{i,\sigma(i')}|}
                {|\hgamma_{i,\sigma(i)}|}
           \log
             \left(
               \frac{|\hgamma_{i,\sigma(i')}|}
                    {|\hgamma_{i,\sigma(i)}|}
             \right)
         \right) \nonumber \\
    &\quad\ 
       +
       O(\temp^2) \nonumber \\
    &\fourstarseq
       C
       -
       \temp
       \sum_{i}
         \sum_{i' \neq i}
           \underbrace{
             \mu_i
             \cdot
             p_{i,i'}
           }_{= \ Q_{i,i'}}
           \cdot
           \big[
             -
             \log(p_{i,i'})
             +
             T_{i,i'}
           \big]
       +
       O(\temp^2),
         \label{eq:FBethe:behavior:1}
\end{align}
where at step~$\onestar$ we have used
Corollary~\ref{cor:FBethe:as:function:of:gamma:1}, where at step~$\twostars$
we have inserted the above expression for $\UBethe(\matrgamma)$ and the
expression for $\HBethe(\matrgamma)$ from
Lemma~\ref{lemma:linear:behavior:HBethe:1}, where at step~$\threestars$ we
have replaced the summations over $j \in \setJ$, $j \neq \sigma(i)$, by
summations over $i' \in \setI$, $\sigma(i') \neq \sigma(i)$, \ie, by
summations over $i' \in \setI$, $i' \neq i$, and where at step~$\fourstars$ we
have introduced the definitions
\begin{align}
  \mu_i
    &\defeq
       |\hgamma_{i,\sigma(i)}|
              \label{eq:mu:def:1} \\
  p_{i,i'}
    &\defeq
       \frac{|\hgamma_{i,\sigma(i')}|}
            {|\hgamma_{i,\sigma(i)}|},
              \label{eq:p:def:1} \\
  Q_{i,i'}
    &\defeq
       \mu_i \cdot p_{i,i'}
     = |\hgamma_{i,\sigma(i')}|,
      \label{eq:Q:def:1} \\
  T_{i,i'}
    &\defeq
       \log
         \left(
           \frac{\theta_{i,\sigma(i')}}
                {\theta_{i,\sigma(i)}}
         \right),
           \label{eq:T:def:1}
\end{align}
for all $(i,i') \in \setI \times \setI$ with $i \neq i'$. One can verify that
the assumptions on $\hmatrgamma$ imply that
\begin{alignat*}{2}
  \sum_{i}
     \mu_i
     &= 1, \\
  \sum_{i' \neq i}
     p_{i,i'}
     &= 1
     \quad
     &&\text{(for all $i \in \setI$)}, \\
  \sum_{i' \neq i}
    Q_{i,i'}
     &= \mu_i
     \quad
     &&\text{(for all $i \in \setI$)}, \\
  \sum_{i \neq i'}
    Q_{i,i'}
     &= \mu_{i'}
     \quad
     &&\text{(for all $i' \in \setI$)}, \\
  \sum_i
    \sum_{i' \neq i}
      Q_{i,i'}
     &= 1.
\end{alignat*}

\begin{figure}
  \begin{center}
    \epsfig{file=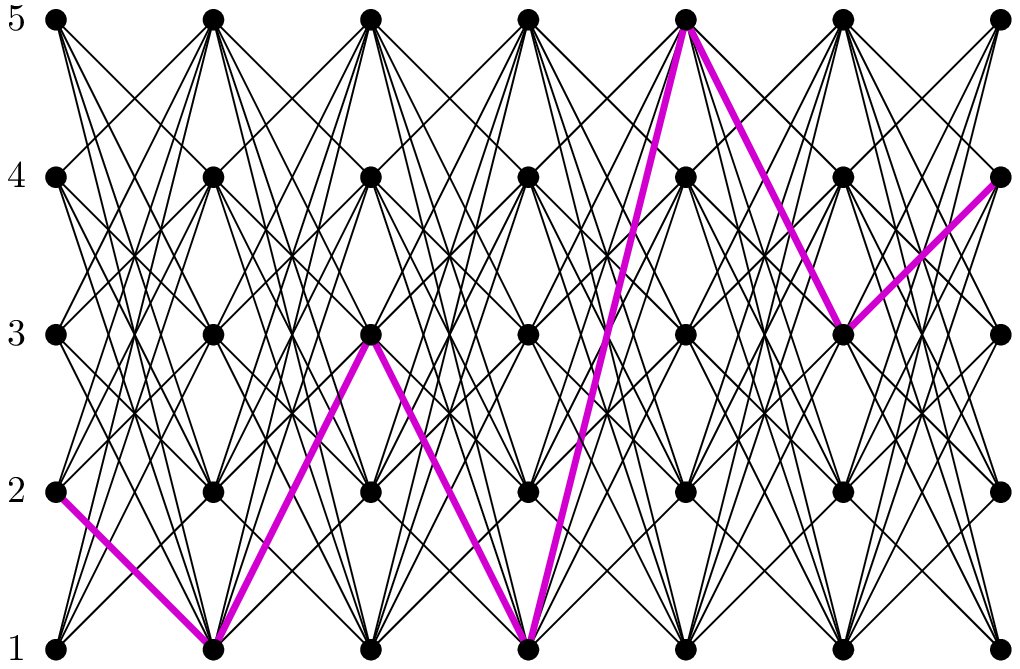, width=0.80\linewidth}
  \end{center}
  \caption{Trellis for the random walk described in
    Appendix~\ref{sec:proof:theorem:FBethe:extremal:behavior:1}. (Here $n =
    5$.) Highlighted is an instance of a possible walk.}
  \label{fig:trellis:random:walk:1}
\end{figure}

In order to obtain the theorem statement, we need to maximize the coefficient
of $(-\temp)$ in~\eqref{eq:FBethe:behavior:1}. Before doing this, let us quickly
discuss the meaning of this coefficient.

Namely, consider the trellis in Fig.~\ref{fig:trellis:random:walk:1} with
state space $\setI$ (\ie, with $n$ states) and where a trellis section has a
branch from state $i \in \setI$ to state $i' \in \setI$ if and only if $i \neq
i'$. It is straightforward to see that there is a bijection between, on the
one hand, the set of all left-to-right walks in the time-invariant trellis
shown in Fig.~\ref{fig:trellis:random:walk:1}, and, on the other hand, the
set of backtrackless walks in $\graphN(\matrtheta)$ (see
Fig.~\ref{fig:ffg:permanent:1}) that were mentioned after
Lemma~\ref{lemma:linear:behavior:HBethe:1}. In particular, going from state $i
\in \setI$ to state $i' \in \setI \setminus \{ i \}$ in the trellis of
Fig.~\ref{fig:trellis:random:walk:1} corresponds to the two half-steps of
going from node $i \in \setI$ to node $\sigma(i') \in \setJ$ and then to node
$i' \in \setI$ in $\graphN(\matrtheta)$. With this, translating
(backtrackless) random walks to left-to-right random walks in the trellis in
Fig.~\ref{fig:trellis:random:walk:1}, we obtain that
\begin{itemize}

\item $\mu_i$ is the probability of being in state $i$,

\item $p_{i,i'}$ is the probability of going to state $i' \neq i$, conditioned
  on being in state $i$,

\item $Q_{i,i'}$ is the probability of being in state $i$ and then going to
  state $i' \neq i$,

\item $-\sum_{i} \sum_{i' \neq i} \mu_i p_{i,i'} \log(p_{i,i'})$ is the
  entropy rate of (the Markov chain corresponding to) the random walk on this
  trellis,

\item $T_{i,i'}$ is a branch metric, 

\item $\sum_{i} \sum_{i' \neq i} \mu_i p_{i,i'} T_{i,i'}$ is the average
  branch metric of the random walk on this trellis,

\item and maximizing the coefficient of $(-\temp)$ in the above expression for
  $\FBethe\big( \matrgamma(\temp) \big)$ (see~\eqref{eq:FBethe:behavior:1})
  means to find the (time-invariant) left-to-right random walk on this trellis
  that maximizes
  \begin{align*}
    \sum_{i}
      \sum_{i' \neq i}
        \mu_i
        \cdot
        p_{i,i'}
        \cdot
        \big[
          -
          \log(p_{i,i'})
          +
          T_{i,i'}
        \big],
  \end{align*}
  \ie, the sum of the entropy rate and the average branch metric of the random
  walk. (In statistical physics terms, this expression can be considered to be
  some negative free energy function.)

\end{itemize}

The purpose of rewriting the above expression in the way we did, was so that
it is very close to the notation used
in~\cite[Lemma~44]{Vontobel:Kavcic:Arnold:Loeliger:08:1} that solved exactly
the above maximization problem. (Note that related problems were also solved
in~\cite{Justesen:Hoeholdt:84:1} and~\cite{Khayrallah:Neuhoff:96:1}.)

As was shown in~\cite[Lemma~44]{Vontobel:Kavcic:Arnold:Loeliger:08:1}, the
maximal value of
\begin{align*}
  \sum_{i}
    \sum_{i' \neq i}
      \underbrace{
        \mu_i
        \cdot
        p_{i,i'}
      }_{= \ Q_{i,i'}}
      \cdot
      \big[
        -
        \log(p_{i,i'})
        +
        T_{i,i'}
      \big]
\end{align*}
is $\log(\rho)$ and is attained by
\begin{align*}
  \mu^*_i
    &= \kappa
       \cdot
       \lefteigvect_i
       \cdot
       \righteigvect_i, \\
  p^*_{i,i'}
    &= \begin{cases}
         \frac{\righteigvect_{i'}}
              {\righteigvect_i}
         \cdot
         \frac{A_{i,i'}}
              {\rho}
           & \text{(if $i \neq i'$)} \\
         0
           & \text{(otherwise)}
       \end{cases}, \\
  Q^*_{i,i'}
    &= \mu^*_i
       \cdot
       p^*_{i,i'}
     = \begin{cases}
         \kappa
         \cdot
         \frac{\lefteigvect_i
               \cdot
               A_{i,i'}
               \cdot
               \righteigvect_{i'}}
              {\rho}
           & \text{(if $i \neq i'$)} \\
         0
           & \text{(otherwise)}
       \end{cases},
\end{align*}
where $\matrA$, $\rho$, $\vlefteigvect$, and $\vrighteigvect$ are defined in
the theorem statement, and where $\kappa$ is a normalization constant such
that $\sum_i \mu^*_i = 1$. Note that $\matrA$, called the noisy adjacency
matrix in~\cite[Lemma~44]{Vontobel:Kavcic:Arnold:Loeliger:08:1}, is such that
$A_{i,i'} = \exp(T_{i,i'})$ for $i \neq i'$ and such that $A_{i,i} = 0$.

Because $\matrA$ contains only non-negative entries, $\rho$ is the so-called
Perron eigenvector of $\matrA$, and $\vlefteigvect$ and $\vrighteigvect$ are
the so-called left and right, respectively, Perron eigenvectors of $\matrA$;
one can show that these two vectors contain only non-negative entries.

Translating this result back using~\eqref{eq:mu:def:1}, \eqref{eq:p:def:1},
and~\eqref{eq:Q:def:1}, we obtain the result given in the theorem statement.

\section{Proof of Lemma~\ref{lemma:sum:product:algorithm:LR:update:rules:1}}
\label{sec:proof:lemma:sum:product:algorithm:LR:update:rules:1}

We start by formulating the SPA message update rule for functions node $g_i$,
$i \in \setI$, at iteration $t \geq
1$. Following~\cite{Kschischang:Frey:Loeliger:01, Forney:01:1, Loeliger:04:1},
we have for every $i \in \setI$, every $j \in \setJ$, and every $\bar a_{i,j}
\in \setA_{i,j}$,
\begin{align*}
  \murightijt(\bar a_{i,j})
    &\defeq
        \frac{1}{C_{i,j}}
        \cdot
        \sum_{\va_i \in \setA_i \atop a_{i,j} = \bar a_{i,j}}
          \!\!\!\!
          f_i(\va_i)
          \cdot
          \prod_{j' \neq j}
            \muleftijptmo(a_{i,j'}),
\end{align*}
where $C_{i,j}$ is some suitable normalization constant. Consequently, the
update of the likelihood ratio reads
\begin{align*}
  \Lambdarightijt
    &\defeq
       \frac{\murightijt(0)}
            {\murightijt(1)}
     = \frac{\sum_{\va_i \in \setA_i \atop a_{i,j} = 0}
               f_i(\va_i)
               \cdot
               \prod_{j'' \neq j}
                 \muleftijtmo(a_{i,j''})}
            {\sum_{\va_i \in \setA_i \atop a_{i,j} = 1}
               f_i(\va_i)
               \cdot
               \prod_{j'' \neq j}
                 \muleftijpptmo(a_{i,j''})} \\
    &\onestareq
       \frac{\sum_{j' \neq j}
               \sqrt{\theta_{i,j'}}
               \cdot
               \muleftijptmo(1)
                 \cdot
                 \prod_{j'' \neq j, j'}
                   \muleftijpptmo(0)}
            {\sqrt{\theta_{i,j}}
             \cdot
             \prod_{j'' \neq j}
               \muleftijpptmo(0)} \\
    &\twostarseq
       \frac{1}
            {\sqrt{\theta_{i,j}}}
       \cdot
       \sum_{j' \neq j}
         \sqrt{\theta_{i,j'}}
         \cdot
         \left(
           \Lambdaleftijptmo
         \right)^{-1},
\end{align*}
where at step~$\onestar$ we have used $\setA_i = \{ \vu_j \ | \ j \in \setJ
\}$ for simplifying the numerator, and where at step~$\twostars$ we have used
the definition of $\Lambdaleftijptmo$, $j' \neq j$. This yields the first
expression in the lemma statement. The second expression is obtained
analogously by considering the SPA message update rule for function nodes
$g_j$, $j \in \setJ$, at iteration $t \geq 1$.

Now we turn our attention to computing the beliefs at the function nodes
$g_i$, $i \in \setI$, at iteration $t \geq
0$. Following~\cite{Kschischang:Frey:Loeliger:01, Forney:01:1, Loeliger:04:1},
we have for every $i \in \setI$ and every $\va_i \in \setA_i$,
\begin{align*}
  \belivait
    &= \frac{1}{C_i}
       \cdot
       f_i(\va_i)
       \cdot
       \prod_{j}
         \muleftijt(a_{i,j}),
\end{align*}
where $C_i$ is chosen such that $\sum_{\va_i} \belivait = 1$. In particular,
for $\va_i = \vu_j$, $j \in \setJ$, we get
\begin{align*}
  \belivait
    &= \frac{1}{C_i}
       \cdot
       f_i(\va_i)
       \cdot
       \prod_{j'}
         \muleftijpt(a_{i,j'}) \\
    &= \frac{1}{C_i}
       \cdot
       f_i(\va_i)
       \cdot
       \left(
         \prod_{j'}
           \muleftijpt(0)
       \right)
       \cdot
       \prod_{j'}
         \frac{\muleftijpt(a_{i,j'})}
              {\muleftijpt(0)} \\
    &= \frac{1}{C_i}
       \cdot
       \sqrt{\theta_{i,j}}
       \cdot
       \left(
         \prod_{j'}
           \muleftijpt(0)
       \right)
       \cdot
       \udLambdaleftijt.
\end{align*}
Because $C_i$ and the expression in the parentheses are independent of $j$, we
have just verified the third expression in the lemma statement. The fourth
expression in the lemma statement is obtained analogously by considering the
beliefs at function nodes $g_j$, $j \in \setJ$, at iteration $t \geq 1$.

\section{Proof of Lemma~\ref{lemma:pseudo:dual:Bethe:free:energy:1}}
\label{sec:proof:lemma:pseudo:dual:Bethe:free:energy:1}

The pseudo-dual function of the Bethe free energy function is given by
evaluating the Lagrangian of the Bethe free energy function at a stationary
point~\cite{Regalia:Walsh:07:1}. Therefore, in a first step, we want to write
down the Lagrangian of the Bethe free energy function. To that end, we take
the Bethe free energy function as in Definition~\ref{def:Bethe:free:energy:1},
\ie,
\begin{align*}
  &
  \FBethe
    \big(
      \{ \vbel_i \}, \{ \vbel_j \}, \{ \vbel_e \}
    \big)
     = \sum_i
         \UBethesub{i}(\vbel_i)
       +
       \sum_j
         \UBethesub{j}(\vbel_j) \\
    &\quad\quad\quad\quad
       -
       \sum_i
         \HBethesub{i}(\vbel_i)
       -
       \sum_j
         \HBethesub{j}(\vbel_j)
       +
       \sum_e
         \HBethesub{e}(\vbel_e).
\end{align*}
(For the purposes of this appendix, the expression for $\FBethe$ in
Definition~\ref{def:Bethe:free:energy:1} is somewhat more convenient than the
one in Lemma~\ref{lemma:Bethe:free:energy:as:function:of:gamma:1}.)

Now, introducing a Lagrange multiplier for the edge consistency constraints
(but not for the other constraints imposed by the local marginal polytope
$\lmpB$, see Definition~\ref{def:permanent:normal:factor:graph:lmp:1}), we
obtain the relevant Lagrangian
\begin{align*}
  &\!\!\!\LBethe\big(\{ \vbel_i \}, \{ \vbel_j \}, \{ \vbel_e \}, 
                     \{ \vlambdalefte \}, \{ \vlambdarighte \} \big) \\
    &= \FBethe(\{ \vbel_i \}, \{ \vbel_j \}, \{ \vbel_e \}) \\
    &\quad\quad
       - \sum_{e = (i,j)}
           \sum_{a_e}
             \lambdalefteae
             \cdot
             \left(
               \sum_{\va_i: \, a_{i,e} = a_e} 
                 \bel_{i, \va_i}
               -
               \bel_{e, a_e}
             \right) \\
    &\quad\quad
       - \sum_{e = (i,j)}
           \sum_{a_e}
             \lambdarighteae
             \cdot
             \left(
               \sum_{\va_j: \, a_{j,e} = a_e} 
                 \bel_{j, \va_j}
               -
               \bel_{e, a_e}
             \right),
\end{align*}
Because $\FBethe$ is convex in $\{ \vbel_i \}_i$ and $\{ \vbel_j \}_j$, but
concave in $\{ \vbel_e \}_e$, the pseudo-dual function of $\FBethe$ is given
by
\begin{align*}
  &
  \FpdualBethe\big( \{ \vlambdalefte \}, \{ \vlambdarighte \} \big) \\
    &\ 
     = \max_{\{ \vbel_e \}} \ \ 
         \min_{\{ \vbel_i \}, \, \{ \vbel_j \}}
           \LBethe\big( \{ \vbel_i \}, \{ \vbel_j \}, \{ \vbel_e \},
                   \{ \vlambdalefte \}, \{ \vlambdarighte \} \big),
\end{align*}
where the maximization/minimization is over all $\{ \vbel_e \}_e$, $\{ \vbel_i
\}_i$, $\{ \vbel_j \}_j$ that satisfy the constraints imposed by the local
marginal polytope $\lmpB$, except for the edge consistency constraints. We
obtain the maximizing $\{ \vbel_e \}_e$ and the minimizing $\{ \vbel_i \}_i$,
$\{ \vbel_j \}_j$ by setting suitable partial derivatives to zero. This
yields,
\begin{align*}
  \bel_{i,\va_i}
    &= \frac{1}{Z_i}
       \cdot
       g_i(\va_i)
       \cdot
       \prod_{e: \, i(e)=i}
         \exp
           \left(
             \lambdaleftext{e,a_{i,j(e)}}
           \right), \\
  \bel_{j,\va_j}
    &= \frac{1}{Z_j}
       \cdot
       g_j(\va_j)
       \cdot
       \prod_{e: \, j(e)=j}
           \exp
             \left(
               \lambdarightext{e,a_{i(e),j}}
             \right), \\
  \bel_{e,a_e}
    &= \frac{1}{Z_e}
       \cdot
       \exp
         \left(
           \lambdalefteae
         \right)
       \cdot
       \exp
         \left(
           \lambdarighteae
         \right),
\end{align*}
where $i(e)$ and $j(e)$ give the label of the, respectively, left and right
vertex to which $e$ is incident, and where $\{ Z_i \}_i$, $\{ Z_j \}_j$, and
$\{ Z_e \}_e$ are suitable normalization constants such that relevant sums are
equal to one.

Now, plugging these beliefs into the Lagrangian, we obtain (after cancelling
several terms) the expression
\begin{align*}
  \FpdualBethe&\big( \{ \vlambdalefte \}, \{ \vlambdarighte \} \big) \\
    &= -
       \sum_i
         \log(Z_i)
       -
       \sum_j
         \log(Z_j)
       +
       \sum_e
         \log(Z_e) \\
    &= -
       \sum_i
         \log
           \left(
             \sum_{\va_i}
               g_i(\va_i)
               \cdot
               \prod_{e: \, i(e)=i}
               \exp
               \left(
                 \lambdaleftext{e,a_{i,j(e)}}
               \right)
           \right) \\
    &\quad\ 
       -
       \sum_j
         \log
           \left(
             \sum_{\va_j}
               g_j(\va_j)
               \cdot
               \prod_{e: \, j(e)=j}
               \exp
               \left(
                 \lambdarightext{e,a_{i(e),j}}
               \right)
           \right) \\
    &\quad\ 
       +
       \sum_e
         \log
           \left(
             \sum_{a_e}
               \exp
               \left(
                 \lambdalefteae
                 +
                 \lambdarighteae
               \right)
           \right).
\end{align*}

We proceed by using some details of the definition of
$\graphN(\matrtheta)$. Namely, using the definition of the local function
nodes and taking advantage of the binary alphabet $\setA_e = \{ 0, 1 \}$, $e
\in \setE$, we obtain (after some simplifications)
\begin{align*}
  &
  \FpdualBethe\big( \{ \vlambdalefte \}, \{ \vlambdarighte \} \big) \\
    &= -
       \sum_i
         \log
           \left(
             \sum_j
               \sqrt{\theta_{i,j}}
               \cdot
               \exp
               \left(
                 \lambdaleftext{(i,j),1}
                 -
                 \lambdaleftext{(i,j),0}
               \right)
           \right) \\
    &\quad\ 
       -
       \sum_j
         \log
           \left(
             \sum_i
               \sqrt{\theta_{i,j}}
               \cdot
               \exp
               \left(
                 \lambdarightext{(i,j),1}
                 -
                 \lambdarightext{(i,j),0}
               \right)
           \right) \\
    &\quad\ 
       +
       \sum_e
         \log
           \left(
             1
             +
             \exp
               \left(
                 \big(
                   \lambdaleftext{e,1}
                   -
                   \lambdaleftext{e,0}
                 \big)
                 +
                 \big(
                   \lambdarightext{e,1}
                   -
                   \lambdarightext{e,0}
                 \big)
               \right)
             \right)
\end{align*}

From the results in~\cite{Yedidia:Freeman:Weiss:05:1} it follows that at a
fixed point of the SPA, the quantity $\lambdaleftext{(i,j),0} -
\lambdaleftext{(i,j),1}$ represents the log-likelihood ratio of the left-going
message along the edge $(i,j)$, and the quantity $\lambdarightext{(i,j),0} -
\lambdarightext{(i,j),1}$ represents the log-likelihood ratio of the
right-going message along the edge $(i,j)$. Clearly, for every edge $(i,j) \in
\setI \times \setJ$, these quantities are related to the inverse likelihood
ratios by
\begin{align*}
  \udLambdaleftij
    &= \exp
         \left(
           \lambdaleftext{(i,j),1}
           -
           \lambdaleftext{(i,j),0}
         \right), \\
  \udLambdarightij
    &= \exp
         \left(
           \lambdarightext{(i,j),1}
           -
           \lambdarightext{(i,j),0}
         \right),
\end{align*}
respectively. Therefore, we get
\begin{align*}
  \FpdualBethe&\big( \{ \udLambdaleftij \}, \{ \udLambdarightij \} \big) \\
    &= -
       \sum_i
         \log
           \left(
             \sqrt{\theta_{i,j}}
             \cdot
             \udLambdaleftij
           \right)
       -
       \sum_j
         \log
           \left(
             \sqrt{\theta_{i,j}}
             \cdot
             \udLambdarightij
           \right) \\
    &\quad\ 
       +
       \sum_{i,j}
         \log
           \left(
             1
             +
             \udLambdaleftij
             \cdot
             \udLambdarightij
           \right),
\end{align*}
which is the expression in the lemma statement.

Although the interpretation of the log-likelihood ratios was given by looking
at fixed points of the SPA, it is not difficult to see that we can evaluate
this last expression for any set of inverse likelihood ratios.

\section{Proof of Theorem~\ref{theorem:convergence:spa:1}}
\label{sec:proof:theorem:convergence:spa:1}

This appendix has two subsections. The first subsection considers the case
where the global minimum of $\FBethe$ is achieved at a vertex of
$\setGamma{n}$, whereas the second subsection considers the case where the
global minimum of $\FBethe$ is achieved in the interior of $\setGamma{n}$.

For ease of reference, we reproduce here the SPA message update rules from
Lemma~\ref{lemma:sum:product:algorithm:LR:update:rules:1}, \ie,
\begin{alignat}{2}
  \hskip-0.3cm
  \udLambdarightijt
    &= \frac{\sqrt{\theta_{i,j}}}
            {\sum_{j' \neq j}
               \sqrt{\theta_{i,j'}}
               \cdot
               \udLambdaleftijptmo},
       &&\ t \geq 1, \ (i,j) \in \setI \times \setJ,
           \label{eq:spa:update:repr:1} \\
  \hskip-0.3cm
  \udLambdaleftijt
    &= \frac{\sqrt{\theta_{i,j}}}
            {\sum_{i' \neq i}
               \sqrt{\theta_{i',j}}
               \cdot
               \udLambdarightipjt
            },
       &&\ t \geq 1, \ (i,j) \in \setI \times \setJ.
           \label{eq:spa:update:repr:2}
\end{alignat}
In both parts of this appendix, the main task will be to exhibit a contraction
operation of a suitably chosen subset of the SPA messages.

\subsection{Global Minimum of $\FBethe$ is Achieved at a Vertex of
                    $\setGamma{n}$}

Let $\matrgamma \in \codeC$ be the vertex of $\setGamma{n}$ that uniquely
minimizes $\FBethe$. This means that $\matrgamma$ corresponds to the
permutation $\sigma_{\matrgamma}$. (In the following, we will use the
short-hands $\sigma \defeq \sigma_{\matrgamma}$ and $\bsigma \defeq
\sigma_{\matrgamma}^{-1}$.)

From~\eqref{eq:spa:update:repr:1} it follows that
$\Lambdarightijextext{\sigma(i)}{t} = 1 /
\udLambdarightijextext{\sigma(i)}{t}$, $i \in \setI$, can be written
as\footnote{For simplicity, because $j$ does not appear on the left-hand side
  of this equation, we use $j$ as a summation variable on the right-hand
  side. This is in contrast to~\eqref{eq:spa:update:repr:1} where $j$ appears
  on the left-hand side and where the summation variable on the right-hand
  side is $j'$.}
\begin{align*}
  \Lambdarightijextext{\sigma(i)}{t}
    &= \frac{1}{\sqrt{\theta_{i,\sigma(i)}}}
       \cdot
       \sum_{j \neq \sigma(i)}
         \sqrt{\theta_{i,j}}
         \cdot
         \udLambdaleftijtmo,
           \quad t \geq 1, \ i \in \setI.
\end{align*}
On the other hand, for $i \in \setI$ and $j \neq \sigma(i)$ the SPA message
update equation in~\eqref{eq:spa:update:repr:2} implies
\begin{align*}
  \udLambdaleftijtmo
    &= \frac{\sqrt{\theta_{i,j}}}
            {\sum_{i' \neq i}
               \sqrt{\theta_{i',j}}
               \cdot
               \udLambdarightipjtmo
            } \\
     &= \frac{\sqrt{\theta_{i,j}}}
             {\sqrt{\theta_{\bsigma(j),j}}
              \cdot
              \udLambdaright_{\bsigma(j),j}^{(t-1)}
             }
        \cdot
        \frac{1}{1
                 +
                 \sum\limits_{i' \neq i, \bsigma(j)}
                   \frac{\sqrt{\theta_{i',j}}
                         \cdot
                         \udLambdarightipjtmo
                        }
                        {\sqrt{\theta_{\bsigma(j),j}}
                         \cdot
                         \udLambdaright_{\bsigma(j),j}^{(t-1)}
                        }
                } \\
     &\leq
        \frac{\sqrt{\theta_{i,j}}}
             {\sqrt{\theta_{\bsigma(j),j}}
              \cdot
              \udLambdaright_{\bsigma(j),j}^{(t-1)}} \\
     &= \frac{\sqrt{\theta_{i,j}}}
             {\sqrt{\theta_{\bsigma(j),j}}}
        \cdot
        \Lambdaright_{\bsigma(j),j}^{(t-1)},
          \quad t \geq 1, \ i \in \setI, \ j \neq \sigma(i),
\end{align*}
where the inequality follows from the fact that all terms in the summation
$\sum_{i' \neq i, \bsigma(j)}$ are non-negative. Then, combining the two above
expressions, we obtain
\begin{align*}
  \Lambdarightijextext{\sigma(i)}{t}
    &\leq
       \sum_{j \neq \sigma(i)}
         \frac{\theta_{i,j}}
              {\sqrt{\theta_{i,\sigma(i)}}
               \sqrt{\theta_{\bsigma(j),j}}}
        \cdot
        \Lambdaright_{\bsigma(j),j}^{(t-1)}, 
          \quad t \geq 1, \ i \in \setI.
\end{align*}
Rearranging terms, we obtain
\begin{align*}
  \frac{\Lambdarightijextext{\sigma(i)}{t}}
       {\sqrt{\theta_{i,\sigma(i)}}}
    &\leq
       \sum_{j \neq \sigma(i)}
         \frac{\theta_{i,j}}
              {\theta_{i,\sigma(i)}}
        \cdot
        \frac{\Lambdaright_{\bsigma(j),j}^{(t-1)}}
             {\sqrt{\theta_{\bsigma(j),j}}} \\
    &= \sum_{i' \neq i}
         \frac{\theta_{i,\sigma(i')}}
              {\theta_{i,\sigma(i)}}
        \cdot
        \frac{\Lambdaright_{i',\sigma(i')}^{(t-1)}}
             {\sqrt{\theta_{i',\sigma(i')}}},
               \quad t \geq 1, \ i \in \setI.
\end{align*}
Now, for every $t \geq 0$, consider the length-$n$ vector
$\overrightarrow{\vm}^{(t)}$ whose $i$th entry is
$\Lambdarightijextext{\sigma(i)}{t} / \sqrt{\theta_{i,\sigma(i)}}$. Grouping
several of the above inequalities together, we obtain the vector inequality
\begin{align}
  \overrightarrow{\vm}^{(t)}
    &\leq
       \matrA
       \cdot
       \overrightarrow{\vm}^{(t-1)},
         \quad t \geq 1,
           \label{eq:SPA:message:subvector:update:1}
\end{align}
where the vector inequality has to be understood component-wise, and where the
$n \times n$ matrix $\matrA$ was defined in
Theorem~\ref{theorem:FBethe:extremal:behavior:1} for the vertex $\matrgamma$
of $\setGamma{n}$. Let $\rho$ be the maximal (real) eigenvalue of
$\matrA$. Then, Corollary~\ref{cor:FBethe:minimum:at:vertex:1} and the
assumption that $\matrgamma$ is the unique minimizer of $\FBethe$ allow us to
conclude that $\rho < 1$. However, because $\rho < 1$ implies that all
eigenvalues of $\matrA$ have magnitude strictly smaller than $1$, the update
equation in~\eqref{eq:SPA:message:subvector:update:1} represents a
contraction, and so
\begin{align*}
  \big\lVert
    \overrightarrow{\vm}^{(t)}
  \big\rVert_2
    &\ \xrightarrow{t \to \infty} \ 
       0.
\end{align*}
Therefore,
\begin{align*}
  \Lambdarightijextext{\sigma(i)}{t}
    &\ \xrightarrow{t \to \infty} \ 
       0, \quad i \in \setI.
\end{align*}
A similar argument shows that
\begin{align*}
  \Lambdaleftijextext{\bsigma(j)}{t}
    &\ \xrightarrow{t \to \infty} \ 
       0, \quad j \in \setJ.
\end{align*}
Finally, from~\eqref{eq:spa:update:repr:1} and~\eqref{eq:spa:update:repr:2}
and the above results it follows that
\begin{align*}
  \udLambdarightijt
    &\ \xrightarrow{t \to \infty} \ 
       0, \quad i \in \setI, \ j \in \setJ, \ j \neq \sigma(i), \\
  \udLambdaleftijt
    &\ \xrightarrow{t \to \infty} \ 
       0, \quad i \in \setI, \ j \in \setJ, \ j \neq \sigma(i).
\end{align*}
All these quantities converge to zero exponentially fast.

When $\FBethe$ achieves its minimum in the interior of $\setGamma{n}$, then we
have equality between $\FBethe$ and $\FpdualBethe$ at stationary points of the
SPA. However, we also have equality in the present case. Namely, evaluating
$\FpdualBethe$ (see Lemma~\ref{lemma:pseudo:dual:Bethe:free:energy:1}) for
the above messages, we obtain
\begin{align*}
  \FpdualBethe
    \big(
      \bigl\{ \udLambdaleftijt \bigr\}, 
      \bigl\{ \udLambdarightijt \bigr\}
    \big)
    &\ \xrightarrow{t \to \infty} \ 
       -
       \sum_i
         \log(\theta_{i,\sigma(i)}),
\end{align*}
which indeed equals $\FBethe(\matrgamma)$. From $\rho < 1$ and
$\FBethe(\matrgamma) = - \log\bigl( \permBethe(\matrtheta) \bigr)$ it also
follows that
\begin{align*}
  \left|\,
    \exp
    \bigg(\!\!\!
      -
      \FpdualBethe
        \Big(
          \bigl\{ \udLambdaleftijt \bigr\}, 
          \bigl\{ \udLambdarightijt \bigr\}
        \Big) \!\!
    \bigg)
    -
    \permBethe(\matrtheta)
  \right|
    &\leq
       C
       \cdot
       \e^{-\nu \cdot t}
\end{align*}
for some suitable constants $C, \nu \in \Rpp$.

\subsection{Global Minimum of $\FBethe$ is Achieved in the Interior of
                    $\setGamma{n}$}

In Corollary~\ref{cor:convexity:Bethe:free:energy:function:1} we established
that the Bethe free energy function of $\graphN(\matrtheta)$ is convex, \ie,
it does not have stationary points besides the global minimum. Therefore,
using a theorem by Yedidia, Freeman, Weiss~\cite{Yedidia:Freeman:Weiss:05:1},
we know that fixed points of the SPA correspond to the global minimum of the
Bethe free energy function.

Let $\bigl\{ \udLambdaleftij \bigr\}_{i,j}$, $\bigl\{ \udLambdarightij
\bigr\}_{i.j}$ be inverse likelihood ratios that constitute a fixed point of
the SPA update rules
in~\eqref{eq:spa:update:repr:1}--\eqref{eq:spa:update:repr:2}. As such, these
inverse likelihoods must satisfy
\begin{align}
  \udLambdarightij
    &= \frac{\sqrt{\theta_{i,j}}}
            {\sum_{j' \neq j}
               \sqrt{\theta_{i,j'}}
               \cdot
               \udLambdaleftijp
            },
                 \label{eq:spa:fixed:point:1:1} \\
  \udLambdaleftij
    &= \frac{\sqrt{\theta_{i,j}}}
            {\sum_{i' \neq i}
               \sqrt{\theta_{i',j}}
               \cdot
               \udLambdarightipj
            },
              \label{eq:spa:fixed:point:1:2}
\end{align}
for every $(i,j) \in \setE$. Note that these SPA fixed point inverse
likelihood ratios satisfy $0 < \udLambdarightij < \infty$ and $0 <
\udLambdaleftij < \infty$, otherwise the assumption that we are dealing with
an interior point of $\setGamma{n}$ would be violated.

It follows from the message gauge invariance mentioned in
Remark~\ref{remark:gauge:invariance:1} that, for any positive real number $C$,
the inverse likelihoods $\bigl\{ C \cdot \udLambdaleftij \bigr\}_{i,j}$,
$\bigl\{ \frac{1}{C} \cdot \udLambdarightij \bigr\}_{i.j}$ also constitute a
fixed point of the SPA update rules. We will use this fact later on.

On the other hand, let $\bigl\{ \udLambdaleftijt \bigr\}_{i,j,t}$, $\bigl\{
\udLambdarightijt \bigr\}_{i,j,t}$ be a set of inverse likelihoods obtained by
running the SPA on $\graphN(\matrtheta)$ according to the SPA update rules
in~\eqref{eq:spa:update:repr:1}--\eqref{eq:spa:update:repr:2}. In the
following, we will not work with $\bigl\{ \udLambdaleftijt \bigr\}_{i,j,t}$,
$\bigl\{ \udLambdarightijt \bigr\}_{i,j,t}$ directly, but with $\bigl\{
\epsleftijt \bigr\}_{i,j,t}$, $\bigl\{ \epsrightijt \bigr\}_{i,j,t}$, which
are implicitly defined by the equations
\begin{align}
  \udLambdarightijt
    &= \udLambdarightij
       \cdot
       \left(
         1
         +
         \epsrightijt
       \right),
         \label{def:eps:right:def:1} \\
  \udLambdaleftijt
    &= \udLambdaleftij
       \cdot
       \left(
         1
         +
         \epsleftijt
       \right).
         \label{def:eps:left:def:1}
\end{align}
(Note that $-1 < \epsleftijt < \infty$ and $-1 < \epsrightijt < \infty$.)
Clearly, $\bigl\{ \epsleftijt \bigr\}_{i,j,t}$, $\bigl\{ \epsrightijt
\bigr\}_{i,j,t}$ can be considered to be a ``measure'' of the distance of the
SPA messages to the fixed-point messages. In particular, we have established
convergence of the SPA if we can show that these values converge to zero for
$t \to \infty$.

In a first step, we express the SPA message update rules in terms of $\bigl\{
\epsleftijt \bigr\}_{i,j,t}$ and $\bigl\{ \epsrightijt \bigr\}_{i,j,t}$.

\begin{Lemma}
  \label{lemma:spa:epsilon:message:update:1}

  For the right-going messages it holds that
  \begin{align}
    \deltarightijt
      &\defeq
         \frac{\sum_{j' \neq j}
                 \sqrt{\theta_{i,j'}}
                 \cdot
                 \udLambdaleftijp
                 \cdot
                 \epsleftijptmo
              }
              {\sum_{j' \neq j}
                 \sqrt{\theta_{i,j'}}
                 \cdot
                 \udLambdaleftijp
              },
                \label{eq:delta:right:def:1} \\
    \epsrightijt
      &= -
         \frac{\deltarightijt}
              {1+\deltarightijt}
           \label{eq:error:mapping:1}.
  \end{align}
  For the left-going messages it holds that
  \begin{align}
    \deltaleftijt
      &\defeq
         \frac{\sum_{i' \neq i}
                 \sqrt{\theta_{i',j}}
                 \cdot
                 \udLambdarightipjt
                 \cdot
                 \epsrightipjt
              }
              {\sum_{i' \neq i}
                 \sqrt{\theta_{i',j}}
                 \cdot
                 \udLambdarightipjt
              },
                \label{eq:delta:left:def:1} \\
    \epsleftijt
      &= -
         \frac{\deltaleftijt}
              {1+\deltaleftijt}.
           \label{eq:error:mapping:2}
  \end{align}
\end{Lemma}

\begin{Proof}
  Let us establish~\eqref{eq:error:mapping:1}. The expression
  in~\eqref{eq:error:mapping:2} then follows analogously. We compute
  \begin{align*}
    \udLambdarightij
    \cdot
    \left(
      1
      +
      \epsrightijt
    \right)
      &\onestareq
          \udLambdarightijt \\
      &\twostarseq
          \frac{\sqrt{\theta_{i,j}}}
              {\sum_{j' \neq j}
                 \sqrt{\theta_{i,j'}}
                 \cdot
                 \udLambdaleftijptmo} \\
      &\threestarseq
          \frac{\sqrt{\theta_{i,j}}}
              {\sum_{j' \neq j}
                 \sqrt{\theta_{i,j'}}
                 \cdot
                 \udLambdaleftijp
                 \cdot
                 \left(
                   1
                   +
                   \epsleftijptmo
                 \right)} \\
      &\fourstarseq
          \frac{\sqrt{\theta_{i,j}}}
              {\left(
                  \sum_{j' \neq j}
                    \sqrt{\theta_{i,j'}}
                    \cdot
                    \udLambdaleftijp
               \right)
               \cdot
               \left(
                 1
                 +
                 \deltarightijt
               \right)} \\
       &\fivestarseq
          \frac{\udLambdarightij}
               {1
                 +
                 \deltarightijt
               },
  \end{align*}
  where at step~$\onestar$ we have used~\eqref{def:eps:right:def:1}, where at
  step~$\twostars$ we have used~\eqref{eq:spa:update:repr:1}, where at
  step~$\threestars$ we have used~\eqref{def:eps:left:def:1}, where at
  step~$\fourstars$ we have used~\eqref{eq:delta:right:def:1}, and where at
  step~$\fivestars$ we have used~\eqref{eq:spa:fixed:point:1:1}. Dividing both
  sides by $\udLambdarightij$, and then subtracting~$1$ from both sides,
  yields the expression in~\eqref{eq:error:mapping:1}.
\end{Proof}

Note that $\deltarightijt$ is a weighted arithmetic average of the error
values $\bigl\{ \epsleftijptmo \bigr\}_{j' \neq j}$, and that $\deltaleftijt$
is a weighted arithmetic average of the error values $\bigl\{ \epsrightipjt
\bigr\}_{i' \neq i}$.

Note also that the expressions in~\eqref{eq:error:mapping:1}
and~\eqref{eq:error:mapping:2} have the following peculiarity. Namely, solving
$\varepsilon = - \delta / (1 + \delta)$ for $\delta$ we obtain $\delta = -
\varepsilon / (1 + \varepsilon)$, which is structurally the same expression as
the first expression but with the roles of $\varepsilon$ and $\delta$
interchanged.

\begin{Lemma}
  \label{lemma:one:step:characterization:1}

  Fix an iteration number $t \geq 1$. Taking advantage of the message gauge
  invariance that was mentioned in Remark~\ref{remark:gauge:invariance:1}, we
  can rescale the left-going and right-going fixed-point messages such that
  all $\{ \epsleftijtmo \}_{i,j}$ are non-negative. With this we define the
  numbers $\epsleftmaxtmo \geq 0$ and $\epsleftmaxt \geq 0$ to be the smallest
  numbers that satisfy
  \begin{alignat*}{2}
    \epsleftijtmo
       &\leq
         \epsleftmaxtmo,
         &\quad (i,j) \in \setE, \\
    \epsleftijt
       &\leq 
         \epsleftmaxt,
         &\quad (i,j) \in \setE.
  \end{alignat*}
  Then
  \begin{align*}
    0 &\leq
         \epsleftijt
       \leq
         \epsleftmaxt
       \leq
         \epsleftmaxtmo
         \quad (i,j) \in \setE.
  \end{align*}
\end{Lemma}

\begin{Proof}
  It follows immediately from~\eqref{eq:delta:right:def:1} that
  \begin{align*}
    0
      &\leq
         \deltarightijt
       \leq
         \epsleftmaxtmo,
           \quad (i,j) \in \setE,
  \end{align*}
  and so, because of~\eqref{eq:error:mapping:1}, we have
  \begin{align}
    -1
      &< -
         \frac{\epsleftmaxtmo}
              {1+\epsleftmaxtmo}
       \leq
         \epsrightijt
       \leq
         0,
         \quad (i,j) \in \setE.
           \label{eq:error:range:assumption:2}
  \end{align}
  Using~\eqref{eq:delta:left:def:1}, this implies
  \begin{align*}
    -1
      &< -
         \frac{\epsleftmaxtmo}
              {1+\epsleftmaxtmo}
       \leq
         \deltaleftijt
       \leq 
       0,
       \quad (i,j) \in \setE,
  \end{align*}
  and so, because of~\eqref{eq:error:mapping:2}, we have
  \begin{align}
    0 &\leq
         \epsleftijt
       \leq
         \epsleftmaxtmo,
         \quad (i,j) \in \setE.
           \label{eq:error:range:assumption:3}
  \end{align}
  This proves the statement in the lemma.
\end{Proof}

This shows that the errors stay bounded but it does not prove convergence
yet. (This result is essentially equivalent to the result that is obtained by
taking the zero-temperature limit of the contraction coefficient that is
computed in the SPA convergence analysis of~\cite{Bayati:Nair:06:1}: the
result is a contraction coefficient of $1$, which is non-trivial, but not good
enough to show that the message update map is a contraction.\footnote{Given
  the difference in the graphical model in~\cite{Bayati:Nair:06:1} and the
  graphical model considered here, some care is required when comparing the
  temperature that is mentioned here and the temperature that is mentioned in
  Sections~\ref{sec:factor:graph:representation:1}
  and~\ref{sec:bethe:entropy:1}.})

It turns out that in order to improve these bounds we have to track the error
values over two iteration, \ie, four half iterations. (We suspect that this is
related to the fact that the girth of $\graphN(\matrtheta)$, \ie, the length
of the shortest cycle of $\graphN(\matrtheta)$, is~$4$.)

\begin{Lemma}
  \label{lemma:two:step:characterization:1}

  Fix an iteration number $t \geq 1$. Taking advantage of the message gauge
  invariance that was mentioned in Remark~\ref{remark:gauge:invariance:1}, we
  can rescale the left-going and right-going fixed-point messages such that
  all $\{ \epsleftijtmo \}_{i,j}$ are non-negative and such that,
  additionally, $\min_{i,j} \epsleftijtmo = 0$. With this, we define the
  numbers $\epsleftmaxtmo \geq 0$ and $\epsleftmaxtpo \geq 0$ to be the
  smallest numbers that satisfy
  \begin{alignat*}{2}
    \epsleftijtmo
       &\leq
         \epsleftmaxtmo,
         &\quad (i,j) \in \setE, \\
    \epsleftijtpo
       &\leq 
         \epsleftmaxtpo,
         &\quad (i,j) \in \setE.
  \end{alignat*}
  Then
  \begin{align*}
    0 &\leq
         \epsleftijtpo
       \leq
         \epsleftmaxtpo
       \leq
         \nu'
         \cdot
         \epsleftmaxtmo
         \quad (i,j) \in \setE,
  \end{align*}
  for some constant $0 \leq \nu' < 1$ that depends only on $\matrtheta$ and
  the fixed-point messages $\{ \udLambdaleftij \}_{i,j}$ and $\{
  \udLambdarightij \}_{i,j}$, \ie, $\nu'$ is independent of $t$.
\end{Lemma}

\begin{Proof}
  The statement $\epsleftijtpo \geq 0$, $(i,j) \in \setE$ follows from
  applying Lemma~\ref{lemma:one:step:characterization:1} twice. Therefore, we
  can focus on the proof of $\epsleftmaxtpo \leq \nu' \cdot \epsleftmaxtmo$.

  For a given edge $(i,j) \in \setE$, we observe that $- \epsleftmaxtmo \big/
  \bigl( 1 + \epsleftmaxtmo \bigr) \leq \epsrightijt$
  in~\eqref{eq:error:range:assumption:2} holds with equality only if
  $\epsleftijptmo = \epsleftmaxtmo$ for all edges $(i,j')$ with $j' \neq
  j$. Similarly, for a given edge $(i,j) \in \setE$ we observe that
  $\epsleftijt \leq \epsleftmaxtmo$ in~\eqref{eq:error:range:assumption:3}
  holds with equality only if $\epsrightipjt = - \epsleftmaxtmo \big/ \bigl( 1
  + \epsleftmaxtmo \bigr)$ for all edges $(i',j)$ with $i' \neq i$. This
  motivates the definition of the following sets where we track the edges for
  which a strict inequality holds w.r.t.\ the inequalities just
  mentioned. Namely, for $t \geq 1$ we define
  \begin{align*}
    \setErighttime{t}
      &\defeq
         \left\{ 
           (i,j) \in \setE
         \ \middle| \ 
           \begin{array}{c}
             \text{there is at least one edge $(i,j')$,} \\ 
             \text{$j' \neq j$, such that $(i,j') \in \setElefttime{t-1}$}
           \end{array}
         \right\}, \\
    \setElefttime{t}
      &\defeq
         \left\{ 
           (i,j) \in \setE
         \ \middle| \ 
           \begin{array}{c}
             \text{there is at least one edge $(i',j)$,} \\ 
             \text{$i' \neq i$, such that $(i',j) \in \setErighttime{t}$}
           \end{array}
         \right\}.
  \end{align*}

  With this, assume that $\setElefttime{t-1}$ contains all the edges for which
  $\epsleftijtime{t-1} < \epsleftmaxtime{t-1}$. Clearly, $\setErighttime{t}$
  then contains all edges $(i,j)$ for which $\epsrightijtime{t} > -
  \epsleftmaxtime{t-1} \big/ \bigl( 1 + \epsleftmaxtime{t-1}
  \bigr)$. Similarly, $\setElefttime{t}$ contains all edges $(i,j)$ for which
  $\epsleftijtime{t} < \epsleftmaxtime{t-1}$.
  
  If $\epsleftmaxtmo = 0$ then the lemma is clearly true. So, assume that
  $\epsleftmaxtmo > 0$. Let $\setElefttime{t-1}$ contain all edges $(i,j)$ for
  which $\epsrightijtmo < \epsleftmaxtmo$. The assumptions in the lemma
  statement guarantee that there is at least one such edge, namely the edge(s)
  $(i,j)$ for which $\epsrightijtmo = 0$, and so the set $\setElefttime{t-1}$
  is non-empty. It can then be verified that four half-iterations later we
  have $\setElefttime{t+1} = \setE$.

  The fact that there is, as mentioned in the lemma statement, a constant
  $\nu'$ that is $t$-independent and strictly smaller than $1$ is then
  established by tracking the differences between the left- and the right-hand
  sides in the above-mentioned strict inequalities. This is done with the help
  of~\eqref{eq:delta:right:def:1} and~\eqref{eq:delta:left:def:1}.
\end{Proof}

The convergence proof is then completed by applying
Lemma~\ref{lemma:two:step:characterization:1} repeatedly. One detail needs to
be mentioned, though. Namely, if $\min_{i,j} \epsleftijtpo > 0$, and a
non-trivial re-gauging occurs at the beginning of the next application of
Lemma~\ref{lemma:two:step:characterization:1}, then in this re-gauging process
the value of $\max_{i,j} \epsleftijtpo > 0$ never increases (in fact, it
always decreases).

Finally, we have
\begin{align*}
  \left|\,
    \exp\!
    \bigg(\!\!\!
      -
      \FpdualBethe
        \Big(
          \bigl\{ \udLambdaleftijt \bigr\}, 
          \bigl\{ \udLambdarightijt \bigr\}
        \Big) \!\!
    \bigg)
    -
    \permBethe(\matrtheta)
  \right|
    &\leq
       C
       \cdot
       \e^{-\nu \cdot t}
\end{align*}
for suitable constants $C, \nu \in \Rpp$. This follows from, on the one hand,
the fact that when $\FBethe$ achieves its minimum in the interior of
$\setGamma{n}$ then we have equality between $\FBethe$ and $\FpdualBethe$ at
stationary points of the SPA~\cite{Yedidia:Freeman:Weiss:05:1}, and, on the
other hand, the above convergence analysis.

\section{Proof of Lemma~\ref{lemma:perm:Bethe:for:all:one:matrix:1}}
\label{sec:proof:lemma:perm:Bethe:for:all:one:matrix:1}

In a first step we evaluate $\perm(\matrallone{n})$. Namely, we obtain
\begin{align}
  \perm(\matrallone{n})
    &= n!
     \onestareq
       \sqrt{2\pi n}
       \cdot
       \left(
         \frac{n}{\e}
       \right)^{n}
       \cdot
       \big(
         1 + o(1)
       \big),
         \label{eq:perm:all:one:matrix:1}
\end{align}
where at step~$\onestar$ we have used Stirling's approximation of $n!$.

In a second step we evaluate $\permBethe(\matrallone{n})$. From
Definitions~\ref{def:Bethe:partition:function:1}
and~\ref{def:Bethe:permanent:1} it follows that
\begin{align*} 
  \permBethe(\matrallone{n})
    &\defeq
       \exp
         \left(
           -
           \min_{\vgamma}
             \FBethe(\vgamma)
         \right).
\end{align*}
From Corollary~\ref{cor:convexity:Bethe:free:energy:function:1} and symmetry
considerations it follows that the minimum in the above expression is achieved
by $\gamma_{i,j} = 1/n$, $(i,j) \in \setI \times \setJ$. Therefore,
\begin{align*}
  &
  \!\!\!\!\!
  \log
    \big(
      \permBethe(\matrallone{n})
    \big) \\
    &= \big.
         -
         \FBethe(\vgamma)
       \big|_{\gamma_{i,j} = 1/n, \ (i,j) \in \setI \times \setJ} \\
    &\onestareq
       \big.
         -
         \UBethe(\vgamma)
         +
         \HBethe(\vgamma)
       \big|_{\gamma_{i,j} = 1/n, \ (i,j) \in \setI \times \setJ} \\
    &\twostarseq
       -
       n^2 \cdot \frac{1}{n} \cdot \log\left( \frac{1}{n} \right)
       +
       n^2
         \cdot \left( 1 \!-\! \frac{1}{n} \right)
         \cdot \log\left( 1 \!-\! \frac{1}{n} \right) \\
    &= n \cdot \log(n)
       +
       n \cdot (n-1) \cdot \log\left( 1 - \frac{1}{n} \right) \\
    &= n \cdot \log(n)
       +
       n \cdot (n-1)
         \cdot
         \left(
           -
           \frac{1}{n}
           -
           \frac{1}{2n^2}
           +
           o
             \left(
               \frac{1}{n^2}
             \right)
         \right) \\
    &= n \cdot \log(n)
       -
       (n-1)
       -
       \frac{n-1}{2n}
       +
       o(1) \\
    &= n \cdot \log(n)
       -
       n
       +
       \frac{1}{2}
       +
       o(1),
\end{align*}
where at steps~$\onestar$ and $\twostars$ we have used
Corollary~\ref{cor:FBethe:as:function:of:gamma:1}. Consequently,
\begin{align}
  \permBethe(\matrallone{n})
    &= \sqrt{\e}
       \cdot
       \left(
         \frac{n}{e}
       \right)^n
       \cdot
       \big(
         1 + o(1)
       \big).
         \label{eq:Bethe:perm:all:one:matrix:1}
\end{align}

Combining~\eqref{eq:perm:all:one:matrix:1}
and~\eqref{eq:Bethe:perm:all:one:matrix:1} we obtain the promised result in
the lemma statement.

\section{Proof of Conjecture~\ref{conj:perm:matrix:lifiting:bounds:1}
               for $\matrtheta = \matrallone{n}$}
\label{sec:proof:cor:Bethe:permanent:all:one:matrix:bound:1}

Let $\matrtheta = \matrallone{n}$. In this appendix we prove that for any $M
\in \Zpp$ and any $\cmatrP \in \csetPsi_M$ it holds that
\begin{align}
  \perm
    \left(
      \matrtheta^{\uparrow\cmatrP}
    \right)
      &\leq
         \big(
           \perm(\matrtheta)
         \big)^M.
           \label{eq:graph:cover:permanent:inequality:1}
\end{align}
Although the proof is somewhat lengthy, the combinatorial idea behind it is
quite straightforward. Moreover, the only inequality that we use is the AM--GM
inequality, which says that the arithmetic mean of a list of non-negative real
numbers is at least as large as the geometric mean of this list of
numbers. Notably, there is no need to use Stirling's approximation of the
factorial function.

Towards showing~\eqref{eq:graph:cover:permanent:inequality:1}, let us fix some
positive integer $M$, fix some collection of permutation matrices $\cmatrP =
\bigl\{ \cmatrP^{(i,j)} \bigr\}_{i \in \setI, j \in \setJ} \in \csetPsi_M$,
define $\cmatrtheta \defeq \matrtheta^{\uparrow \cmatrP}$ as in
Definition~\ref{def:maththeta:lifting:1}, and let the row and column index
sets of $\matrtheta^{\uparrow \cmatrP}$ be $\setI \times [M]$ and $\setJ
\times [M]$, respectively. With this, it follows from
Definition~\ref{def:permanent:1} that
\begin{align}
  \perm(\matrtheta)
    &= \sum_{\sigma}
         \prod_{i \in \setI}
           \theta_{i,\sigma(i)},
           \label{eq:perm:theta:def:1} \\
  \perm(\cmatrtheta)
    &= \sum_{\csigma}
         \prod_{(i,m) \in \setI \times [M]}
           \cover{\theta}_{(i,m),\csigma((i,m))},
           \label{eq:perm:thetaP:def:1}
\end{align}
where $\sigma$ ranges over all permutations of the set $\setI$ and where
$\csigma$ ranges over all permutations of the set $\setI \times [M]$.

Note that, because all entries of $\cmatrtheta$ are either equal to zero or to
one, the products in~\eqref{eq:perm:thetaP:def:1} evaluate either to zero or
to one. Computing $\perm(\cmatrtheta)$ is therefore equivalent to counting the
$\csigma$'s for which these products evaluate to one. Equivalently,
$\perm(\cmatrtheta)$ equals the number of perfect matchings in the
NFG~$\graphN(\cmatrtheta)$.

\begin{Example}
  Some of the steps of the proof will be illustrated with the help of the NFGs
  in Fig.~\ref{fig:gc:graph:cover:interpretation:3:by:3:1} (which are
  reproduced in
  Fig.~\ref{fig:gc:graph:cover:interpretation:3:by:3:1:partition:1} for ease
  of reference), where $n = 3$ and $M = 4$.
  \begin{itemize}

  \item Fig.~\ref{fig:gc:graph:cover:interpretation:3:by:3:1:partition:1}(a)
    shows the NFG $\graphN(\matrtheta)$; $\perm(\matrtheta)$ equals the number
    of perfect matchings in
    Fig.~\ref{fig:gc:graph:cover:interpretation:3:by:3:1:partition:1}(a).
    Note: $\perm(\matrtheta) = n!$\hskip0.05cm.

  \item If $\cmatrP = \bigl\{ \cmatrP^{(i,j)} \bigr\}_{i \in \setI, j \in
      \setJ} = \bigl\{ \cmatrI \bigr\}_{i \in \setI, j \in \setJ}$, where
    $\cmatrI$ is the identity matrix of size $M \times M$, then we obtain the
    $M$-cover shown in
    Fig.~\ref{fig:gc:graph:cover:interpretation:3:by:3:1:partition:1}(b),
    which is a ``trivial'' $M$-cover of $\graphN(\matrtheta)$; $\perm\big(
    \matrtheta^{\uparrow \cmatrP} \big)$ equals the number of perfect
    matchings in
    Fig.~\ref{fig:gc:graph:cover:interpretation:3:by:3:1:partition:1}(b).
    Note: $\perm\big( \matrtheta^{\uparrow \cmatrP} \big) = \bigl( \perm(
    \matrtheta) \bigr)^M = (n!)^M$.
  
  \item For a ``non-trivial'' collection of permutation matrices $\cmatrP =
    \bigl\{ \cmatrP^{(i,j)} \bigr\}_{i \in \setI, j \in \setJ}$ we obtain an
    $M$-cover like in
    Fig.~\ref{fig:gc:graph:cover:interpretation:3:by:3:1:partition:1}(c);
    $\perm\big( \matrtheta^{\uparrow \cmatrP} \big)$ equals the number of
    perfect matchings in
    Fig.~\ref{fig:gc:graph:cover:interpretation:3:by:3:1:partition:1}(c).
    \exampleend
    
  \end{itemize}
\end{Example}

Let us therefore count the number of perfect matchings in
$\graphN(\cmatrtheta)$,
see Fig.~\ref{fig:gc:graph:cover:interpretation:3:by:3:1:partition:1}(c).
Before continuing, we define $\cdel((i,m))$, $(i,m) \in \setI \times [M]$, to
be the set of neighbors of the vertex $(i,m)$ in $\graphN(\cmatrtheta)$, \ie,
\begin{align*}
  \cdel((i,m))
    \defeq
      \left\{
        (j,m') \in \setJ \times [M]
      \ \middle| \ 
        \cover{P}^{(i,j)}_{m,m'} = 1
      \right\}.
\end{align*}
One can easily verify that for every $i \in \setI$, the sets $\cdel((i,m))$,
$m \in [M]$, form a partition of $\setJ \times [M]$. (See
Figs.~\ref{fig:gc:graph:cover:interpretation:3:by:3:1:partition:1}(b)--(c)
that highlight this partitioning for $i = 1$.) This observation will be the
crucial ingredient of the following steps.

We count the number of perfect matchings in $\graphN(\cmatrtheta)$ by
considering the vertices $\big\{ (i,m) \big\}_{m \in [M]}$ for $i = 1$, $i =
2$, up to $i = n$, thereby counting in how many ways we can specify $\csigma$
such that the product in~\eqref{eq:perm:thetaP:def:1} equals one.  Note that
because of the above partitioning observation, we can, conditioned on the
selection of a perfect matching up to and including step~$i-1$ (which we shall
symbolically denote by $\csigma_1^{i-1}$), consider the vertices $\big\{ (i,m)
\big\}_{m \in [M]}$ independently. Then we define
\begin{align*}
  \cover{d}_{i,m|\csigma_1^{i-1}}, \ (i,m) \in \setI \times [M],
\end{align*}
to be the number of possibilities of choosing $\csigma((i,m))$, \ie, the
number of ways that the edge of the perfect matching of $\graphN(\cmatrtheta)$
that is incident on $(i,m)$ can be chosen.
\begin{itemize}

\item Let $i = 1$. Then $\cover{d}_{i,m|\csigma_1^{i-1}}$, $m \in [M]$, is the
  number of possibilities of choosing the edge of the perfect matching of
  $\graphN(\cmatrtheta)$ that is incident on $(i,m)$. Because the $i$th row
  of $\matrtheta$ contains only ones, and because of the above partitioning
  observation, we find that $\cover{d}_{i,m} = n$ for all $m \in [M]$, and so,
  \begin{align*}
    \sum_{m \in [M]}
      \cover{d}_{i,m|\csigma_1^{i-1}}
      &= Mn.
  \end{align*}
  We observe that, whatever the selection of these $M$ edges is, $M$
  vertices on the right-hand side will be incident on a selected edge, and
  therefore be ``not available anymore'' in the following steps. This
  reduces the number of ``available'' right-hand side vertices to $Mn - M =
  M \cdot (n-1)$.

\item Let $i = 2$. Then $\cover{d}_{i,m|\csigma_1^{i-1}}$, $m \in [M]$, is the
  number of possibilities of choosing the edge of the perfect matching of
  $\graphN(\cmatrtheta)$ that is incident on $(i,m)$. Because the $i$th row
  of $\matrtheta$ contains only ones, because of the above partitioning
  observation, and because of the observation at the end of the above step, we
  find that
  \begin{align}
    \sum_{m \in [M]}
      \cover{d}_{i,m|\csigma_1^{i-1}}
      &\leq M \cdot (n-1).
         \label{eq:matrallones:Bethe:perm:2:1}
  \end{align}
  (If all permutation matrices in $\cmatrP$ are identity matrices, then it
  can be verified that the inequality
  in~\eqref{eq:matrallones:Bethe:perm:2:1} is an equality. However, for
  general $\cmatrP$, equality in~\eqref{eq:matrallones:Bethe:perm:2:1} does
  not need to hold.) Similar to the end of the above step, we observe that
  whatever the selection of these $M$ edges is, $M$ vertices on the
  right-hand side will be incident on a selected edge, and therefore be
  ``not available anymore'' in the following steps. This reduces the number
  of ``available'' right-hand side vertices to $M \cdot (n-1) - M = M \cdot
  (n-2)$.

\begin{figure}
  \begin{center}
    \begin{tabular}{ccc}
    \begin{minipage}[c]{0.25\linewidth}
      \begin{center}
        \subfigure[]
        {\epsfig{file=ffg_permanent_3by3_base_graph1_1.fig.eps,
                       scale=0.5}}
    \label{fig:gc:graph:cover:interpretation:3:by:3:base:partition:1}
      \end{center}
    \end{minipage}
    &
    \ 
    \begin{minipage}[c]{0.30\linewidth}
      \begin{center}
        \subfigure[]
        {\epsfig{file=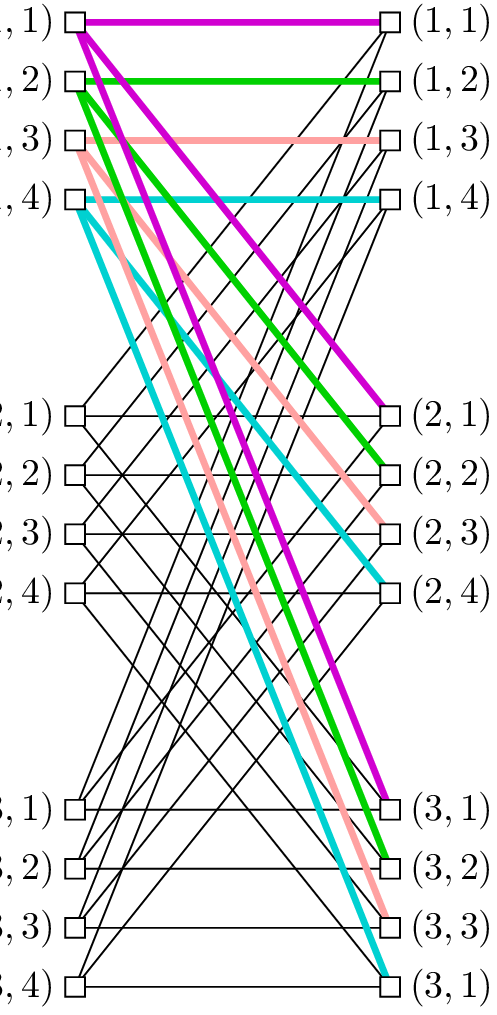,
                       scale=0.5}}
    \label{fig:gc:graph:cover:interpretation:3:by:3:fourfold:cover:1:partition:1}
      \end{center}
    \end{minipage}
    &
    \ 
    \begin{minipage}[c]{0.30\linewidth}
      \begin{center}
        \subfigure[]
        {\epsfig{file=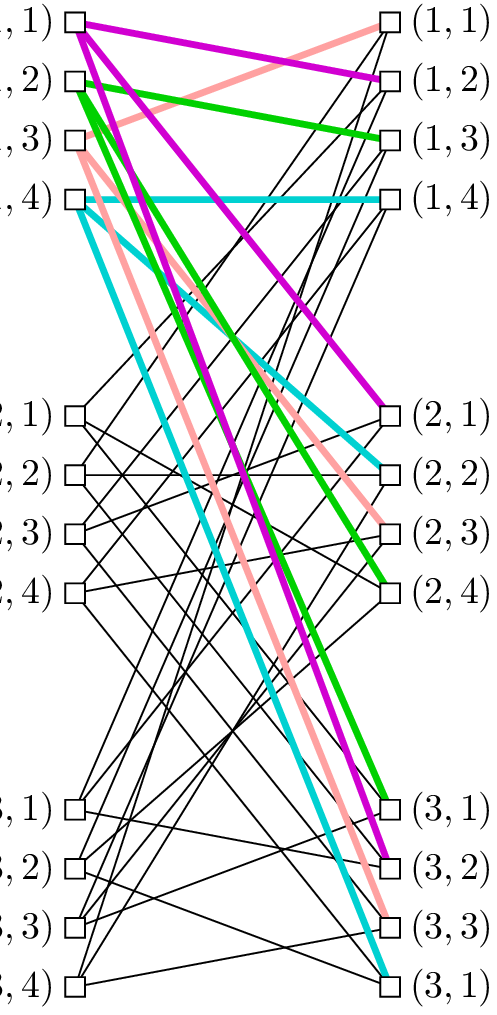,
                       scale=0.5}}
      \end{center}
    \end{minipage}
    \end{tabular}
  \end{center}
  \caption{(a) NFG $\graphN(\matrtheta)$ for $n = 3$. (b) ``Trivial''
    $4$-cover of $\graphN(\matrtheta)$ (c) A possible $4$-cover of
    $\graphN(\matrtheta)$.  The coloring of the edges in (b) and (c) show
    visually the fact that he sets $\cdel((i,m))$, $m \in [M]$, form a
    partition of $\setJ \times [M]$ (here for $i = 1$). (For more details, see
    the text in
    Appendix~\ref{sec:proof:cor:Bethe:permanent:all:one:matrix:bound:1}).}
  \label{fig:gc:graph:cover:interpretation:3:by:3:1:partition:1}
\end{figure}

\item Continuing as above, we observe that for general $i \in \setI$ it
  holds that
  \begin{align}
    \sum_{m \in [M]}
      \cover{d}_{i,m|\csigma_1^{i-1}}
      &\leq M \cdot (n-i+1).
         \label{eq:matrallones:Bethe:perm:i:1}
  \end{align}

\end{itemize}
Note that for $i \in \setI$ we have
\begin{align}
  \prod_{m \in [M]}
    \cover{d}_{i,m|\csigma_1^{i-1}}
    &= \left(
         \prod_{m \in [M]}
           \cover{d}_{i,m|\csigma_1^{i-1}}^{\ 1/M}
       \right)^M \nonumber \\
    &\onestarleq
       \left(
         \frac{1}{M}
         \sum_{m \in [M]}
           \cover{d}_{i,m|\csigma_1^{i-1}}
       \right)^M \nonumber \\
    &\twostarsleq
       \left(
         \frac{1}{M}
         \cdot
         M
         \cdot
         (n-i+1)
       \right)^M \nonumber \\
    &= (n-i+1)^M,
         \label{eq:matrallones:Bethe:perm:i:2}
\end{align}
where at step~$\onestar$ we have used the fact that the geometric mean of a
collection of non-negative numbers is upper bounded by the arithmetic mean of
the same collection of numbers, and where at step~$\twostars$ we have
used~\eqref{eq:matrallones:Bethe:perm:i:1}.

With this, we obtain the following upper bound on
$\perm(\cmatrtheta)$. Namely,
\begin{align}
  \perm(\cmatrtheta)
    &\onestareq
       \sum_{\csigma_1^1}
       \sum_{\csigma_1^2|\csigma_1^1}
       \cdots
       \sum_{\csigma_1^{n-1}|\csigma_1^{n-2}}
       \sum_{\csigma_1^n|\csigma_1^{n-1}}
         1 \nonumber \\
    &\twostarseq
       \sum_{\csigma_1^1}
       \sum_{\csigma_1^2|\csigma_1^1}
       \cdots
       \sum_{\csigma_1^{n-1}|\csigma_1^{n-2}}
       \prod_{m_n \in [M]}
         \cover{d}_{n,m_n|\csigma_1^{n-1}} \nonumber \\
    &\threestarsleq
       \sum_{\csigma_1^1}
       \sum_{\csigma_1^2|\csigma_1^1}
       \cdots
       \sum_{\csigma_1^{n-1}|\csigma_1^{n-2}}
         (n-n+1)^M \nonumber \\
    &\fourstarsleq
       (n-n+1)^M
       \cdot
       \sum_{\csigma_1^1}
       \sum_{\csigma_1^2|\csigma_1^1}
       \cdots
       \sum_{\csigma_1^{n-1}|\csigma_1^{n-2}}
         1 \nonumber \\
    &\ \, \vdots \nonumber \\
    &\fivestarsleq
       \prod_{i \in \setI}
         (n-i+1)^M \nonumber \\
    &= (n!)^M \nonumber \\
    &\sixstarseq
       \perm(\matrtheta)^M, \nonumber
\end{align}
where at step~$\onestar$ we have used the fact that $\perm(\cmatrtheta)$
equals the number of perfect matchings in $\graphN(\cmatrtheta)$, where at
step~$\twostars$ we have used the definition of
$\cover{d}_{n,m|\csigma_1^{n-1}}$, where at step~$\threestars$ we have
used~\eqref{eq:matrallones:Bethe:perm:i:2} for $i = n$, where at
step~$\fourstars$ we take advantage of the fact that $(n-n+1)^M$ is
independent of $\csigma_1^{n-1}$, where at step~$\fivestars$ we apply similar
results as at steps~$\twostars$--$\fourstars$ (note that for all $i$, the
quantity $(n-i+1)^M$ is independent of $\csigma_1^{i-1}$), and where at
step~$\sixstars$ we have used the observation $\perm(\matrtheta) = n!$. This
shows that the desired
inequality~\eqref{eq:graph:cover:permanent:inequality:1} indeed holds for
arbitrary positive integer $M$ and $\cmatrP \in \csetPsi_M$.

\section{Proof of Lemma~\ref{lemma:Bethe:permanent:of:cover:matrix:1}}
\label{sec:proof:lemma:Bethe:permanent:of:cover:matrix:1}

We first prove $\permBethe \bigl( \matrtheta^{\uparrow\cmatrP} \bigr) \geq
\bigl( \permBethe(\matrtheta) \bigr)^M$ and then $\permBethe \bigl(
\matrtheta^{\uparrow\cmatrP} \bigr) \leq \bigl( \permBethe(\matrtheta)
\bigr)^M$, from which the promised equality follows.

For the rest of the proof, we will use the short-hand $\cmatrtheta$ for
$\matrtheta^{\uparrow\cmatrP}$. We remind the reader of
Assumption~\ref{assumption:non:negative:matrix:theta:1}, \ie, we will assume
that there is at least one permutation $\sigma: [n] \to [n]$ such that
$\prod_i \theta_{i,\sigma(i)} > 0$ (otherwise, $\permBethe(\cmatrtheta) =
\permBethe(\matrtheta) = 0$). Moreover, $\graphN(\cmatrtheta)$ will be the NFG
associated with $\cmatrtheta$.\footnote{Let $\cgraph{N}$ be the $M$-cover of
  $\graphN(\matrtheta)$ corresponding to $\cmatrP$. Note that, strictly
  speaking, $\cgraph{N}$ and $\graphN(\cmatrtheta)$ are not the same NFG. The
  former is an $M$-cover of $\graphN(\matrtheta)$ (therefore it has two times
  $Mn$ function nodes, all of them with degree $n$), whereas the latter is a
  complete bipartite graph with two times $Mn$ function nodes. However, with
  the above condition on $\matrtheta$, for all practical purposes they are the
  same because $\FBethesub{\graphN(\cmatrtheta)}(\cmatrgamma) < \infty$ only
  for matrices $\cmatrgamma \in \setGamma{(Mn)}$ for which
  $\cgamma_{(i,m),(j,m')} = 0$ whenever $\cover{P}^{(i,j)}_{m,m'} = 0$,
  $(i,m,j,m') \in \setI \times [M] \times \setJ \times [M]$.}

Towards proving the first inequality, let $\matrgamma \in \setGamma{n}$ be a
matrix that minimizes $\FBethesub{\graphN(\matrtheta)}$. Based on
$\matrgamma$, we define the $(Mn) \times (Mn)$ matrix $\cmatrgamma$ with
entries
\begin{align*}
  \cgamma_{(i,m),(j,m')}
    &\defeq
       \gamma_{i,j}
       \cdot
       \cover{P}^{(i,j)}_{m,m'}
\end{align*}
for all $(i,m,j,m') \in \setI \times [M] \times \setJ \times [M]$. One can
easily verify that $\cmatrgamma \in \setGamma{(Mn)}$ and that
$\FBethesub{\graphN(\cmatrtheta)}(\cmatrgamma) = M \cdot
\FBethesub{\graphN(\matrtheta)}(\matrgamma)$. From this and
Corollary~\ref{cor:FBethe:as:function:of:gamma:1} it then follows that
\begin{align*}
  \permBethe(\cmatrtheta)
      &\geq \big(\!
           \permBethe(\matrtheta)
         \big)^M.
\end{align*}

Towards proving the second inequality, let $\cmatrgamma \in \setGamma{(Mn)}$
be a matrix that minimizes $\FBethesub{\graphN(\cmatrtheta)}$. One can easily
verify that $\cgamma_{(i,m),(j,m')} = 0$ whenever $\cover{P}^{(i,j)}_{m,m'} =
0$, $(i,m,j,m') \in \setI \times [M] \times \setJ \times [M]$. Based on
$\cmatrgamma$, we define the $n \times n$ matrix $\matrgamma$ with entries
\begin{align*}
  \gamma_{i,j}
    &\defeq
       \frac{1}{M}
       \sum_{m}
         \sum_{m'}
           \cgamma_{(i,m),(j,m')}
           \cdot
           \cover{P}^{(i,j)}_{m,m'}
\end{align*}
for all $(i,j) \in \setI \times \setJ$. One can easily verify that $\matrgamma
\in \setGamma{n}$. Let $\cmatrgamma_{(i,m)}$ be the length-$n$ vector based on
the $(i,m)$th row of $\cmatrgamma$, where we include an entry only if
$\cover{P}^{(i,j)}_{m,m'} = 1$. Similarly, define the length-$n$ vector
$\cmatrgamma_{(j,m')}$ based on the $(j,m')$th column of $\cmatrgamma$. One
can verify that the $i$th row of $\matrgamma$, \ie, $\matrgamma_i$, equals
$\frac{1}{M} \sum_m \cmatrgamma_{(i,m)}$. Similarly, the $j$th column of
$\matrgamma$, \ie, $\matrgamma_j$, equals $\frac{1}{M} \sum_{m'}
\cmatrgamma_{(j,m')}$. Then
\begin{align*}
  \HBethesub{\graphN(\cmatrtheta)}(\cmatrgamma)
    &\onestareq
       \frac{1}{2}
         \sum_i
           \sum_{m}
           S(\cmatrgamma_{(i,m)})
         +
         \frac{1}{2}
         \sum_j
           \sum_{m'}
           S(\cmatrgamma_{(j,m')}) \\
     &\twostarsleq
        \frac{M}{2}
         \sum_i
           S(\cmatrgamma_{i})
         +
         \frac{M}{2}
         \sum_j
           S(\cmatrgamma_{j}) \\
     &\threestarseq
        M
        \cdot
        \HBethesub{\graphN(\matrtheta)}(\matrgamma),
\end{align*}
where at step~$\onestar$ we have used
Lemma~\ref{lemma:Bethe:entropy:function:in:terms:of:S:1}, where at
step~$\twostars$ we have used the concavity of the $S$-function
(see Theorem~\ref{theorem:md:ess:function:properties:1}), and where at
step~$\threestars$ we have used once again
Lemma~\ref{lemma:Bethe:entropy:function:in:terms:of:S:1}. Moreover, one can
easily show that $\UBethesub{\graphN(\cmatrtheta)}(\cmatrgamma) = M \cdot
\UBethesub{\graphN(\matrtheta)}(\matrgamma)$, and so
$\FBethesub{\graphN(\cmatrtheta)}(\cmatrgamma) \geq M \cdot
\FBethesub{\graphN(\matrtheta)}(\matrgamma)$. From this and
Corollary~\ref{cor:FBethe:as:function:of:gamma:1} it then follows that
\begin{align*}
  \permBethe(\cmatrtheta)
    &\leq
       \big(\!
         \permBethe(\matrtheta)
       \big)^M.
\end{align*}

\section{Proof of Lemma~\ref{lemma:frac:perm:Bethe:for:all:one:matrix:1}}
\label{sec:proof:lemma:frac:perm:Bethe:for:all:one:matrix:1}

Because $\vkappa$ satisfies the conditions listed in
Theorem~\ref{theorem:concavity:fractional:Bethe:entropy:1}, the concavity
statement for the Bethe entropy function and the convexity statement for the
Bethe free energy function follow immediately.

Therefore, let us turn our attention to evaluating the ratio
$\perm(\matrallone{n}) / \fracpermBethe{\vkappa}(\matrallone{n})$. In a first
step we evaluate $\perm(\matrallone{n})$. Namely, as in the proof of
Lemma~\ref{lemma:perm:Bethe:for:all:one:matrix:1} in
Appendix~\ref{sec:proof:lemma:perm:Bethe:for:all:one:matrix:1} we have
\begin{align}
  \perm(\matrallone{n})
    &= n!
     = \sqrt{2\pi n}
       \cdot
       \left(
         \frac{n}{\e}
       \right)^{n}
       \cdot
       \big(
         1 + o(1)
       \big).
         \label{eq:perm:all:one:matrix:1:2}
\end{align}

In a second step we evaluate $\fracpermBethe{\vkappa}(\matrallone{n})$. From
Theorem~\ref{theorem:concavity:fractional:Bethe:entropy:1} and symmetry
considerations it follows that the minimum in the above expression is achieved
by $\gamma_{i,j} = 1/n$, $(i,j) \in \setI \times \setJ$. Therefore,
\begin{align*}
  &
  \!\!\!\!\!
  \log
    \big(
      \permBethe(\matrallone{n})
    \big) \\
    &\onestareq
       -
       \UBethe(\vgamma)
       +
       \fracHBethe{\vkappa}(\vgamma) \\
    &\twostarseq
       - \,
       n^2
       \cdot
       \left(
         1
         +
         \frac{1}{2n}
       \right)
       \cdot
       \frac{1}{n} \cdot \log\left( \frac{1}{n} \right) \\
     &\quad\ 
       +
       n^2
       \cdot
       \left(
         1
         -
         \frac{1}{2n}
       \right)
         \cdot \left( 1 \!-\! \frac{1}{n} \right)
         \cdot \log\left( 1 \!-\! \frac{1}{n} \right) \\
    &= \left(
         n
         +
         \frac{1}{2}
       \right)
       \cdot \log(n)
       +
       \left(
         n
         -
         \frac{1}{2}
       \right)
       \cdot (n-1) \cdot \log\left( 1 - \frac{1}{n} \right) \\
    &= \left(
         n
         +
         \frac{1}{2}
       \right)
       \cdot \log(n) \\
    &\quad\,
       +
       \left(
         n
         -
         \frac{1}{2}
       \right)
         \cdot
         (n-1)
         \cdot
         \left(
           -
           \frac{1}{n}
           -
           \frac{1}{2n^2}
           +
           o
             \left(
               \frac{1}{n^2}
             \right)
         \right) \\
    &= \left(
         n
         +
         \frac{1}{2}
       \right)
       \cdot \log(n)
       -
       n
       +
       1
       +
       o(1),
\end{align*}
where at step~$\onestar$ we have used $\fracFBethe{\vkappa}(\vgamma) =
\UBethe(\vgamma) - \fracHBethe{\vkappa}(\vgamma)$, where at $\twostars$ we
have used $\UBethe(\vgamma) = - \sum_{i,j} \gamma_{i,j} \log(\theta_{i,j}) =
0$ and the expression for $\fracHBethe{\vkappa}(\vgamma)$ from
Lemma~\ref{lemma:fractional:Bethe:entropy:function:reexpressed:1}. Therefore,
\begin{align}
  \fracpermBethe{\vkappa}(\matrallone{n})
    &= \e
       \cdot
       \sqrt{n}
       \cdot
       \left(
         \frac{n}{e}
       \right)^n
       \cdot
       \big(
         1 + o(1)
       \big).
         \label{eq:frac:Bethe:perm:all:one:matrix:1}
\end{align}
By combining~\eqref{eq:perm:all:one:matrix:1:2}
and~\eqref{eq:frac:Bethe:perm:all:one:matrix:1} we obtain the promised result.

\bibliographystyle{IEEEtran}
\bibliography{/home/vontobel/references/references}

\end{document}